\documentclass[reprint,nofootinbib,aps,prd]{revtex4-2}

\usepackage{amssymb}
\usepackage{amsmath}
\usepackage{times}
\usepackage{bm}
\usepackage[colorlinks]{hyperref}
\usepackage{graphicx}
\usepackage{dcolumn}
\usepackage{hhline}
\usepackage[american]{babel}
\usepackage{booktabs}
\usepackage{dcolumn}
\usepackage[T1]{fontenc}
\usepackage[utf8]{inputenc}
\usepackage{multirow}
\usepackage{url}
\usepackage[dvipsnames]{xcolor}
\usepackage{float}
\usepackage{resizegather}
\usepackage{appendix}

\newcolumntype{d}[1]{D{.}{.}{#1}}
\newcolumntype{v}[1]{D{,}{,\ }{#1}}

\begin{document}

\title{Second-order corrections to Starobinsky inflation}

\author{G. Rodrigues-da-Silva}
\email{gesiel.neto.090@ufrn.edu.br}
\affiliation{Departamento de Física, Universidade Federal do Rio Grande do Norte,\\
	Campus Universitário, s/n - Lagoa Nova, CEP 59072-970, Natal, Rio Grande do Norte, Brazil}

\author{L. G. Medeiros}
\email{leo.medeiros@ufrn.br}
\affiliation{Escola de Ciências e Tecnologia, 	Universidade Federal do Rio Grande do Norte,\\
	Campus Universitário, s/n - Lagoa Nova, CEP 59072-970, Natal, Rio Grande do Norte, Brazil}

\date{\today }

\begin{abstract}
	Higher-order theories of gravity are extensions to general relativity (GR)
	motivated mainly by high-energy physics searching for GR ultraviolet
	completeness. They are characterized by the inclusion of correction terms in
	the Einstein-Hilbert action that leads to higher-order field equations. In
	this paper, we propose investigating inflation due to the GR extension built
	with all correction terms up to the second-order involving only the scalar
	curvature $R$, namely, $R^{2}$, $R^{3}$, $R\square R$. We investigate
	inflation within the Friedmann cosmological background, where we study the
	phase space of the model, as well as explore inflation in slow-roll
	leading-order. Furthermore, we describe the evolution of scalar
	perturbations and properly establish the curvature perturbation. Finally, we
	confront the proposed model with recent observations from Planck,
	BICEP3/Keck, and BAO data.
\end{abstract}

\maketitle


\section{Introduction} \label{Introduction}

Despite the immense predictive power of general relativity (GR), extensions
to it have been motivated by several areas. In high-energy physics, which
aims for the ultraviolet completeness of GR, quantum gravity and inflation
models are included. On the other hand, models involving low-energy physics
include, among others, the phenomenology of the dark sector of the universe
and spherically symmetric solutions in a weak-field regime.

According to Lovelock's theorem, fundamentally, GR is constructed based on
some hypotheses: it is a $4$-dimensional Riemannian metric gravity theory,
containing the metric $g_{\mu \nu }$ as the only fundamental field,
invariant by diffeomorphism and with second-order field equations. In this
sense, extensions to GR are achieved by violating any of these hypotheses
\cite{Clifton:2011jh}. By violating the first hypothesis, we can allow a
higher-dimensional spacetime or even consider a gravitational action
constructed with curvature and torsion invariants due to a Riemann-Cartan
spacetime \cite{RevModPhys.48.393}. If we violate the hypothesis that the
theory of gravity has the metric as the only fundamental field, we can
obtain, for example, the Horndeski theories. These, in turn, are the more
general $4$-dimensional theories of gravity whose action, constructed with
the metric and a scalar field, leads to second-order field equations \cite%
{Horndeski:1974wa}. On the other hand, by allowing field equations above
the second-order and preserving all other assumptions, we find the
higher-order gravities.

Higher-order theories of gravity are characterized by the inclusion of
correction terms in the Einstein-Hilbert (EH) action that lead to
higher-order field equations. Such corrections can be conveniently
classified according to their mass (energy) scale. In this scenario, EH plus
the cosmological constant represents the usual zero-order term. First-order
corrections to EH are fourth mass terms constructed from the $4$ possible
invariants\footnote{%
	Although we present all of the following terms, not all are relevant to
	field equations. The $\square R$ is a surface term, and it does not
	contribute. Also, due to the Gauss-Bonnet invariant $G=R_{\mu \nu \alpha
		\beta }R^{\mu \nu \alpha \beta }-4R_{\mu \nu }R^{\mu \nu }+R^{2}$, the
	contraction of the Riemann tensor can be written in terms of the other two
	terms.}%
\begin{equation*}
R^{2}\text{, }R_{\mu \nu }R^{\mu \nu }\text{, }R_{\mu \nu \alpha \beta
}R^{\mu \nu \alpha \beta }\text{ \ and }\square R.
\end{equation*}%
In turn, the second-order corrections to EH are sixth mass terms, built with
the invariants\footnote{%
	In this case, since the number of terms grows vastly, we present only those
	that contribute to the field equations.}
\begin{gather*}
R\square R\text{, }R_{\mu \nu }\square R^{\mu \nu }\text{,} \\
R^{3}\text{, }RR_{\mu \nu }R^{\mu \nu }\text{, }R_{\mu \nu }R_{\text{ \ }%
	\alpha }^{\nu }R^{\alpha \mu }\text{,} \\
RR_{\mu \nu \alpha \beta }R^{\mu \nu \alpha \beta }\text{, }R_{\mu \alpha
}R_{\nu \beta }R^{\mu \nu \alpha \beta }\text{ \ and }R_{\mu \nu \alpha
	\beta }R_{\text{ \ \ }\kappa \rho }^{\alpha \beta }R^{\kappa \rho \mu \nu }.
\end{gather*}%
And so on, we will have more higher-order correction terms as we increase
the energy scales.

Models involving higher-order gravities have been explored in various
contexts. There are papers in the literature whose purpose is to show the
equivalence between different classes of gravity theories, in particular,
between $f\left( R\right) $ or $f\left( R,\square ^{k}R\right) $ and
scalar-tensor theories \cite{Gottlober:1989ww,RevModPhys.82.451,Nojiri:2010wj,DeFelice:2010aj,Capozziello:2011et,Cuzinatto:2016ehv,Nojiri:2017ncd}. In some contexts, it becomes
more convenient to pass from the original frame to the Jordan or Einstein
frames, through a conformal transformation, in order to handle equations for
scalar fields rather than higher-order equations for the metric. Another
topic of great interest is the investigation of spherically symmetric and
static solutions in higher-order gravities, with Stelle's paper \cite%
{Stelle:1977ry} being one of those responsible for shedding light on this
line of research. In particular, the study of the possibility of
non-Schwarzschild black hole solutions through different approaches is
addressed in Refs. \cite{Nelson:2010ig,PhysRevD.92.124019,Lu:2015cqa,Kokkotas:2017zwt,Bueno:2017sui,Goldstein:2017rxn,Podolsky:2018pfe,Rodrigues-da-Silva:2020cpd},
whereas researches involving weak-field regime solutions are covered in
Refs. \cite{Accioly:2016qeb,GIACCHINI2017306,Giacchini:2018gxp,Rodrigues-da-Silva:2020cpd}. There are also models that study the
generation and properties of gravitational waves \cite{Berry:2011pb,PhysRevD.96.064044,Zinhailo:2018ska,Holscher:2018jhm,Datta:2019npq,Kim:2019sqk,Yamada:2019zrb,Gogoi:2020ypn,Faria:2020kbv,Ezquiaga:2021ler,PhysRevD.104.084061}. The
latter is a topic of great current appeal due to the direct detections \cite%
{Abbott:2016blz,TheLIGOScientific:2016src,PhysRevLett.119.161101} that allow the rising of gravitational wave astrophysics.

Regarding inflation, it is well known in the literature that the Starobinsky
model \cite{doi.org/10.1016/0370-2693(80)90670-X,Starobinsky:1983zz} has a good fit for recent
observational data from Planck, BICEP3/Keck and BAO \cite{Akrami:2018odb,BICEPKeck:2021gln}. Furthermore, the fact that it has a well-grounded
theoretical motivation makes it one of the strongest inflationary
candidates, despite the immense plethora of inflation models \cite%
{MARTIN201475}. Such reasons motivate the investigation of models based on
extensions to the Starobinsky model via higher-order gravity theories. There
is a large amount of research in this context. Some of them are based on $%
f\left( R\right) $ theories, as in Refs. \cite{Huang:2013hsb,Cheong:2020rao,PhysRevD.105.063504,Sebastiani:2013eqa,Odintsov:2017fnc,Ivanov:2021chn}, others consider the introduction
of Weyl's term \cite{Salvio:2017xul,Salvio:2019wcp,Anselmi:2020lpp,Anselmi:2021dag,Anselmi:2021rye}. There are
those that consider local gravitational actions involving a finite number of
curvature derivative terms \cite{Berkin:1990nu,Asorey:1996hz,Iihoshi:2010pf,Modesto:2016ofr,Cuzinatto:2018chu,Cuzinatto:2018vjt,Castellanos:2018dub}, while others are nonlocal, involving infinite
derivatives \cite{Koshelev:2016xqb,Edholm:2016seu,Diamandis:2017ems,SravanKumar:2018dlo,Bezerra-Sobrinho:2022dkv}.

In this paper, we propose to investigate the
extension to the Starobinsky model due to the inclusion of all correction
terms up to the second-order involving only the scalar curvature $R$. In
this sense, we have the following gravitational action

\begin{equation}
S=\frac{M_{Pl}^{2}}{2}\int d^{4}x\sqrt{-g}\left( R+\frac{1}{2\kappa _{0}}%
R^{2}+\frac{\alpha _{0}}{3\kappa _{0}^{2}}R^{3}-\frac{\beta _{0}}{2\kappa
	_{0}^{2}}R\square R\right) ,  \label{Acao do Trabalho}
\end{equation}%
where $\kappa _{0}$ has squared mass unit and parameters $\alpha _{0}$ and $%
\beta _{0}$ are dimensionless quantities. Furthermore, $M_{Pl}$ is the
reduced Planck mass, such that $M_{Pl}^{2}\equiv \left( 8\pi G\right) ^{-1}$
and $\square \equiv \nabla _{\sigma }\nabla ^{\sigma }$ represents the
covariant d'Alembertian operator. In this scenario, where we only address
the scalar sector of corrections, $R^{2}$ represents the first-order
correction, while the last two terms correspond to the second-order
corrections to EH. Note that the parameter $\kappa _{0}$ is responsible for
establishing the energy scale of inflation, while the parameters $\alpha _{0}
$ and $\beta _{0}$ give us a measure of the Starobinsky deviation. Since $R^{3}$ and $R\square R$ are
both second-order correction terms on energy scales, they must contribute
similarly to inflation, so there is a joint effect that must be considered.
In that regard, it is
worth noting that our paper goes a step further in recent researches
developed in \cite{PhysRevD.105.063504} and \cite{Cuzinatto:2018vjt,Castellanos:2018dub}, which address the models Starobinsky$+R^{3}$ and
Starobinsky$+R\square R$, respectively. In this paper, the multi-field treatment associated with the $R\square
R$ term is different from that used in Ref. \cite{Cuzinatto:2018vjt}. While in that paper,
inflation is described by a scalar and a vector field, here, inflation is
driven through the dynamics of two scalar fields. Furthermore, by properly
constructing the curvature perturbation, we can obtain observational
constraints different from those obtained in Ref. \cite{Cuzinatto:2018vjt} for
the tensor-to-scalar ratio. In turn, by assuming the $R\square
R$ sixth-derivative term as a small perturbation to Starobinsky inflation, Ref. \cite{Castellanos:2018dub} uses a somewhat different approach, being able to map the model into a \textit{one-scalar theory}.

It is important to comment that the discussed model (\ref{Acao do Trabalho})
is not seen as a fundamental theory of gravity. On the other hand, it is
seen as a classical model of gravity in a context of effective theory. One could legitimately worry about the ghost-type instabilities introduced with the $R\square R$ sixth-derivative term\footnote{%
	This occurs for $\beta _{0}>0$.}. Nevertheles, as previously pointed out by Refs. \cite{Salles:2014rua,Peter:2017xxf}, the complications of the growing up explosive behaviour of the ghost-type perturbations will not take place only until the initial seeds of such perturbations do not have sufficiently high frequencies. Usually, as long as the energy scales involved are close to the Planck order of magnitude, cosmological solutions are stable.

The paper is structured as follows. In section \ref{sec:eqs-campo}, we start
from the original frame for the action (\ref{Acao do Trabalho}) and rewrite
it in the scalar-tensor representation in the Einstein frame, where the
theory is described through a metric and two auxiliary scalar fields, only
one of which is associated with a canonical kinetic term. Then we write the
field equations for each of the fields. Section \ref{sec:inflacao-background}
is responsible for making the full description of inflation in the
cosmological background. In section \ref{sec - EspFase}, we study the
critical points and the $4$-dimensional phase space of the model. Next, we
explore inflation in the slow-roll leading order regime by defining the
slow-roll factor, and thus, we obtain the slow-roll parameters and the
number of $e$-folds. In section \ref{sec:eqs-perturbadas}, we give a
complete description of the evolution of scalar perturbations. In addition
to writing the perturbed field equations in the slow-roll leading order
regime, we define the adiabatic and isocurvature perturbations by separating
of the background phase space trajectories in the tangent (adiabatic
perturbation) and orthogonal (isocurvature perturbation) directions. This
allows us to properly establish the curvature perturbation, which is
essential to connect our model with the observations. In section \ref%
{sec:vinculos-observacionais}, we confront the proposed model with the
recent observations of Ref. \cite{BICEPKeck:2021gln}, where by using the
constraint for the number of inflation $e$-folds found in \cite%
{PhysRevD.105.063504}, we build the usual $n_{s}\times r_{0.002}$ plane and
the Plot for the parameter space $\alpha _{0}\times \beta _{0}$. In section %
\ref{sec:comentarios-finais}, we make some final comments.

\section{Field Equations \label{sec:eqs-campo}}

The first step is to rewrite the action (\ref{Acao do Trabalho}%
) in the Einstein frame. Performing this calculation, we get

\begin{align}
	\bar{S}& =\frac{M_{Pl}^{2}}{2}\int d^{4}x\sqrt{-\bar{g}}\left[ \bar{R}%
	-3\left( \frac{1}{2}\bar{\nabla}_{\rho }\chi \bar{\nabla}^{\rho }\chi
	+\right. \right.  \notag \\
	& \left. \left. -\frac{\beta _{0}}{6}e^{-\chi }\bar{\nabla}_{\rho }\lambda 
	\bar{\nabla}^{\rho }\lambda +V\left( \chi ,\lambda \right) \right) \right] ,
	\label{Acao Frame de Einstein Adim}
\end{align}
with%
\begin{equation}
V\left( \chi ,\lambda \right) =\frac{\kappa _{0}}{3}e^{-2\chi }\lambda
\left( e^{\chi }-1-\frac{1}{2}\lambda -\frac{\alpha _{0}}{3}\lambda
^{2}\right) ,  \label{eq:potencial-modelo}
\end{equation}%
the potential associated with the model. The quantities with bar are defined
from the metric as $\bar{g}_{\mu \nu }=e^{\chi }g_{\mu \nu }$ and the
dimensionless fields $\chi $ and $\lambda $ are defined as%
\begin{equation}
\lambda =\frac{R}{\kappa _{0}}\text{ \ and \ }\mu =e^{\chi }=1+\lambda
+\alpha _{0}\lambda ^{2}-\frac{\beta _{0}}{\kappa _{0}}\square \lambda \text{%
	,}
\end{equation}
where in the Einstein frame,
$\square\lambda=e^{\chi}\left(  \bar{\square}\lambda-\bar{\partial}^{\mu
}\lambda\bar{\partial}_{\mu}\chi\right)  $.

\bigskip

\subparagraph{Addendum}

By recovering the usual notation and the dimensions of the scalar fields,
and the potential, we must take%
\begin{gather}
	\chi =\sqrt{\frac{2}{3}}\frac{\phi }{M_{Pl}},\text{ \ }\lambda =\sqrt{2}%
	\frac{\psi }{M_{Pl}},\text{ \ and}  \notag \\
	\tilde{V}\left( \phi ,\psi \right) =\frac{3M_{Pl}^{2}}{2}V\left( \chi
	,\lambda \right) .  \label{eq:campos-massivos}
\end{gather}%
This way, we can rewrite the action (\ref{Acao Frame de Einstein Adim}) as%
\begin{align}
	\bar{S}& =\int d^{4}x\sqrt{-\bar{g}}\left( \frac{M_{Pl}^{2}}{2}\bar{R}-\frac{%
		1}{2}\bar{\nabla}_{\rho }\phi \bar{\nabla}^{\rho }\phi +\right.  \notag \\
	& \left. +\frac{\beta _{0}e^{-\sqrt{\frac{2}{3}}\frac{\phi }{M_{Pl}}}}{2}%
	\bar{\nabla}_{\rho }\psi \bar{\nabla}^{\rho }\psi -\tilde{V}\left( \phi
	,\psi \right) \right) .  \label{eq:acao-frame-einstein-dim}
\end{align}

\bigskip

By starting from the action (\ref{Acao Frame de Einstein Adim}), we obtain
three field equations: one for $\bar{g}_{\mu \nu }$ and another two for each
of the scalar fields $\chi $ and $\lambda $. Taking the variation concerning
the metric $\bar{g}_{\mu \nu }$, we find%
\begin{equation}
\bar{R}_{\mu \nu }-\frac{1}{2}\bar{g}_{\mu \nu }\bar{R}=\frac{1}{M_{Pl}^{2}}%
\bar{T}_{\mu \nu }^{\left( \text{eff}\right) },  \label{eq:campo-metrica}
\end{equation}%
where we define an effective energy-momentum tensor as%
\begin{gather}
	\frac{1}{M_{Pl}^{2}}\bar{T}_{\mu \nu }^{\left( \text{eff}\right) }=\frac{3}{2%
	}\left( \bar{\nabla}_{\mu }\chi \bar{\nabla}_{\nu }\chi -\frac{1}{2}\bar{g}%
	_{\mu \nu }\bar{\nabla}^{\rho }\chi \bar{\nabla}_{\rho }\chi \right) + 
	\notag \\
	-\frac{\beta _{0}e^{-\chi }}{2}\left( \bar{\nabla}_{\mu }\lambda \bar{\nabla}%
	_{\nu }\lambda -\frac{1}{2}\bar{g}_{\mu \nu }\bar{\nabla}^{\rho }\lambda 
	\bar{\nabla}_{\rho }\lambda \right) -\frac{3}{2}\bar{g}_{\mu \nu }V\left(
	\chi ,\lambda \right) .  \label{T_munu efetivo}
\end{gather}%
The variation concerning the $\chi $ and $\lambda $ fields results in%
\begin{gather}
\bar{\square}\chi -\frac{\beta _{0}}{6}e^{-\chi }\bar{\nabla}_{\rho }\lambda
\bar{\nabla}^{\rho }\lambda -V_{\chi }=0,  \label{eq:eq-campo-chi} \\
\beta _{0}e^{-\chi }\left( \bar{\nabla}^{\rho }\chi \bar{\nabla}_{\rho
}\lambda -\bar{\square}\lambda \right) -3V_{\lambda }=0.
\label{eq:eq-campo-lambda}
\end{gather}%
where $V_{\chi }=\partial _{\chi }V$ and $V_{\lambda }=\partial _{\lambda }V$
represent derivatives concerning the fields $\chi $ and $\lambda $,
respectively.

\section{Inflation in Friedmann cosmological background \label{sec:inflacao-background}}

On large scales ($\gtrsim 100$ Mpc), we can consider the universe to be
homogeneous and isotropic. Furthermore, for a spatially flat universe, the
line element that describes the evolution of a comoving frame of reference
is given by%
\begin{equation}
ds^{2}=-dt^{2}+a^{2}\left( t\right) \left( dx^{2}+dy^{2}+dz^{2}\right) ,
\label{eq:metrica-friedmann}
\end{equation}%
where $a\left( t\right) $ is the scale factor.

By obtaining the field equations in Friedmann background is to write the
field equations (\ref{eq:campo-metrica}), (\ref{eq:eq-campo-chi}) and (\ref%
{eq:eq-campo-lambda}) for the metric (\ref{eq:metrica-friedmann}). From the
field equation for the metric, we get two independent ones, namely the
Friedmann equations
\begin{gather}
H^{2}=\frac{1}{2}\left( \frac{1}{2}\dot{\chi}^{2}-\frac{\beta _{0}}{6}%
e^{-\chi }\dot{\lambda}^{2}+V\left( \chi ,\lambda \right) \right) ,
\label{eq:H} \\
\dot{H}=-\frac{3}{4}\dot{\chi}^{2}+\frac{1}{4}\beta _{0}e^{-\chi }\dot{%
	\lambda}^{2},  \label{eq:H-dot}
\end{gather}%
where $H=\dot{a}/a$. In addition to these equations, we also have the
equations for the $\chi $ and $\lambda $ fields. Since, for a scalar field $%
\Phi $,%
\begin{equation*}
\bar{\square}\Phi =\bar{\nabla}_{\sigma }\bar{\nabla}^{\sigma }\Phi =-3H\dot{%
	\Phi}-\ddot{\Phi},
\end{equation*}%
for the equation of $\chi $ given in (\ref{eq:eq-campo-chi}), we have%
\begin{equation}
\ddot{\chi}+3H\dot{\chi}-\frac{\beta _{0}}{6}e^{-\chi }\dot{\lambda}%
^{2}+V_{\chi }=0.  \label{eq:Chi}
\end{equation}%
In turn, for the equation of $\lambda $ given in (\ref{eq:eq-campo-lambda}),
we have%
\begin{equation}
\beta _{0}e^{-\chi }\left[ \ddot{\lambda}-\left( \dot{\chi}-3H\right) \dot{%
	\lambda}\right] -3V_{\lambda }=0.  \label{eq:Lambda}
\end{equation}

\subsection{Phase space \label{sec - EspFase}}

In this section, we will analyze the phase space of the model. Therefore, it
becomes convenient to rewrite the field equations in a dimensionless way: we
define the dimensionless time derivative%
\begin{equation*}
A_{t}\equiv \frac{1}{\sqrt{\kappa _{0}}}\dot{A},
\end{equation*}%
the dimensionless Hubble parameter $h$%
\begin{equation*}
h\equiv \frac{1}{\sqrt{\kappa _{0}}}H,
\end{equation*}%
and the dimensionless potential $\bar{V}$ as%
\begin{equation*}
\bar{V}\left( \chi ,\lambda \right) =\frac{1}{\kappa _{0}}V\left( \chi
,\lambda \right) .
\end{equation*}%
With that, it is possible to rewrite the equations of cosmological dynamics (%
\ref{eq:H}), (\ref{eq:H-dot}), (\ref{eq:Chi}) and (\ref{eq:Lambda}) as
follows:%
\begin{gather}
h^{2}=\frac{1}{2}\left( \frac{1}{2}\chi _{t}^{2}-\frac{\beta _{0}}{6}%
e^{-\chi }\lambda _{t}^{2}+\bar{V}\left( \chi ,\lambda \right) \right) ,
\label{eq:H adm} \\
h_{t}=-\frac{3}{4}\chi _{t}^{2}+\frac{1}{4}\beta _{0}e^{-\chi }\lambda
_{t}^{2},  \label{eq:H-dot adm}
\end{gather}%
and%
\begin{gather}
\chi _{tt}+3h\chi _{t}-\frac{\beta _{0}}{6}e^{-\chi }\lambda _{t}{}^{2}+\bar{%
	V}_{\chi }=0,  \label{eq:Chi adm} \\
\beta _{0}e^{-\chi }\left[ \lambda _{tt}-\left( \chi _{t}-3h\right) \lambda
_{t}\right] -3\bar{V}_{\lambda }=0.  \label{eq:Lambda adm}
\end{gather}

We already know the inflationary dynamics of Starobinsky$+R^{3}$ model,
which in the scalar-tensor approach in the Einstein frame is characterized
by its specific potential $V\left( \chi \right) $ \cite{PhysRevD.105.063504}%
, as well as the dynamics inflation of Starobinsky$+R\square R$ model,
explored in Ref. \cite{Cuzinatto:2018vjt} via a scalar-vector approach. A
first step in order to understand the dynamics of our current case is
through the study of its phase space, having as reference the known
particular cases mentioned above. In this first part, we will investigate
the existence of an attracting inflationary regime in some region of the
phase space.

Since the dimensionless equations governing the dynamics of the $\chi $ and $%
\lambda $ fields are written as in (\ref{eq:Chi adm}) and (\ref{eq:Lambda
	adm}), that is, two autonomous second-order differential equations
concerning time, we can rewrite them as a system of four first-order
differential equations. Taking $\chi _{t}=\psi $ and $\lambda _{t}=\phi $,
we have
\begin{align}
\chi _{t}& =\psi , \label{Autonomo-aux01} \\
\psi _{t}& =-3h\psi +\frac{\beta _{0}}{6}e^{-\chi }\phi ^{2}-\bar{V}_{\chi },
\label{Autonomo} \\
\lambda _{t}& =\phi , \label{Autonomo-aux03} \\
\beta _{0}\phi _{t}& =\beta _{0}\left( \psi -3h\right) \phi +3e^{\chi }\bar{V%
}_{\lambda },  \label{Autonomo-aux04}
\end{align}%
where%
\begin{equation*}
h=\sqrt{\frac{1}{2}\left( \frac{1}{2}\psi ^{2}-\frac{\beta _{0}}{6}e^{-\chi
	}\phi ^{2}+\bar{V}\right) },
\end{equation*}
which is associated with a physically consistent system when its root argument is positive\footnote{We refer to a physically consistent system that one with real $h(t)$ and $a(t)$.}.

From that point on, we will study the approximate behavior of the solutions
of the system at critical points. Critical points are equilibrium points of
the system, and it is our interest to investigate their stability, which is
directly related to the necessary conditions for the occurrence of a
physical inflationary regime\footnote{%
	A physical inflationary regime is understood to be a regime that has a
	sufficient number of $e$-folds to solve the flatness, horizon, and
	perturbations generation problems and that has a graceful exit.}. The
analysis of the previous system allows us to conclude that there are two
critical points:%
\begin{align}
P_{0}& =\left( \chi _{0},\lambda _{0},\psi _{0},\phi _{0}\right) =\left(
0,0,0,0\right)  \label{P0} \\
P_{c}& =\left( \chi _{c},\lambda _{c},\psi _{c},\phi _{c}\right) =\left( \ln
\left( 4+\sqrt{\frac{3}{\alpha _{0}}}\right) ,\sqrt{\frac{3}{\alpha _{0}}}%
,0,0\right)  \label{Pc}
\end{align}

The study on the stability of these critical points is done through the
linearization of the $4$-dimensional autonomous system $\left( \chi ,\lambda
,\psi ,\phi \right) $. Linearizing the system given by Eqs. (\ref{Autonomo-aux01}), (\ref{Autonomo}), (\ref{Autonomo-aux03}) and (\ref{Autonomo-aux04}) around $P_{0}$%
, we verify that the Lyapunov exponents $r_{0}$, associated with the
stability of the critical point, satisfy the fourth-order characteristic
equation%
\begin{equation}
\beta _{0}r_{0}^{4}+r_{0}^{2}+\frac{1}{3}=0,  \label{eq Lyapunov P0}
\end{equation}%
whose solution is%
\begin{equation*}
r_{0}=\pm \sqrt{\frac{-1\pm \sqrt{1-\frac{4\beta _{0}}{3}}}{2\beta _{0}}.}
\end{equation*}%
A center or spiral point occurs when we obtain pure imaginary roots. Looking
at the previous expression, we see that this occurs whenever the condition%
\begin{equation}
0\leq \beta _{0}\leq \frac{3}{4},  \label{cond beta zero}
\end{equation}%
is satisfied. Any value of $\beta _{0}$ outside this range contains at least
one Lyapunov exponent with positive real part. That is, outside the range (%
\ref{cond beta zero}) the point $P_{0}$ is unstable.\footnote{%
	An identical result was obtained in Ref. \cite{Cuzinatto:2018vjt}.} A
numerical analysis of the system (\ref{Autonomo}) shows that within the
interval (\ref{cond beta zero}) the point $P_{0}$ is an attracting spiral
point and therefore stable (see figure 1). This behavior is essential for the
existence of a graceful exit. In fact, the spiral dynamics around $P_{0}$
constitute the period of coherent oscillations consistent with the initial
phases of reheating. It is also worth noting that Eq. (\ref{eq Lyapunov P0})
is independent of $\alpha _{0}$, and therefore the term $R^{3}$ plays no
role at the end of the inflationary period.

In turn, linearizing the system (\ref{Autonomo}) around $P_{c}$, we verify
that the Lyapunov exponents $r_{c}$ satisfy the characteristic fourth-order
equation%
\begin{gather}
\beta _{0}\left[ r_{c}\left( r_{c}-G\right) -\frac{4}{9}G^{2}\right]
r_{c}\left( r_{c}-G\right) +  \notag \\
+\frac{4}{9}G^{2}\left[ \left( \sqrt{3\alpha _{0}}+6\alpha _{0}\right)
r_{c}\left( r_{c}-G\right) +\frac{1}{3}\sqrt{3\alpha _{0}}-\frac{4}{9}\left(
\sqrt{3\alpha _{0}}+6\alpha _{0}\right) G^{2}\right] =0,
\label{Eq Lyapunov Pc}
\end{gather}%
where%
\begin{equation*}
G=\frac{-3}{2\sqrt{4\sqrt{3\alpha _{0}}+3}}.
\end{equation*}%
A numerical study of this characteristic equation, considering $\alpha
_{0}>0 $ and $\beta _{0}>0$, shows that at least two of the four roots of
Eq. (\ref{Eq Lyapunov Pc}) are real and have opposite signs. This shows that
$P_{c}$ is a saddle point and therefore unstable. This conclusion also
remains valid for $\beta _{0}=0$ and $\alpha _{0}>0$, in which case we have
only two roots.\footnote{%
	In Ref. \cite{PhysRevD.105.063504}, it was shown that the potential $V$ of
	the real and well-behaved model occurs
	for an $\alpha _{0} \geq 0$. Thus, in this paper, we assume a restricted
	parameter $\alpha _{0}$ in this range.} See Ref. \cite{PhysRevD.105.063504}
for details.

To better understand the dynamics of the $\chi $ and $\lambda $ fields, we
will numerically study the $4$-dimensional phase space. In this study, we
will analyze two $2$-dimensional slices of this space given by $\chi
_{t}\times \chi $ and $\lambda _{t}\times \lambda $. For that, we manipulate
Eqs. (\ref{eq:Chi adm}) and (\ref{eq:Lambda adm}) writing them as%
\begin{gather*}
\frac{d\chi _{t}}{d\chi }=\frac{-3h\chi _{t}+\frac{\beta _{0}}{6}e^{-\chi
	}\lambda _{t}^{2}{}-\bar{V}_{\chi }}{\chi _{t}}, \\
\frac{d\lambda _{t}}{d\lambda }=\left( \chi _{t}-3h\right) +\frac{3e^{\chi }%
}{\beta _{0}\lambda _{t}}\bar{V}_{\lambda },
\end{gather*}%
where $h$ is given by (\ref{eq:H adm}).

Numerical analysis of the equation $d\chi _{t}/d\chi $ is more easily
performed if we write $\lambda =\lambda \left( \chi ,\chi _{t},\lambda
_{t},\right. $ $\left. \lambda _{tt},\alpha _{0},\beta _{0}\right)$. For that, it is
necessary to work with the equations (\ref{eq:Lambda adm}) and (\ref{eq:H
	adm}). Solving the quadratic equation for $\lambda $ in Eq. (\ref{eq:Lambda
	adm}), we get%
\begin{equation*}
\lambda =\frac{-1+\sqrt{1-4\alpha _{0}\left\{ 1-e^{\chi }+\beta _{0}e^{\chi }%
		\left[ \lambda _{tt}-\left( \chi _{t}-3h\right) \lambda _{t}\right] \right\}
}}{2\alpha _{0}},
\end{equation*}%
where we choose the positive sign to guarantee the Starobinsky limit. In
principle, we can substitute (\ref{eq:H adm}) in the previous expression,
obtain a third-degree algebraic equation for $\lambda $ and solve it to
obtain $\lambda =\lambda \left( \chi ,\chi _{t},\lambda _{t},\lambda
_{tt},\alpha _{0},\beta _{0}\right) $. However, we will see in Sec. \ref%
{sec:vinculos-observacionais} that the values of interest for $\alpha _{0}$
and $\beta _{0}$ are such that $\alpha _{0}<10^{-3}$ and $\beta _{0}<3\times
10^{-2}$.\footnote{%
	See also Refs. \cite{PhysRevD.105.063504} and \cite{Cuzinatto:2018vjt}.} In
this case, it is licit to consider only linearized corrections of $\alpha
_{0}$ and disregard terms of the type $\alpha _{0}\beta _{0}$. Performing
these approximations, we obtain the functional forms%

\begin{widetext}
\begin{gather}
F_{\chi }\equiv \frac{d\chi _{t}}{d\chi }\simeq \frac{-3\bar{h}\chi _{t}+%
	\frac{\beta _{0}}{6}e^{-\chi }\lambda _{t}^{2}{}-\frac{1}{9}e^{-2\chi }\bar{%
		\lambda}\left\{ 4-e^{\chi }+\bar{\lambda}-2\beta _{0}e^{\chi }\left[ \lambda
	_{tt}-\left( \chi _{t}-3\bar{h}\right) \lambda _{t}\right] \right\} }{\chi
	_{t}},  \label{Chi x Chi t} \\
F_{\lambda }\equiv \frac{d\lambda _{t}}{d\lambda }=\left( \chi
_{t}-3h\right) +\frac{1}{\beta _{0}\lambda _{t}}\left[ 1-e^{-\chi }\left(
1+\lambda +\alpha _{0}\lambda ^{2}\right) \right] ,  \label{lam x lam t}
\end{gather}%
where
\begin{gather}
h\simeq \bar{h}\equiv \frac{-3\beta _{0}^{2}\lambda _{t}+\sqrt{\left(
		3\beta _{0}^{2}\lambda _{t}\right) ^{2}+\left( 12+9\beta _{0}^{2}\lambda
		_{t}^{2}\right) \left[ 3\chi _{t}^{2}-\beta _{0}e^{-\chi }\lambda
		_{t}^{2}-\left( \lambda _{tt}-\chi _{t}\lambda _{t}\right) ^{2}\beta
		_{0}^{2}+A\right] }}{12+9\beta _{0}^{2}\lambda _{t}^{2}},  \label{h barra} \\
\lambda \simeq \bar{\lambda}\equiv \left( e^{\chi }-1\right) \left[
1-\alpha _{0}\left( e^{\chi }-1\right) \right] -\beta _{0}e^{\chi }\left[
\lambda _{tt}-\left( \chi _{t}-3h\right) \lambda _{t}\right] ,
\label{lambda barra}
\end{gather}%
\end{widetext}

with%
\begin{equation*}
A=\left( 1-e^{-\chi }\right) ^{2}\left[ 1-\frac{2}{3}\alpha _{0}\left(
e^{\chi }-1\right) \right] .
\end{equation*}

\begin{figure}[ht]
	\begin{center}
		\includegraphics[height=7.0cm]{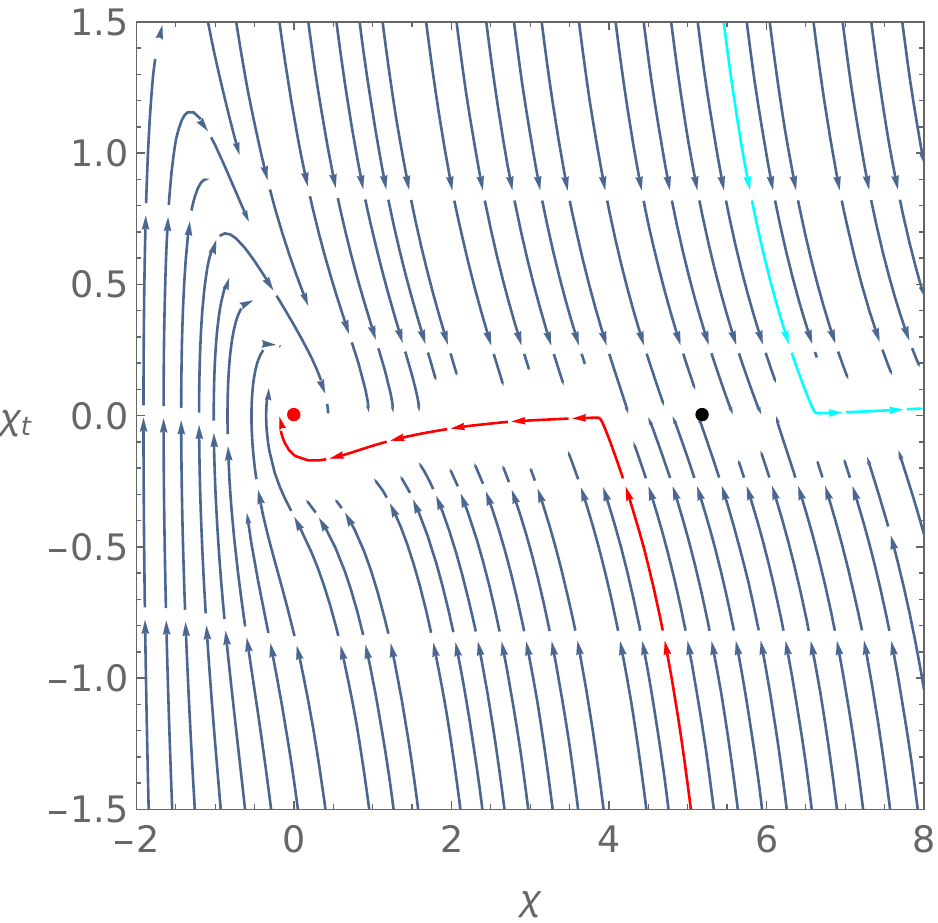} \quad
		\includegraphics[height=7.0cm]{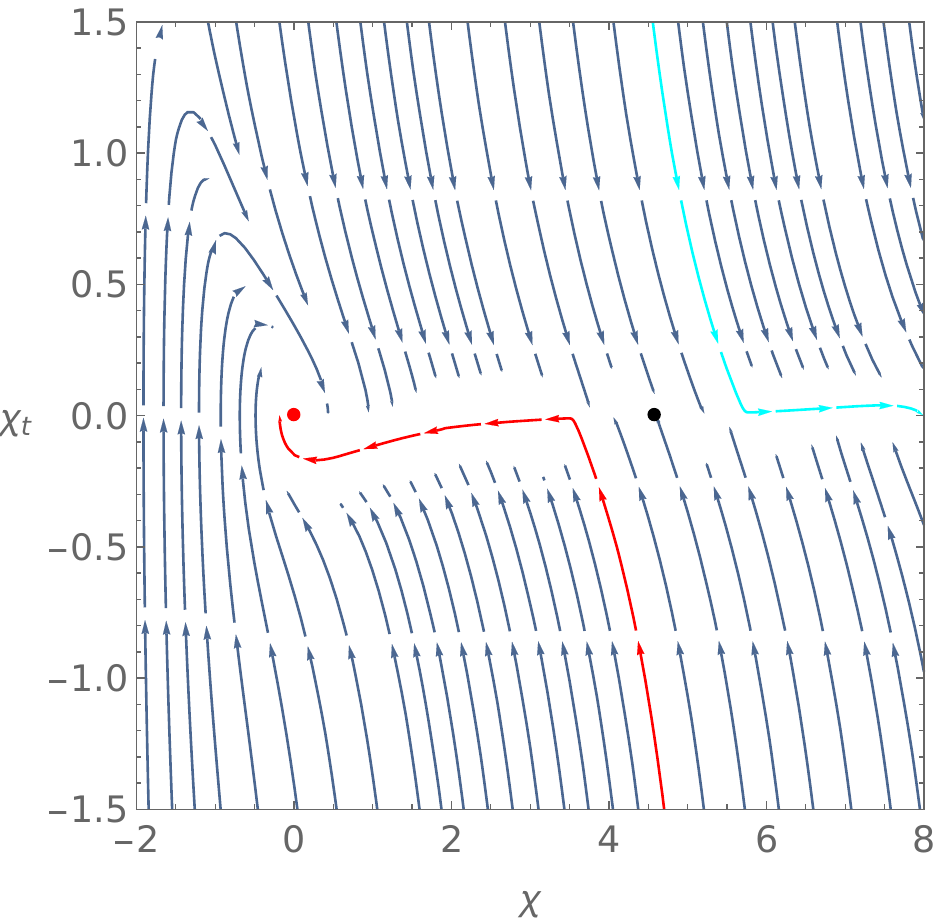}
		\caption{\label{fig:phase-space-chi} The $\chi _{t}\times \chi $ graphs considering phase space cuts $%
			\left( \chi ,\lambda ,\chi _{t},\lambda _{t}\right) $ fixing $\lambda
			_{t}=\lambda _{tt}=0$ and $\beta _{0}=0.001$ with $\left( \lambda ,\alpha
			_{0}\right) =\left( 173,0.0001\right) $ (top graph) and $\left( \lambda
			,\alpha _{0}\right) =\left( 94,0.00034\right) $ (bottom graph). The red and
			black points correspond to the critical points $P_{0}$ and $P_{c}$,
			respectively. For $\alpha _{0}=0.0001$, we have $P_{c}=\left(
			5.18,173,0,0\right) $ and for $\alpha _{0}=0.00034$, we have $P_{c}=\left(
			4.58,94,0,0\right) $. The red (cyan) trajectories represent trajectories
			that, when reaching the attractor line close to $\dot{\chi}=0$, approach
			(depart) from the origin. Details on the interpretation of the graphics are
			presented in the body of the text. \qquad \qquad \qquad \qquad \qquad \qquad \qquad \qquad}
		
	\end{center}
\end{figure}

\begin{figure}[ht]
	\begin{center}
		\includegraphics[height=7.0cm]{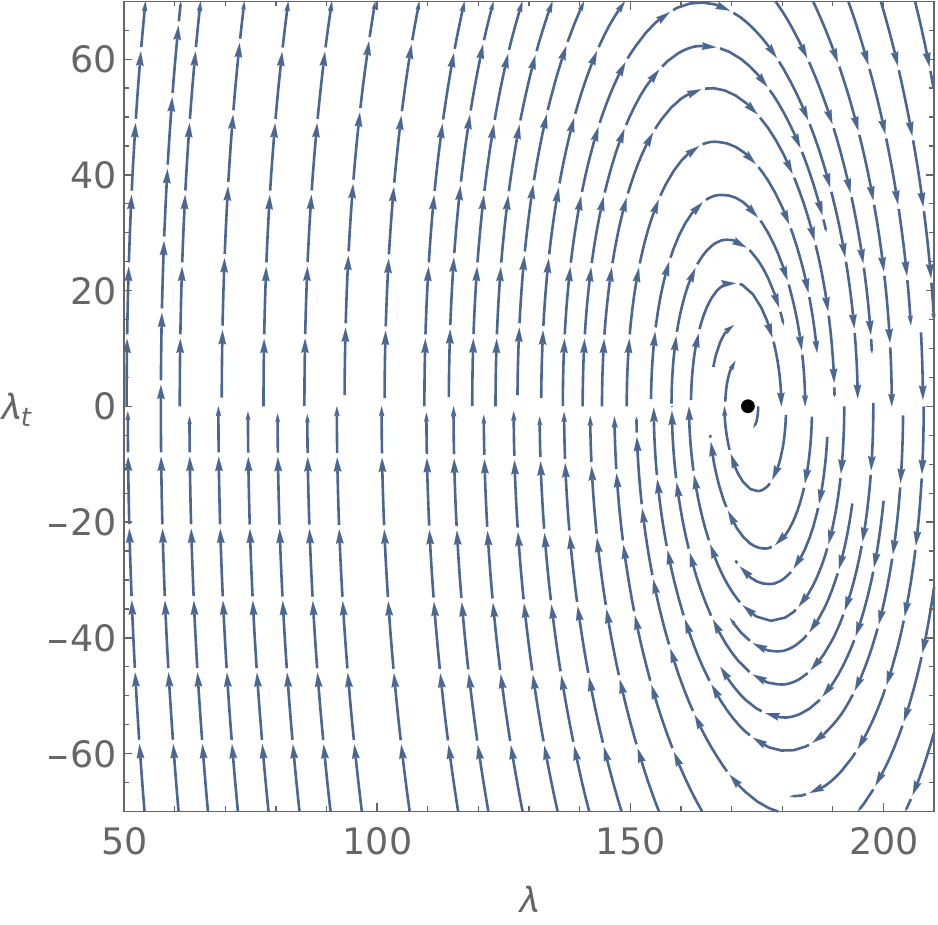} \quad
		\includegraphics[height=7.0cm]{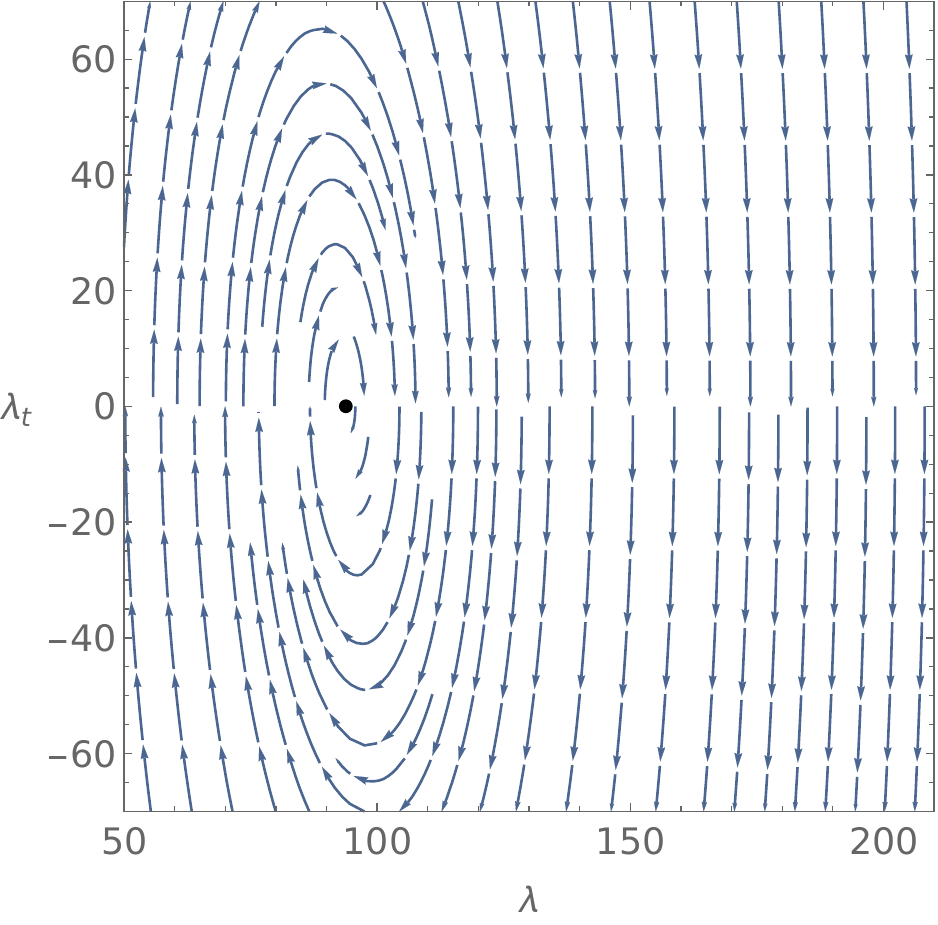}
		\caption{\label{fig:phase-space-lambda} The $\lambda _{t}\times \lambda $ graphs considering phase space cuts $%
			\left( \chi ,\chi _{t},\lambda \text{,}\lambda _{t}\right) $ setting $\chi
			_{t}=0$ and $\beta _{0}=0.001$ with $\left( \chi ,\alpha _{0}\right) =\left(
			5.18,0.0001\right) $ and (top graph) and $\left( \chi ,\alpha _{0}\right)
			=\left( 4.58,0.00034\right) $ (bottom graph). The black points correspond to
			the critical points $P_{c}=\left( 5.18,173,0,0\right) $ (top graph) and $%
			P_{c}=\left( 4.58,94,0,0\right) $ (bottom graph). Details on the
			interpretation of the graphics are presented in the body of the text.  \qquad \qquad \qquad \qquad \qquad \qquad}
	\end{center}
\end{figure}

In figures \ref{fig:phase-space-chi} and \ref{fig:phase-space-lambda}, we show direction fields associated with equations (%
\ref{Chi x Chi t}) and (\ref{lam x lam t}).

The first (and most relevant) point that can be seen in figure \ref{fig:phase-space-chi} is that
there is an attractor line close to $\chi _{t}\simeq 0$. The existence of
this region is consistent with any value of $\alpha _{0}<10^{-3}$ and $\beta
_{0}<3\times 10^{-2}$ and for any interval of $\lambda _{t}$ and $\lambda
_{tt}$ that yields real results in the region of interest $\chi \in \lbrack
0,8]$.\footnote{%
	These ranges are typically between $-10$ e $10$.} At the same time that the $%
\chi $ field tends to the attracting line ($\chi _{t}\simeq 0$), figure \ref{fig:phase-space-lambda}
indicates that $\lambda $ tends to a finite value and $\lambda
_{t}\rightarrow 0$. This finite value of $\lambda $ essentially depends on
the value of $\chi $ with variations on a smaller scale due to changes in
the parameter $\alpha _{0}$. The other fixed parameters $\chi _{t}$ and $%
\beta _{0}$ in figure \ref{fig:phase-space-lambda} change how $\lambda $ approaches the accumulation
point but does not change its value. We will see in the Sec. \ref%
{sec - inflacao SR} that this attractor region in the $4$-dimensional phase
space where $\left( \chi ,\lambda ,\chi _{t}\text{,}\lambda _{t}\right)
\simeq \left( \chi ,\lambda \left( \chi \right) ,0,0\right) $ corresponds to
a slow-roll inflationary regime.

Once the attractor region is reached, we must ask ourselves if inflation
occurs enough, i.e., if it generates a sufficient number of $e$-folds and if
it ends in a reheating phase. The answer to this question essentially
depends on the position where the $\chi $ field hits the attractor line in
figure \ref{fig:phase-space-chi}. If the $\chi $ field is to the left of the critical point $P_{c}$
(black dots in the graphs of figure \ref{fig:phase-space-chi}), inflation proceeds normally and ends
in a phase of coherent oscillations associated with the beginning of
reheating. On the other hand, if $\chi $ is to the right of $P_{c}$ the
value of $\chi $ increases indefinitely, and inflation never ends (see Ref.
\cite{PhysRevD.105.063504} for details). Thus, a physical inflationary
regime, i.e., consistent with a graceful exit, only occurs if $\chi <\chi
_{c}\Rightarrow \alpha _{0}<3\left( e^{\chi }-4\right) ^{-2}$ which for
sufficiently large $\chi $ corresponds to $\alpha _{0}<3e^{-2\chi }$.

In the next section, we will see how to describe the dynamics of the $\chi $
and $\lambda $ fields during the slow-roll inflationary phase.

\subsection{Inflation in the slow-roll leading order regime \label{sec -
		inflacao SR}}

This section aims to describe the dynamics of $\chi $, $\lambda $, and their
derivatives during the inflationary regime considering the slow-roll
approximation. In the region associated with physical inflation, the
parameter $\chi $ is a monotonic decreasing function of time, so we can
parameterize the various quantities in terms of $\chi $. For the case of
Starobinsky model, we know that in the slow-roll leading order regime $\chi
_{t}\sim \delta $ and $\chi _{tt}\sim \delta ^{2}$, where $\delta $ is the
slow-roll factor defined as $\delta \equiv e^{-\chi }$. And for Starobinsky
plus $R^{3}$ model (i.e., $\beta _{0}=0$ and $\alpha _{0}\neq 0$), we have
\cite{PhysRevD.105.063504}
\begin{equation}
\chi _{t}\sim \left( \delta -\frac{\alpha _{0}}{3}\delta ^{-1}\right) .
\label{chi t exemplo}
\end{equation}%
Note that since $\alpha _{0}<3\delta ^{2}$, the second term of the previous
expression is of the same order or less than $\delta $.

The previous discussion allows us to associate the factor $\delta $ as a
parameter that controls the slow-roll approximation order, i.e., a quantity $%
f\sim \delta ^{n}$ will be an $n$th-order slow-roll quantity. In this case, $%
\chi _{t}$ present in (\ref{chi t exemplo}) is first-order in slow-roll,
since both $\delta $ and $\alpha _{0}\delta ^{-1}$ are first-order. To apply
this reasoning in our model, it is also necessary to establish what is the
maximum slow-roll order of the parameter $\beta _{0}$. A more detailed
analysis of the field equations in the attractor region shows us that, for
slow-roll inflation, $\beta _{0}\lesssim \delta $, i.e., $\beta _{0}$ is a
(at most) first-order slow-roll parameter (for details see Ref. \cite%
{Cuzinatto:2018vjt}).

Once the slow-roll (maximum) orders of the parameters $\alpha _{0}$ and $%
\beta _{0}$ are known, we can propose the following ansatz for $\chi _{t}$:

\begin{equation}
\chi _{t}\simeq c_{1}\delta +\beta _{0}\sum\limits_{n=0}^{\infty
}b_{n}\left( \beta _{0}\delta ^{-1}\right) ^{n}+\alpha _{0}\delta
^{-1}\sum\limits_{n=0}^{\infty }d_{n}\left( \beta _{0}\delta ^{-1}\right)
^{n}.  \label{chi t ansatz}
\end{equation}%
This ansatz has the following properties:

\begin{itemize}
	\item All terms are first-order in slow-roll, this being the leading order
	of $\chi _{t}$;
	
	\item In the limit of $\beta _{0}\rightarrow 0$, we recover the result (\ref%
	{chi t exemplo});
	
	\item Derivating Eq. (\ref{chi t ansatz}) with respect to $t$, we increase
	the slow-roll order, i.e., $\chi _{tt}$ is second order, $\chi _{ttt}$ is
	third order, etc.
\end{itemize}

By following similar reasoning, we propose the following ansatz for $%
\lambda $:%
\begin{equation}
\lambda \simeq \delta ^{-1}+\sum\limits_{n=0}^{\infty }g_{n}\left( \beta
_{0}\delta ^{-1}\right) ^{n}+\alpha _{0}\delta ^{-2}\sum_{n=0}^{\infty
}j_{n}\left( \beta _{0}\delta ^{-1}\right) ^{n}.  \label{lambda ansatz}
\end{equation}%
In this case, the first term is of order $\mathcal{O}\left( -1\right) $ in
slow-roll, and the others are zero-order terms. The $\mathcal{O}\left( -1\right) $ order term is
necessary as we know that in the case of Starobinsky $\lambda =\delta
^{-1}-1 $ (see Eq. (\ref{eq:Lambda adm}) with $\alpha _{0}=\beta _{0}=0$).
Analogously to $\chi _{t}$, each derivative of $\lambda $ with respect to $t$
increases by one the slow-roll order.

The next step is substituting these two ansatzes and their derivatives into
Eqs. (\ref{eq:H adm}), (\ref{eq:Chi adm}), and (\ref{eq:Lambda adm}) taking
into account only the slow-roll leading order. In this situation, we get%
\begin{align}
3h\chi _{t}-\frac{1}{3}\delta \lambda \left( 1-\delta \lambda -2\delta -%
\frac{2}{3}\alpha _{0}\delta \lambda ^{2}\right) & \simeq 0,
\label{eq:Chi adm aprox} \\
3\beta _{0}h\lambda _{t}-\left[ 1-\delta \left( 1+\lambda +\alpha
_{0}\lambda ^{2}\right) \right] & \simeq 0,  \label{eq:Lambda adm aprox}
\end{align}%
where%
\begin{equation}
h^{2}\simeq \frac{1}{6}\delta \lambda \left( 1-\frac{1}{2}\delta \lambda
\right) .  \label{eq:H adm aprox}
\end{equation}%
By explicitly substituting Eqs. (\ref{chi t ansatz}) and (\ref{lambda ansatz}%
) in these last three expressions, we get after a long calculation%
\begin{gather}
\chi _{t}\simeq -\frac{2\sqrt{3}}{3\left( 3-\beta _{0}\delta ^{-1}\right) }%
\delta \left( 1-\frac{\alpha _{0}}{3}\delta ^{-2}\right) ,
\label{chi t aprox} \\
\lambda \simeq \delta ^{-1}-\frac{3-2\beta _{0}\delta ^{-1}}{3-\beta
	_{0}\delta ^{-1}}-\alpha _{0}\delta ^{-2}\left( 1+\frac{\frac{1}{3}\beta
	_{0}\delta ^{-1}}{3-\beta _{0}\delta ^{-1}}\right) ,  \label{lambda aprox}
\end{gather}%
with%
\begin{equation}
h^{2}\simeq \frac{1}{12}\left( 1-2\delta -\frac{2}{3}\alpha _{0}\delta
^{-1}\right) .  \label{h2 aprox}
\end{equation}%
For details see appendix \ref{Ap6}. It is worth noting that the previous
expressions are well defined only for $\beta _{0}\delta ^{-1}<3$. Note in
Eq. (\ref{chi t aprox}) the existence of two terms that, in the slow-roll
leading order, are first-order terms. In Eq. (\ref{lambda aprox}), we have the $%
\mathcal{O}\left( -1\right) $ order term in addition to the zero-order
corrections. Finally, $h^{2}$, related to the Hubble parameter, is given by
the zero-order slow-roll leading term plus first-order corrections
(independent of $\beta _{0}$).

\subsubsection{Calculation of slow-roll parameters and number of $e$-folds}

The characterization of the inflationary regime is done through the
slow-roll parameters%
\begin{gather}
\epsilon \equiv -\frac{\dot{H}}{H^{2}}=-\frac{h_{t}}{h^{2}},  \label{epsilon}
\\
\eta \equiv -\frac{1}{H}\frac{\dot{\epsilon}}{\epsilon }=-\frac{1}{h}\frac{%
	\epsilon _{t}}{\epsilon }.  \label{eta}
\end{gather}%
By substituting (\ref{chi t aprox}) and (\ref{lambda aprox}) in (\ref%
{eq:H-dot adm}), we get%
\begin{equation*}
h_{t}\simeq -\frac{\delta ^{2}}{3\left( 3-\beta _{0}\delta ^{-1}\right) }%
\left( 1-\frac{\alpha _{0}}{3}\delta ^{-2}\right) ^{2}.
\end{equation*}%
Thus, in the slow-roll leading order, we have%
\begin{equation}
\epsilon \simeq \frac{4\delta ^{2}}{\left( 3-\beta _{0}\delta ^{-1}\right) }%
\left( 1-\frac{\alpha _{0}}{3}\delta ^{-2}\right) ^{2}.  \label{epsilon SR}
\end{equation}%
The next step is calculating $\eta $. Differentiating $\epsilon $ and using
this result together with $\epsilon $ itself in Eq. (\ref{eta}), we get%
\begin{equation}
\eta \simeq -\frac{4\delta }{\left( 3-\beta _{0}\delta ^{-1}\right) ^{2}}%
\left[ 3\left( 2-\beta _{0}\delta ^{-1}\right) +\alpha _{0}\delta
^{-2}\left( 2-\frac{1}{3}\beta _{0}\delta ^{-1}\right) \right] .
\label{eta SR}
\end{equation}%
Note that by construction $\alpha _{0}\delta ^{-2}<3$ and $\beta _{0}\delta
^{-1}<3$.

In order to have robust inflation, i.e., with enough number of e-folds, we
must have $\epsilon \ll 1$ and $\eta \ll 1$. Thus, from the equations (\ref%
{epsilon SR}) and (\ref{eta SR}), we see that this occurs for $\delta \ll 1$
(typically $\chi \gtrsim 4$). However, unlike the Starobinsky case, we also
have lower bounds for $\delta $. In fact, the slow-roll inflationary regime
only occurs if
\begin{equation}
\delta >\frac{\beta _{0}}{3}\text{ \ and \ }\delta >\sqrt{\frac{\alpha _{0}}{%
		3}}.  \label{condicoes}
\end{equation}%
The first condition does not represent a real difficulty for the existence
of slow-roll inflation, because even if at some point we have $\delta <\beta
_{0}$,\footnote{%
	In this situation, we have no guarantee that an inflationary regime exists.}
the dynamics of the phase space guarantees that $\chi $ decreases
monotonically so that at some point $\delta $ becomes greater than $\beta
_{0}$ (see figure \ref{fig:phase-space-chi}). The second condition represents a real constraint for
carrying out a physical inflation (see discussion at the end of Sec. \ref%
{sec - EspFase}). For a discussion of the implications of this second
constraint and the initial conditions of inflation see Ref. \cite%
{PhysRevD.105.063504}.

Next we will calculate the number of $e$-folds $N$ in the slow-roll leading
order. By the definition of $N$, we have

\begin{equation*}
N=\int_{t}^{t_{e}}Hdt\simeq \frac{1}{4}\int_{\delta }^{\delta _{e}}\frac{%
	\left( 1-2\delta -\frac{2}{3}\alpha _{0}\delta ^{-1}\right) \left( 3-\beta
	_{0}\delta ^{-1}\right) }{\delta ^{2}\left( 1-\frac{\alpha _{0}}{3}\delta
	^{-2}\right) }d\delta ,
\end{equation*}%
where the index $e$ corresponds to the end of inflation. To integrate this
expression, it is convenient to perform the following change of variable:%
\begin{equation}
x=\frac{\delta _{m}}{\delta }\text{ \ where }\delta _{m}=\sqrt{\frac{\alpha
		_{0}}{3}}.  \label{variavel x}
\end{equation}%
In this case, we get%
\begin{equation*}
N\simeq -\frac{1}{4\delta _{m}}\int_{x}^{x_{e}}\frac{\left( x-2\delta
	_{m}-2x^{2}\delta _{m}\right) \left( 3-\beta _{0}\delta _{m}^{-1}x\right) }{%
	1-x^{2}}\frac{dx}{x},
\end{equation*}%
whose solution is%
\begin{gather*}
	N\simeq -\frac{1}{4\delta _{m}}\left\{ -2x\beta _{0}-6\delta _{m}\ln
	x+\right. \\
	\left. +\frac{\beta _{0}+12\delta _{m}^{2}}{2\delta _{m}}\ln \left[ \left(
	1-x\right) \left( 1+x\right) \right] +\frac{3+4\beta _{0}}{2}\ln \left( 
	\frac{1+x}{1-x}\right) \right\} _{x}^{x_{e}}.
\end{gather*}%
By considering only leading terms and taking into account that $x_{e}\ll x$,
we finally get%
\begin{equation}
N\simeq \frac{3}{8}\sqrt{\frac{3}{\alpha _{0}}}\ln \left[ \left( 1-x\right)
^{\gamma -1}\left( 1+x\right) ^{\gamma +1}\right] \text{ \ where \ }\gamma
^{2}=\frac{\beta _{0}^{2}}{3\alpha _{0}}\text{.\ }  \label{efolds aprox}
\end{equation}

By construction, physical inflation occurs in the interval $0\leq x<1$. In
fact, when $x\rightarrow 1$ we have $\delta \rightarrow \delta _{m}$ which
corresponds approximately to $\chi \rightarrow \chi _{c}$ (see eq. (\ref{Pc}%
)). However, the expression (\ref{efolds aprox}) has an extra restriction
due to the presence of the $\beta _{0}$ term. For $\beta _{0}^{2}>3\alpha
_{0}\Rightarrow \gamma >1$, we have that when $x\rightarrow 1$, the value of
$N$ diverges to $-\infty $, and this is clearly not physical. What happens
is that for $\gamma >1$, the function $N$ presents a maximum point within
the interval $0\leq x<1$. Differentiating $N$ with respect to time, we have%
\begin{equation*}
N_{t}\simeq \frac{3x}{8}\sqrt{\frac{3}{\alpha _{0}}}\left[ \frac{-\left(
	\gamma -1\right) \left( 1+x\right) +\left( \gamma +1\right) \left(
	1-x\right) }{\left( 1-x\right) \left( 1+x\right) }\right] \chi _{t}.
\end{equation*}%
So, for $\gamma >1$, we have%
\begin{eqnarray}
	N_{t} &=&0\Rightarrow -\left( \gamma -1\right) \left( 1+x_{\max }\right)
	+\left( \gamma +1\right) \left( 1-x_{\max }\right) =0  \notag \\
	&\Rightarrow &x_{\max }=\frac{1}{\gamma }<1\text{.}  \label{x max}
\end{eqnarray}%
On the other hand, for values of $x$ such that $x_{\max }\leq x<1$, we get
\begin{equation*}
\beta _{0}\delta ^{-1}=3\gamma x\geq 3\Rightarrow \delta \leq \frac{\beta
	_{0}}{3}
\end{equation*}%
which violates the first condition of Eq. (\ref{condicoes}).

Therefore, based on the previous analysis, we conclude that Eqs. (\ref%
{epsilon SR}), (\ref{eta SR}) and (\ref{efolds aprox}) referring to the
quantities $\epsilon $, $\eta $ and $N$ are valid in the following intervals:%
\begin{equation}
\left\{
\begin{array}{c}
\gamma \leq 1\Rightarrow x<1\Rightarrow \chi <\ln \left( \sqrt{\frac{3}{%
		\alpha _{0}}}\right) \\
\gamma >1\Rightarrow x<x_{\max }\Rightarrow \chi <\ln \left( \frac{3}{\beta
	_{0}}\right)%
\end{array}%
\right. .  \label{condicoes 2}
\end{equation}

In the next section, we will study the inflationary regime from the
perturbative point of view.

\section{Inflation via cosmological perturbation theory \label{sec:eqs-perturbadas}}

In this section, we investigate inflation of the model (\ref{Acao Frame de
	Einstein Adim}) via cosmological perturbations. Recall that its background
dynamic equations are the Friedmann ones, given by Eqs. (\ref%
{eq:H}) and (\ref{eq:H-dot}), and the equations of motion
for the scalar fields $\chi $ and $\lambda $, given by Eqs. (\ref{eq:Chi})
and (\ref{eq:Lambda}).

Before proceeding with our developments, it is worth commenting on scalar
perturbations. In addition to the perturbations of the two scalar fields,
which we will denote by $\delta \chi $ and $\delta \lambda $, we have the
scalar perturbations of the metric. The line element in the perturbed
Friedmann-Lema\^{\i}tre-Robertson-Walker\ (FLRW) metric is given by%
\begin{eqnarray}
	ds^{2} &=&-\left( 1+2A\right) dt^{2}+2a\partial _{i}Bdx^{i}dt+  \notag \\
	&&+a^{2}\left[ \left( 1-2\psi \right) \delta _{ij}+2\partial _{ij}E+h_{ij}%
	\right] dx^{i}dx^{j},  \label{eq:metrica-flrw-perturbada}
\end{eqnarray}%
with $A$, $B$, $\psi $ and $E$ being the scalar perturbations of the metric
\cite{MUKHANOV1992203,RevModPhys.78.537}. In order to obtain the
perturbative field equations through a perturbation directly in the action,
we need to write it up to the second order in the perturbations. In this
case, we must consider second-order terms for perturbations involving only
scalar field perturbations (e.g., $\delta \chi ^{2}$), second-order terms
involving only metric scalar perturbations (e.g., $A^{2}$) and cross terms,
that is, a product of first-order terms (e.g., $A\delta \chi $). In the
following subsection, by following a perturbative procedure directly in the
action, along the lines of that found in Refs. \cite{Baumann:2018muz,Wands:2007bd,RevModPhys.78.537}, and
assuming the spatially flat gauge,\footnote{%
	In the spatially flat gauge, the perturbations $\psi =E=0$, in order to kill
	the spatial part of the metric.} we obtain and discuss the equations of
motion for the perturbations.

\subsection{Equations for scalar perturbations \label{subsec:perturbacao-na-acao}}

The first step in order to perturbate the action is defining the
perturbations of the scalar fields. For an inhomogeneous distribution of
matter, we write%
\begin{align}
\chi \left( t,x\right) & =\chi \left( t\right) +\delta \chi \left(
t,x\right) ,  \label{eq:def-pert-chi} \\
\lambda \left( t,x\right) & =\lambda \left( t\right) +\delta \lambda \left(
t,x\right) .  \label{eq:def-pert-l}
\end{align}%
In turn, the metric in Eq. (\ref{eq:metrica-flrw-perturbada}) is written as%
\begin{equation}
g^{\rho \sigma }\left( t,x\right) =g^{\rho \sigma }\left( t\right) +\delta
g^{\rho \sigma }\left( t,x\right) .
\end{equation}%
By writing Eq. (\ref{Acao Frame de Einstein Adim}) up to the second order in
the perturbations and taking their variations with respect to each one of
the perturbations, we are able to obtain the following equations of motion
for the perturbations $\delta \chi $ and $\delta \lambda $
\begin{gather}
	\delta \ddot{\chi}+3H\delta \dot{\chi}-\frac{1}{a^{2}}\nabla ^{2}\left(
	\delta \chi \right) +  \notag \\
	+\frac{\beta _{0}}{6}e^{-\chi }\dot{\lambda}\left( \dot{\lambda}\delta \chi
	-2\delta \dot{\lambda}\right) +V_{\chi \chi }\delta \chi +V_{\chi \lambda
	}\delta \lambda =  \notag \\
	\dot{\chi}\dot{A}+\frac{1}{a}\dot{\chi}\nabla ^{2}B-2V_{\chi }A,
	\label{eq:perturbacao-chi-comp}
\end{gather}%
and%
\begin{gather}
	\beta _{0}e^{-\chi }\left[ \delta \ddot{\lambda}+\left( 3H-\dot{\chi}%
	\right) \delta \dot{\lambda}-\frac{1}{a^{2}}\nabla ^{2}\left( \delta \lambda
	\right) +\right.   \notag \\
	\left. -\dot{\lambda}\delta \dot{\chi}-3 V_{\lambda} \delta \chi \right] -3\left( V_{\chi
		\lambda }\delta \chi +V_{\lambda \lambda }\delta \lambda \right) =  \notag \\
	\beta _{0}e^{-\chi }\left( \dot{\lambda}\dot{A}+\dot{\lambda}\frac{1}{a}%
	\nabla ^{2}B+\dot{\chi}\dot{\lambda}A\right) +6V_{\lambda }A,
	\label{eq:perturbacao-lambda-comp}
\end{gather}%
as well as the Einstein equations%
\begin{gather}
	H\left( 3HA-\frac{k^{2}}{a}B\right) =-\frac{1}{4}\left[ 3\dot{\chi}\delta 
	\dot{\chi}-\beta _{0}e^{-\chi }\dot{\lambda}\delta \dot{\lambda}+\right.  
	\notag \\
	\left. -\left( 3\dot{\chi}^{2}-\beta _{0}e^{-\chi }\dot{\lambda}^{2}\right)
	A+\frac{1}{2}\beta _{0}e^{-\chi }\dot{\lambda}^{2}\delta \chi +V_{\chi
	}\delta \chi +V_{\lambda }\delta \lambda \right] ,  \label{eq:einstein-1}
\end{gather}%
and%
\begin{equation}
HA=\frac{1}{4}\left( 3\dot{\chi}\delta \chi -\beta _{0}e^{-\chi }\dot{\lambda%
}\delta \lambda \right) .  \label{eq:einstein-2}
\end{equation}%
The double subscript in potential $V$ represents second-order
differentiation with respect to the corresponding scalar fields.

\subsection{Equations in the slow-roll leading order regime \label{subsec:pert-slow-roll}}

Once we obtain Eqs. (\ref{eq:perturbacao-chi-comp}), (\ref%
{eq:perturbacao-lambda-comp}), (\ref{eq:einstein-1}) and (\ref{eq:einstein-2}%
), which completely describe the evolution of scalar perturbations, the next
step is to write them in the slow-roll leading order regime. This is not a
trivial task and therefore, we will initially present the particular case of
the Starobinsky model ($\alpha _{0}=\beta _{0}=0$). In this case, Eqs. (\ref%
{eq:perturbacao-chi-comp}), (\ref{eq:perturbacao-lambda-comp}), (\ref%
{eq:einstein-1}) and (\ref{eq:einstein-2}) reduce to%
\begin{gather}
	\delta \ddot{\chi}+3H\delta \dot{\chi}-\frac{1}{a^{2}}\nabla ^{2}\left(
	\delta \chi \right) +\hat{V}_{\chi \chi }\delta \chi +\hat{V}_{\chi \lambda
	}\delta \lambda =  \notag \\
	\dot{\chi}\dot{A}+\frac{1}{a}\dot{\chi}\nabla ^{2}B-2\hat{V}_{\chi }A,
	\label{eq:pert-chi-sta}
\end{gather}%
\begin{equation}
\hat{V}_{\chi \lambda }\delta \chi +\hat{V}_{\lambda \lambda }\delta \lambda
=-2\hat{V}_{\lambda }A,  \label{eq:pert-lambda-sta}
\end{equation}%
\begin{equation}
H\left( 3HA-\frac{k^{2}}{a}B\right) =-\frac{1}{4}\left( 3\dot{\chi}\delta
\dot{\chi}-3\dot{\chi}^{2}A+\hat{V}_{\chi }\delta \chi +\hat{V}_{\lambda
}\delta \lambda \right)  \label{eq:einstein-1-sta}
\end{equation}%
and%
\begin{equation}
HA=\frac{3}{4}\dot{\chi}\delta \chi ,  \label{eq:einstein-2-sta}
\end{equation}%
with%
\begin{equation}
\hat{V}\left( \chi ,\lambda \right) =\frac{1}{3}\kappa _{0}e^{-2\chi
}\lambda \left( e^{\chi }-1-\frac{1}{2}\lambda \right) .
\end{equation}

In Sec. \ref{sec - inflacao SR}, we saw, in the context of the background,
the behavior of the scalar fields, their derivatives and the relationships
they keep between them. Once the slow-roll factor $\delta $ was established,
we recall that in the slow-roll leading order regime, we obtain%
\begin{equation*}
\dot{\chi}\sim \delta \text{, \ }\lambda \sim \delta ^{-1}\text{, \ }H\sim
\delta ^{0}\text{, \ }\beta _{0}\sim \delta \text{ \ and \ }\alpha _{0}\sim
\delta ^{2}\,\text{,}
\end{equation*}%
and that with each differentiation with respect to time in the scalar
fields, an order of slow-roll is increased, that is, $\ddot{\chi}\sim \delta
^{2}$ and $\dot{\lambda}\sim \delta ^{0}$. By making the constructions in
this section, some assumption is necessary, namely, to find the slow-roll
orders of the perturbations, we need to establish the slow-roll order of one
of them. In this sense, we take the perturbation $\delta \chi $ as a
zero-order slow-roll quantity. Also, we keep in mind that derivatives do not
change the slow-roll order of perturbations. We are now able to write the
equations of motion in the slow-roll leading order. Analyzing Eq. (\ref%
{eq:einstein-2-sta}), note that since $\dot{\chi}\delta \chi \sim \delta $,
perturbation $A$ must be at most first-order in slow-roll. Regarding Eq. (%
\ref{eq:einstein-1-sta}), in its right member, we have the first and third
terms, which are of first-order, the second term, which is subdominant $3%
\dot{\chi}^{2}A$ of third-order in slow-roll, and the last one is null,
since $\hat{V}_{\lambda }=0$. Furthermore, since in its left member we have $%
3H^{2}A\sim \delta $, we conclude that perturbation $B$ is at most first-order
in slow-roll. In turn, as $\hat{V}_{\chi \lambda }\sim \delta $ and $%
\hat{V}_{\lambda \lambda }\sim \delta ^{2}$, we see that Eq. (\ref%
{eq:pert-lambda-sta}) establish the perturbation leading order of $\delta
\lambda $, namely, $\delta \lambda \sim \delta ^{-1}$. With Eq. (\ref%
{eq:pert-lambda-sta}), we can still write $\delta \lambda $ in terms of $%
\delta \chi $ and substitute in Eq. (\ref{eq:pert-chi-sta}). Thus, on the
left side of Eq. (\ref{eq:pert-chi-sta}), the first three terms are of zero
order in slow-roll, while all other terms of the equation give us
subdominant contributions. So we can write%
\begin{equation}
\delta \ddot{\chi}+3H\delta \dot{\chi}-\frac{1}{a^{2}}\nabla ^{2}\left(
\delta \chi \right) \simeq 0.
\end{equation}

When developing the previous analysis now for the case of the complete
equations, we find that the perturbations evolve, in the slow-roll leading
order regime, with the same orders obtained previously. In short, the scalar
perturbations evolve in the form $A\sim B\sim \delta $ and $\delta \lambda
\sim \delta ^{-1}$. It is interesting to note that the perturbation $\delta
\lambda $ goes in leading order regime with $\delta ^{-1}$, and that if it
were otherwise, it would seriously compromise the slow-roll dynamics.

By applying all the discussion raised above, we find, in the slow-roll
leading order regime, the following equations of motion for the
perturbations of the scalar fields%
\begin{equation}
\delta \ddot{\chi}+3H\delta \dot{\chi}-\frac{1}{a^{2}}\nabla ^{2}\left(
\delta \chi \right) \simeq \frac{1}{3}\kappa _{0}\left( \delta \chi
-e^{-\chi }\delta \lambda \right) ,  \label{eq:perturbacao-chi-sr-1.2}
\end{equation}%
and%
\begin{equation}
\beta _{0}\left[ \delta \ddot{\lambda}+3H\delta \dot{\lambda}-\frac{1}{a^{2}}%
\nabla ^{2}\left( \delta \lambda \right) \right] \simeq \kappa _{0}\left(
\delta \chi -e^{-\chi }\delta \lambda \right) .
\label{eq:perturbacao-lambda-sr-1.2}
\end{equation}%
These results are in agreement with those obtained in Ref. \cite%
{Cuzinatto:2018vjt}, where the Starobinsky$+R\square R$ model is explored.

\subsection{Adiabatic and isocurvature perturbations \label{subsec:pert-adiab-iso}}

In this subsection, we define adiabatic and isocurvature perturbations,
obtain expressions that describe their dynamics, and study their solutions.

The action (\ref{Acao Frame de Einstein Adim}) can be rewritten, along the lines of
Ref. \cite{Gundhi:2018wyz}, compactly as%
\begin{equation}
S=\frac{M_{Pl}^{2}}{2}\int d^{4}x\sqrt{-g}\left( -\frac{1}{2}g^{\mu \nu
}G_{IJ}\left( \Phi \right) \partial _{\mu }\Phi ^{I}\partial _{\nu }\Phi
^{J}-3V\right) ,
\end{equation}%
where the scalars $\Phi ^{I}\left( x\right) $ are seen as local coordinates
of the scalar field space with metric $G_{IJ}\left( \Phi \right) $%
\begin{equation}
\Phi ^{I}=%
\begin{pmatrix}
\chi \\
\lambda%
\end{pmatrix}%
,\text{ \ }G_{IJ}\left( \Phi \right) =%
\begin{pmatrix}
3 & 0 \\
0 & -\beta _{0}e^{-\chi }%
\end{pmatrix}%
,
\end{equation}%
and $V$ represents the potential of the model, Eq. (\ref{eq:potencial-modelo}). In
a two-field scalar model, the field space is $2$-dimensional and
characterized by $G_{IJ}\left( \Phi \right) $. To conveniently describe the
evolution of perturbations, we can define a basis having a tangent
direction, which we will denote by $\hat{\sigma}^{I}$, and another
orthogonal, $\hat{s}^{I}$, to the background trajectories. Tangent
directions to background trajectories are associated with adiabatic
perturbation, while orthogonal directions are associated with isocurvature
perturbation. In this sense, we build the basis through the definitions,
respectively, of the module of the velocity vector, the unit velocity vector
in the tangent direction and the normalization rule%
\begin{equation}
\dot{\sigma}=\sqrt{G_{IJ}\dot{\Phi}^{I}\dot{\Phi}^{J}},\text{\ \ }\hat{\sigma%
}^{I}=\frac{\dot{\Phi}^{I}}{\dot{\sigma}}\text{ \ and \ }G_{IJ}\hat{\sigma}%
^{I}\hat{\sigma}^{J}=1,  \label{eq:vetor-velocidade}
\end{equation}%
and for the orthogonal direction, the normalization\footnote{%
	This proposed normalization condition for $\hat{s}^{I}$ is necessary due to
	the negative sign in the metric $G_{IJ}\left( \Phi \right) $.} and
orthogonality rules%
\begin{equation}
G_{IJ}\hat{s}^{I}\hat{s}^{J}=-1\text{ \ and \ }G_{IJ}\hat{s}^{I}\hat{\sigma}%
^{J}=0.
\end{equation}%
For our case, we have for the velocity module $\dot{\sigma}$%
\begin{equation}
\dot{\sigma}=\sqrt{3\dot{\chi}^{2}-\beta _{0}e^{-\chi }\dot{\lambda}^{2}}.
\label{eq:modulo-veloc}
\end{equation}%
Note that it is directly related to the Friedmann equation (\ref%
{eq:H-dot}). For the unit velocity vectors, we write%
\begin{equation}
\hat{\sigma}^{\chi }=\frac{\dot{\chi}}{\dot{\sigma}}\text{ \ and \ }\hat{%
	\sigma}^{\lambda }=\frac{\dot{\lambda}}{\dot{\sigma}}.
\end{equation}%
In turn, for the unit vectors in the orthogonal direction to the background
trajectories, we have%
\begin{equation}
\hat{s}^{\chi }=\sqrt{\frac{\beta _{0}e^{-\chi }}{3}}\frac{\dot{\lambda}}{%
	\dot{\sigma}}\text{ \ and \ }\hat{s}^{\lambda }=\sqrt{\frac{3}{\beta
		_{0}e^{-\chi }}}\frac{\dot{\chi}}{\dot{\sigma}}.
\end{equation}

Continuing our study on the evolution of scalar perturbations, we point out
that the quantity $\delta \Phi _{g}^{I}$ given by%
\begin{equation}
\delta \Phi _{g}^{I}=\delta \Phi ^{I}+\frac{\dot{\Phi}^{I}}{H}\psi ,
\end{equation}%
is gauge invariant. It turns out that $\psi =0$ when working on a spatially
flat gauge, so that $\delta \Phi _{g}^{I}=\delta \Phi ^{I}$. That said, by
projecting $\delta \Phi ^{I}$ in the $\hat{\sigma}$ and $\hat{s}$
directions, we construct the adiabatic $Q_{\sigma }$ and isocurvature $Q_{s}$
perturbations, respectively. In that sense, we have%
\begin{equation}
Q_{\sigma } =\hat{\sigma}^{J}G_{IJ}\delta \Phi ^{I}
=\frac{3\dot{\chi}\delta \chi -\beta _{0}e^{-\chi }\dot{\lambda}%
	\delta \lambda }{\dot{\sigma}},  \label{eq:pert-q-sigma}
\end{equation}%
and%
\begin{equation}
Q_{s} =\hat{s}^{J}G_{IJ}\delta \Phi ^{I}
=\frac{\sqrt{3\beta _{0}e^{-\chi }}\left( \dot{\lambda}\delta \chi -%
	\dot{\chi}\delta \lambda \right) }{\dot{\sigma}}.  \label{eq:pert-q-s}
\end{equation}

It is worth noting that from the point of view of the slow-roll
approximation both Eq. (\ref{eq:pert-q-sigma}) and Eq. (\ref{eq:pert-q-s})
are zero-order. By having written the expressions for the adiabatic and
isocurvature perturbations, the next step is to invert the relations in
order to obtain $\delta \Phi ^{I}=\delta \Phi ^{I}\left( Q\right) $. We
obtain these relations by solving the linear system given by Eqs. (\ref%
{eq:pert-q-sigma}) and (\ref{eq:pert-q-s}). Thus, we find the expressions%
\begin{equation}
\delta \chi =\frac{1}{\dot{\sigma}}\left( \dot{\chi}Q_{\sigma }-\sqrt{\frac{%
		\beta _{0}e^{-\chi }}{3}}\dot{\lambda}Q_{s}\right) ,  \label{eq:d-chi-q}
\end{equation}%
and%
\begin{equation}
\delta \lambda =\frac{1}{\dot{\sigma}}\left( \dot{\lambda}Q_{\sigma }-\sqrt{%
	\frac{3}{\beta _{0}e^{-\chi }}}\dot{\chi}Q_{s}\right) .
\label{eq:d-lambda-q}
\end{equation}


Once the field perturbations in terms of the adiabatic and isocurvature
perturbations were obtained, we can write the second order perturbed action
for $Q_{\sigma }$ and $Q_{s}$. Such an action is fundamentally constituted
by quadratic terms involving $Q_{\sigma }$ and $Q_{s}$ (e.g., $Q_{\sigma
}^{2}$), cross terms involving a $Q_{\sigma }$ or $Q_{s}$ and a metric
perturbation (e.g., $AQ_{\sigma }$), and quadratic terms for metric
perturbations (e.g., $A^{2}$). On the other hand, we can express the cross
term only in terms of the perturbations $Q_{\sigma }$ and $Q_{s}$ by making
use of the constraints from the Einstein equations, Eqs. (\ref{eq:einstein-1}) and (\ref{eq:einstein-2}). By taking these considerations into account, we find
the following structure for the part of the action that depends only on $%
Q_{\sigma }$ and $Q_{s}$
\begin{gather}
S^{\left( 2\right) }=\frac{M_{Pl}^{2}}{2}\int d^{4}x\sqrt{-g}\left( -\frac{1}{2}\partial _{\kappa }Q_{\sigma }\partial
^{\kappa }Q_{\sigma }+\frac{1}{2}\partial _{\kappa }Q_{s}\partial ^{\kappa
}Q_{s}\right. +  \notag \\
+C_{Q_{\sigma }^{2}}Q_{\sigma }^{2}+C_{Q_{\sigma }Q_{s}}Q_{\sigma
}Q_{s}+C_{Q_{s}^{2}}Q_{s}^{2}+  \notag \\
+\left. C_{Q_{\sigma }\dot{Q}_{\sigma }}Q_{\sigma }\dot{Q}_{\sigma }+C_{\dot{%
		Q}_{\sigma }Q_{s}}\dot{Q}_{\sigma }Q_{s}+C_{Q_{\sigma }\dot{Q}_{s}}Q_{\sigma
}\dot{Q}_{s}+C_{Q_{s}\dot{Q}_{s}}Q_{s}\dot{Q}_{s}\right) ,
\label{eq:acao-qs-geral}
\end{gather}
where the coefficient of the cross kinetic term $\partial _{\kappa }Q_{\sigma }\partial
^{\kappa }Q_{s}$ is zero, and the others can be found in Appendix \ref{Ap:coefficients}. It is interesting to analyze the behavior of kinetic terms and their possible contribution to the
emergence of ghost-type instabilities.
Note that the kinetic terms are canonical, equal
and with reversed signs. This characteristic irremediably indicates
that the existence of ghost-type instabilities is something intrinsic to the
model and that it is essential to take this into account when performing the
perturbation quantization process.\footnote{%
	This conclusion is valid for the case of a positive $\beta _{0}$.} Furthermore, the fact of the non-existence of the cross kinetic term is something
expected and is directly related to our approach of making a consistent
decomposition of the perturbations in the tangent and orthogonal directions
to the trajectories of the background phase space.

When writing Eq. (\ref{eq:acao-qs-geral}) considering a slow-roll leading order regime, we obtain
\begin{eqnarray}
	S^{\left( 2\right) } &\simeq &\frac{M_{Pl}^{2}}{2}\int d^{4}x\sqrt{-g}%
	\left\{ -\frac{1}{2}\partial _{\kappa }Q_{\sigma }\partial ^{\kappa
	}Q_{\sigma }+\frac{1}{2}\partial _{\kappa }Q_{s}\partial ^{\kappa
	}Q_{s}+\right.  \notag \\
	&&\left. -\frac{\kappa _{0}}{\dot{\sigma}^{2}}\left[ 1-\frac{1}{2}\left( 
	\frac{3}{\beta _{0}e^{\chi }}+\frac{\beta _{0}e^{\chi }}{3}\right) \right] 
	\dot{\chi}^{2}Q_{s}^{2}\right\} .  \label{eq:acao-pert-qs}
\end{eqnarray}%


We now turn our attention to the task of writing the equations for the
evolution of adiabatic and isocurvature perturbations. This is done by
substituting the expressions (\ref{eq:d-chi-q}) and (\ref{eq:d-lambda-q}) in
the dynamic equations (\ref{eq:perturbacao-chi-sr-1.2}) and (\ref%
{eq:perturbacao-lambda-sr-1.2}). In this case, taking the first derivatives
of Eqs. (\ref{eq:d-chi-q}) and (\ref{eq:d-lambda-q}), remembering that the
background quantities can be considered constant, we are able to write%
\begin{equation}
\ddot{Q}_{\sigma }+3H\dot{Q}_{\sigma }-\frac{1}{a^{2}}\nabla ^{2}Q_{\sigma
}\simeq 0,  \label{eq:din-pert-adiab-sig}
\end{equation}%
and%
\begin{equation}
\ddot{Q}_{s}+3H\dot{Q}_{s}+m^{2}Q_{s}\simeq 0,  \label{eq:din-pert-adiab-s}
\end{equation}%
where%
\begin{equation}
m^{2}=-\left[ \frac{1}{a^{2}}\nabla ^{2}+\frac{\kappa _{0}}{3}\left( 1-\frac{%
	3}{\beta _{0}e^{\chi }}\right) \right] .
\end{equation}%
Note that relations (\ref{eq:din-pert-adiab-sig}) and (\ref%
{eq:din-pert-adiab-s}) indicate that the adiabatic $Q_{\sigma }$ and
isocurvature $Q_{s}$ perturbations are decoupled, that is, they evolve
independently in our model. That is an interesting result since such a decoupling
usually does not occur. Generally, the isocurvature perturbation enter as source
of the adiabatic one. \cite{Gundhi:2018wyz}.

From this point on, it becomes convenient to treat the field equations for the
perturbations in a Mukhanov-Sasaki form. By making a redefinition of the perturbations
and assuming conformal time and Fourier space, we get
\begin{equation}
\delta \varphi _{\sigma }^{\prime \prime }+\left( k^{2}-\frac{a^{\prime
		\prime }}{a}\right) \delta \varphi _{\sigma }\simeq 0,
\end{equation}%
\begin{equation}
\delta \varphi _{s}^{\prime \prime }+\left[ k^{2}-\frac{a^{\prime \prime }}{a%
}-\frac{a^{2}\kappa _{0}}{3}\left( 1-\frac{3}{\beta _{0}e^{\chi }}\right) %
\right] \delta \varphi _{s}\simeq 0,
\end{equation}%
with the prime representing derivative with respect to conformal time and
where%
\begin{equation}
\delta \varphi _{\sigma }\equiv aQ_{\sigma }\text{ \ and \ }\delta \varphi
_{s}\equiv aQ_{s}\text{.}
\end{equation}%
In a de Sitter background (slow-roll zero order), we have%
\begin{equation}
\frac{a^{\prime \prime }}{a}\simeq \frac{2}{\eta ^{2}}\text{,\ \ }a\simeq -%
\frac{1}{H\eta }\text{\ \ and \ }H^{2}\simeq \frac{\kappa _{0}}{12}.
\label{eq:relacao-background-dS}
\end{equation}%
Furthermore, in the slow-roll zero order regime, we have
\begin{equation}
\dot{\chi}\simeq -\frac{1}{3}\frac{H^{-1}}{3-\beta _{0}\delta ^{-1}}\delta
\left( 1-\frac{\alpha _{0}}{3}\delta ^{-2}\right) \simeq 0\Rightarrow \chi
=cte\text{.}
\end{equation}

For consistency with several previous results, we have $\beta _{0}e^{\chi
}<3 $ so that we can define a quantity%
\begin{equation}
M\equiv \frac{3}{\beta _{0}e^{\chi }}-1>0,
\end{equation}%
and in this way we write the expressions%
\begin{equation}
\delta \varphi _{\sigma }^{\prime \prime }+k^{2}\left( 1-\frac{2}{k^{2}\eta
	^{2}}\right) \delta \varphi _{\sigma }\simeq 0,
\label{eq:din-pert-adiab-ms2}
\end{equation}%
\begin{equation}
\delta \varphi _{s}^{\prime \prime }+k^{2}\left[ 1-\frac{2}{k^{2}\eta ^{2}}%
\left( 1-2M\right) \right] \delta \varphi _{s}\simeq 0.
\label{eq:din-pert-iso-ms2}
\end{equation}%
Next, we will explore the solutions of Eqs. (\ref{eq:din-pert-adiab-ms2})
and (\ref{eq:din-pert-iso-ms2}).

\subsection{Solutions to the perturbations \label{subsec:solutions-pert}}

Once the equations for the dynamics of adiabatic and isocurvature
perturbations have been established in the appropriate form, given by Eqs. (%
\ref{eq:din-pert-adiab-ms2}) and (\ref{eq:din-pert-iso-ms2}), we can write
and analyze their solutions.

In a subhorizon regime, $k\eta \gg 1$, equations to the perturbations are
approximated by%
\begin{equation*}
\delta \varphi _{\sigma }^{\prime \prime }+k^{2}\delta \varphi _{\sigma
}\simeq 0,\ \ k\eta \gg 1,
\end{equation*}%
\begin{equation*}
\delta \varphi _{s}^{\prime \prime }+k^{2}\delta \varphi _{s}\simeq 0,\text{
	\ \ }k\eta \gg 1.
\end{equation*}%
The quantization process in de Sitter takes the following initial conditions
\cite{Baumann:2009ds,Piattella:2018hvi}
\begin{equation}
\delta \varphi _{\sigma }\simeq \delta \varphi _{s}\simeq \frac{1}{\sqrt{2k%
}}e^{-ik\eta },\text{\ \ }k\eta \gg 1,  \label{eqs:cond-inic}
\end{equation}%
or%
\begin{equation}
Q_{\sigma }\simeq Q_{s}\simeq \frac{1}{\sqrt{2k}a}e^{-ik\eta }\simeq -%
\frac{H\eta }{\sqrt{2k}}e^{-ik\eta }.
\end{equation}%
Note that due to the ghost-type behavior of isocurvature perturbation, the
quantization of the $Q_{s}$ field was performed in the same way as in Refs.
\cite{Ivanov:2016hcm,Salvio:2017xul}. In principle, this behavior can
raise questions about the unitarity of the theory \cite{Sbisa:2014pzo} (see also the discussions in Refs. \cite{Salles:2014rua,Peter:2017xxf}). However, as we will see below, the $Q_{s}$ field decays rapidly after
crossing the horizon, suppressing any observable effects associated with
isocurvature perturbation.\footnote{%
	Note also that in the slow-roll leading order regime, the adiabatic and
	isocurvature perturbations of the model evolve independently.}

The exact solution of Eqs. (\ref{eq:din-pert-adiab-ms2}) and (\ref%
{eq:din-pert-iso-ms2}) can be written using a combination of Hankel's
functions as \cite{Piattella:2018hvi,gradshteyn2007}

\begin{equation}
\delta \varphi \left( \eta ,k\right) =C_{1}\left( k\right) \sqrt{-\eta }%
H_{\nu }^{\left( 1\right) }\left( -k\eta \right) +C_{2}\left( k\right) \sqrt{%
	-\eta }H_{\nu }^{\left( 2\right) }\left( -k\eta \right) ,
\end{equation}%
where for the adiabatic perturbation $\delta \varphi _{\sigma }$, we have $%
\nu _{\sigma }=3/2$, and for the isocurvature perturbation $\delta \varphi
_{s}$, we have%
\begin{equation}
\nu _{s}=\frac{3}{2}\sqrt{1-\frac{16M}{9}}.  \label{eq:nu_s}
\end{equation}%
To determine the constants, we compare the general solution with the initial
conditions in Eq. (\ref{eqs:cond-inic}). For the adiabatic case ($\nu
_{\sigma }=3/2$) in the subhorizon regime, we find%
\begin{equation}
C_{1\sigma }=-\frac{\sqrt{\pi }}{2}\text{ \ and \ }C_{2\sigma }=0,
\end{equation}%
so that%
\begin{equation}
\delta \varphi _{\sigma }\left( \eta ,k\right) =-\frac{\sqrt{-\pi \eta }}{2}%
H_{3/2}^{\left( 1\right) }\left( -k\eta \right) .
\label{eq:solucao-pert-adiab}
\end{equation}%
In turn, for the case of isocurvature perturbation, where $\nu _{s}$ is
given by Eq. (\ref{eq:nu_s}), for a subhorizon regime, we get\footnote{%
	Note that in the limit where $M\rightarrow 0$ (adiabatic case $\nu _{\sigma
	}=3/2$), we have $C_{1s}=C_{1\sigma }=-\sqrt{\pi }/2$.}

\begin{equation}
C_{1s}=\frac{\sqrt{\pi }}{2}e^{i\frac{\pi }{2}\left( \nu _{s}+\frac{1}{2}%
	\right) }\text{ \ and \ }C_{2s}=0.
\end{equation}%
Thus, the solution to the isocurvature perturbation is written as%
\begin{equation}
\delta \varphi _{s}\left( \eta ,k\right) =\frac{\sqrt{-\pi \eta }}{2}e^{i%
	\frac{\pi }{2}\left( \nu _{s}+\frac{1}{2}\right) }H_{\nu _{s}}^{\left(
	1\right) }\left( -k\eta \right) .  \label{eq:solucao-pert-iso}
\end{equation}

Since the solutions (\ref{eq:solucao-pert-adiab}) and (\ref%
{eq:solucao-pert-iso}) for the perturbations have been established, we can
analyze their behavior in the superhorizon regime ($k\eta \ll 1$). Taking into
account Eq. (\ref{eq:relacao-background-dS}), which gives us the relation
between $\eta $ and $a$ in a de Sitter background, for adiabatic
perturbation, we find%
\begin{equation}
\delta \varphi _{\sigma }\left( \eta ,k\right) \simeq -\frac{i}{4\eta }%
\left( \frac{k}{2}\right) ^{-3/2}\text{ \ or \ }Q_{\sigma }\simeq \frac{iH}{%
	4}\left( \frac{k}{2}\right) ^{-3/2},\text{ \ }k\eta \ll 1,
\label{eq:solucao-Q-sigma-super-h}
\end{equation}%
while for isocurvature perturbation, we have%
\begin{equation}
\delta \varphi _{s}\left( \eta ,k\right) \simeq -\frac{i\sqrt{\pi }e^{i%
		\frac{\pi }{2}\left( \nu _{s}+\frac{1}{2}\right) }}{2\sin \left( \nu _{s}\pi
	\right) \Gamma \left( 1-\nu _{s}\right) }\left( \frac{k}{2}\right) ^{-\nu
	_{s}}\left( -\eta \right) ^{\frac{1}{2}-\nu _{s}},\text{ \ }k\eta \ll 1,
\end{equation}%
or%
\begin{equation}
Q_{s}\simeq -\frac{i\sqrt{\pi }e^{i\frac{\pi }{2}\left( \nu _{s}+\frac{1}{2}%
		\right) }H^{\nu _{s}-\frac{1}{2}}}{2\sin \left( \nu _{s}\pi \right) \Gamma
	\left( 1-\nu _{s}\right) }\left( \frac{k}{2}\right) ^{-\nu _{s}}a^{\nu _{s}-%
	\frac{3}{2}}.  \label{eq:solucao-Q-s-super-h}
\end{equation}

The result in Eq. (\ref{eq:solucao-Q-sigma-super-h}) tells us that the
adiabatic perturbation is constant in the superhorizon limit, whereas Eq. (%
\ref{eq:solucao-Q-s-super-h}) reveals a decaying behavior for the isocurvature
one. Remembering that $M>0$, we have the following situations:

\begin{itemize}
	\item If the quantity $\nu _{s}$ is real, we have
	\begin{equation}
	0<M\leq \frac{9}{16}\Rightarrow 0\leq \nu _{s}<\frac{3}{2},
	\end{equation}%
	representing a solution to $Q_{s}$ that decays with $a$.
	
	\item If the quantity $\nu _{s}$ is imaginary,
	\begin{equation}
	M>\frac{9}{16}\Rightarrow \nu _{s}=i\frac{3}{2}\sqrt{\left\vert 1-\frac{16M}{%
			9}\right\vert },
	\end{equation}%
	also providing a decaying solution, since we have the product of an
	oscillatory term and the term decaying with $a^{-\frac{3}{2}}$.
\end{itemize}

We can provide a quantitative measure of isocurvature perturbation by
considering scales of interest during inflation (measured in CMB
anisotropies). These ones are within the range $%
10^{-3}Mpc^{-1}<k<10^{4}Mpc^{-1}$, where the pivot scale is $k_{\ast }=0.002$
with $50<N_{\ast }<60$. The point is that during the inflationary regime, a
given scale $k$ crosses the horizon at a specific value of the number of $e$%
-folds $N$. The smaller/larger the scale $k$ is, the smaller/larger the
number of $e$-folds $N$ it experiences after crossing the horizon. Taking
the smallest scale\footnote{%
	Since all others decay further.} $k_{sm}=10^{4}Mpc^{-1}$, we find $%
N_{sm}=N_{\ast }-15.4$. In this sense, for $N_{\ast }=50$ and $\nu _{s}=1/2$%
, we get%
\begin{equation}
Q_{s}\simeq \frac{1}{2}\left( \frac{k_{sm}}{2}\right) ^{-\frac{1}{2}%
}e^{-N_{sm}}\sim 10^{-18},
\end{equation}%
which shows us that isocurvature perturbation is negligible after
inflation. Since, in addition to this, they do not enter as a source of
adiabatic perturbation, we can consider them negligible. All the previous analysis was carried out considering the slow-roll approximation.

In the next section, we will analyze the connection of our model with the observations.

\section{Observational constraints \label{sec:vinculos-observacionais}}

At the end of the previous section, we show why isocurvature
perturbation is negligible after inflation. Furthermore, since adiabatic perturbation
has the same behavior as in the case of a single scalar field, we easily recognize the
power spectrum and its connection with observational parameters.
To make the connection with the observations, specifically to
write the power spectrum, it is interesting to recover the mass units of the
fields. In this sense, equations such as (\ref{eq:H}), (\ref{eq:H-dot}), (%
\ref{eq:vetor-velocidade}) and (\ref{eq:pert-q-sigma}) need to be written in
terms of massive fields given in Eq. (\ref{eq:campos-massivos}). Since the
curvature perturbation is given by $\mathcal{R=}\frac{H}{\dot{\sigma}}\left(
\frac{\sqrt{2}}{M_{Pl}}Q_{\sigma }\right) $,\footnote{%
	The quantity $\sqrt{2}/M_{Pl}$ is introduced in the definition of the
	curvature perturbation $\mathcal{R}$ to recover the conventional units of $%
	Q_{\sigma }$.} the power spectrum of adiabatic perturbation is written as%
\begin{equation}
\mathcal{P}_{\mathcal{R}}^{2}=\left. \frac{k^{3}}{2\pi ^{2}}\left\vert
\mathcal{R}\right\vert ^{2}\right\vert _{k=Ha}=\left. \frac{1}{8\pi
	^{2}M_{Pl}^{2}}\frac{H^{2}}{\epsilon }\right\vert _{k=Ha},  \label{eq:P_R}
\end{equation}%
where we evaluate it to $k=Ha$ at the instant when $k$ crosses the horizon.
The result in Eq. (\ref{eq:P_R}) is identical to the power spectrum for
single-field inflationary models. Thus, in the slow-roll leading order
regime, the scalar spectral index $n_{s}$ and the tensor-to-scalar ratio $r$
are, respectively,%
\begin{equation}
n_{s}=1+\eta -2\epsilon \text{ \ and \ }r=16\epsilon ,
\label{eq:parametros-ns-r}
\end{equation}%
where $\epsilon $ and $\eta $ are the slow-roll parameters of the model given by the
Eqs. (\ref{epsilon SR}) and (\ref{eta SR}). These equations depend on $\alpha _{0}$, $%
\beta _{0}$, and the number of $e$-folds $N$ through Eq. (\ref{efolds aprox}%
) which carries the dependency between $N$ and $\delta $.

In our paper, there are two types of Plots where we compare our model with
observational data \cite%
{BICEPKeck:2021gln}, built from Eq. (\ref{eq:parametros-ns-r}) taking the
three independent parameters $\alpha _{0}$, $\beta _{0}$, and $N$: the usual
$n_{s}\times r_{0.002}$ plane and the parameter space $\alpha _{0}\times
\beta _{0}$. The Plots are constructed by setting one of the parameters and varying the others.
We use the range $52\leq N\leq 59$ for the number of
inflation $e$-folds $N$ based on a reheating modeling. For details, see
appendix \ref{Ap:vinculo-reaquecimento}.

The figure \ref{fig:nsxr-plane} shows the $n_{s}\times r_{0.002}$ plane
containing the observational constraints (in blue) obtained from Ref. \cite%
{BICEPKeck:2021gln} and the theoretical evolution of the model in two
different situations.

In the top graph of figure \ref{fig:nsxr-plane}, we fixed the parameter $\beta _{0}$
and varied the others. In it, the light red region represents Starobinsky$%
+R^{3}$ model, which starts at the light red points. In turn, the light
yellow region represents the complete model with $\beta _{0}=1.5\times
10^{-2}$, starting at the yellow points. As we increase the values of the
parameter $\alpha _{0}$, the region predicted by the model shifts to the
left and slightly downwards, until it crosses the region of $95\%$ C.L..
This behavior can also be seen in Ref. \cite{PhysRevD.105.063504}, and it is consistent with the results obtained in Ref. \cite{Huang:2013hsb}, where $\beta _{0}=0$. These constraints establish, in the most
conservative way, a maximum value for $\alpha _{0}\sim 10^{-4}$.

In the bottom graph of figure \ref{fig:nsxr-plane}, on the other hand, we fixed the parameter $\alpha _{0}$
and varied the others. The light red region represents the Starobinsky$%
+R\square R$ model, which starts at the light red points. In turn, the light
green region represents the complete model with $\alpha _{0}=10^{-5}$,
starting at the light green points. As the values of $\beta _{0}$ increase,
the region predicted by the model moves to the right and slightly upwards, until it
crosses the region of $95\%$ C.L.. These
constraints establish a maximum value for $\beta _{0}\sim 10^{-2}$. Similar results were obtained in Refs.
\cite{Cuzinatto:2018vjt,Castellanos:2018dub} for the Starobinsky$+R\square R$
case. However, a considerable difference between our results and those in
Ref. \cite{Cuzinatto:2018vjt} is checked for the constraint on the
tensor-to-scalar ratio $r_{0.002}$. There, $r_{0.002}$ can assume larger
values, so that the growth of the region predicted by the model is more
accentuated. This difference is due to the fact that in Ref. \cite%
{Cuzinatto:2018vjt}, the definition for the curvature perturbation was not
established properly by not making a separation of the background phase
space trajectories in the tangent (adiabatic perturbation) and orthogonal
(isocurvature perturbation) directions. On the other hand, our results are closer to those in Ref. \cite{Castellanos:2018dub}, indicating that the approach of treating the $R\square R$ term as a small perturbation is relevant and consistent.
\begin{figure}[ht]
	\begin{center}
		\includegraphics[height=4.76cm]{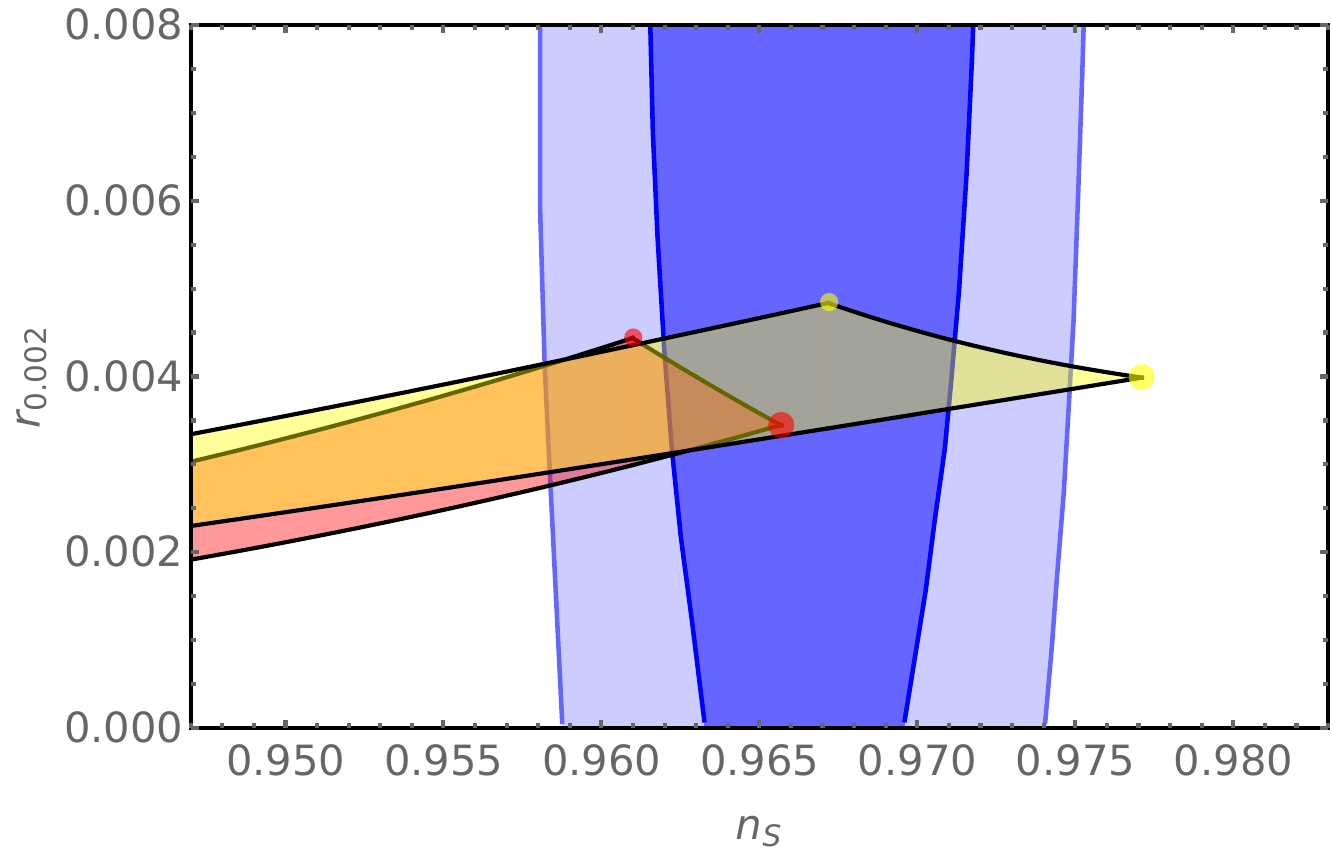} \quad
		\includegraphics[height=4.76cm]{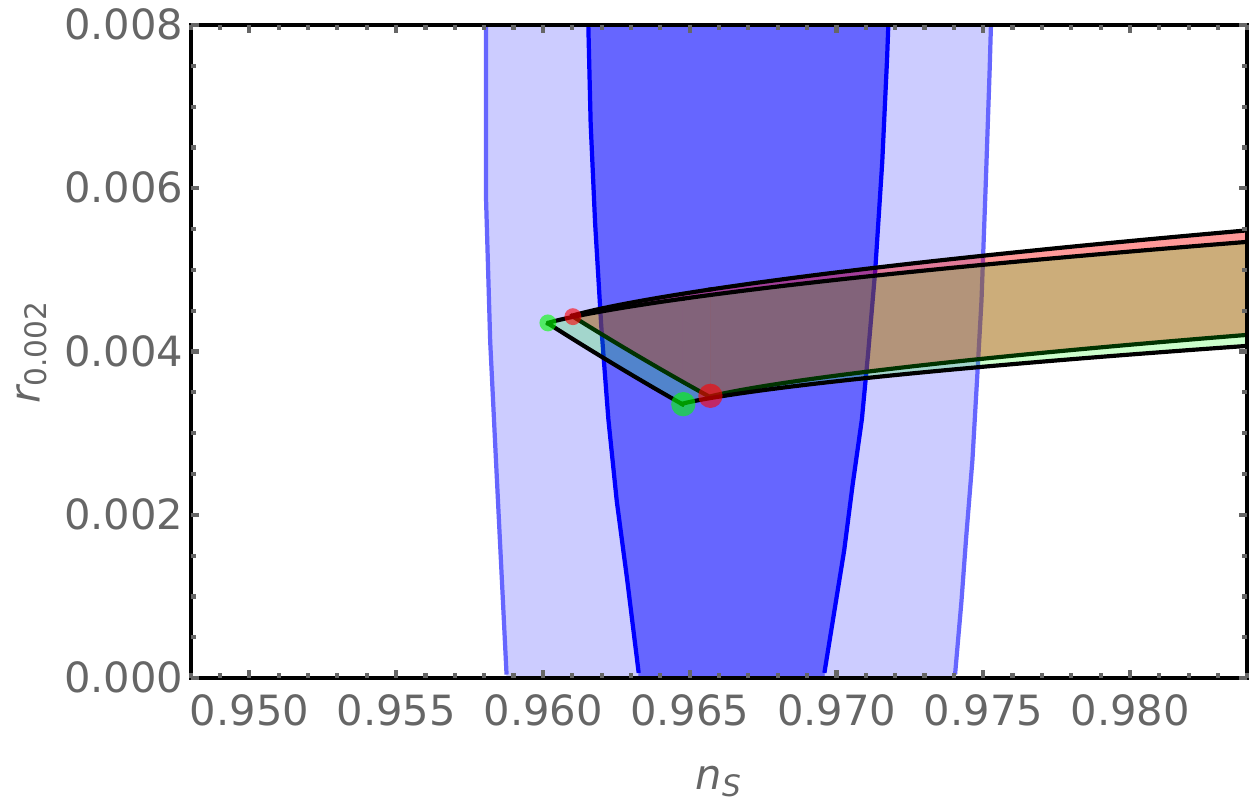}
		\caption{\label{fig:nsxr-plane} The contours in blue represent the constraints of
			the $n_{s}\times r_{0.002}$ plane in $68\%$ and $95\%$ C.L. due to
			observational data from Planck plus BICEP3/Keck plus BAO \cite%
			{BICEPKeck:2021gln}. In the top graph, we set the parameter $\beta _{0}$ and vary
			the others. The light red circles represent Starobinsky$+R^{3}$ model for $%
			N=52$ (smaller one) and $N=59$ (bigger one). The yellow circles represent
			the complete model with $\beta _{0}=1.5\times 10^{-2}$ for $N=52$ and $N=59$%
			. As the values of $\alpha _{0}$ increase, the region predicted by the model
			shifts to the left and downwards, until it crosses the region of $95\%$
			C.L.. When it crosses, the curves for Starobinsky$+R^{3}$ and the complete
			model with $\beta _{0}=1.5\times 10^{-2}$ for $N=52$ correspond to $\alpha
			_{0}=3.5\times 10^{-5}$ and $r_{0.002}=4.1\times 10^{-3}$ and $\alpha
			_{0}=4\times 10^{-5}$ and $r_{0.002}=4.1\times 10^{-3}$, respectively; for $%
			N=59$ they correspond to $\alpha _{0}=8.2\times 10^{-5}$ and $%
			r_{0.002}=2.8\times 10^{-3}$ and $\alpha _{0}=5.4\times 10^{-5}$ and $%
			r_{0.002}=2.9\times 10^{-3}$, respectively. In the bottom graph, in turn, we set
			the parameter $\alpha _{0}$ and vary the others. The light red circles
			represent Starobinsky$+R\square R$ model for $N=52$ (smaller one) and $N=59$
			(bigger one). The green circles represent the complete model with $\alpha
			_{0}=10^{-5}$ for $N=52$ and $N=59$. As the values of $\beta _{0}$ increase,
			the region predicted by the model shifts to the right and slightly upwards,
			until they cross the $95\%$ C.L. region. As it crosses, the curves for $N=52$
			correspond approximately to $\beta _{0}=1.7\times 10^{-2}$ and $%
			r_{0.002}=5.2\times 10^{-3}$; for $N=59$ they correspond approximately to $%
			\beta _{0}=1.5\times 10^{-2}$ and $r_{0.002}=3.9\times 10^{-3}$. \qquad \qquad \qquad \qquad \qquad \qquad \qquad \qquad \qquad \qquad}		
	\end{center}
\end{figure}

\begin{figure}[ht]
	\begin{center}
		\includegraphics[height=4.59cm]{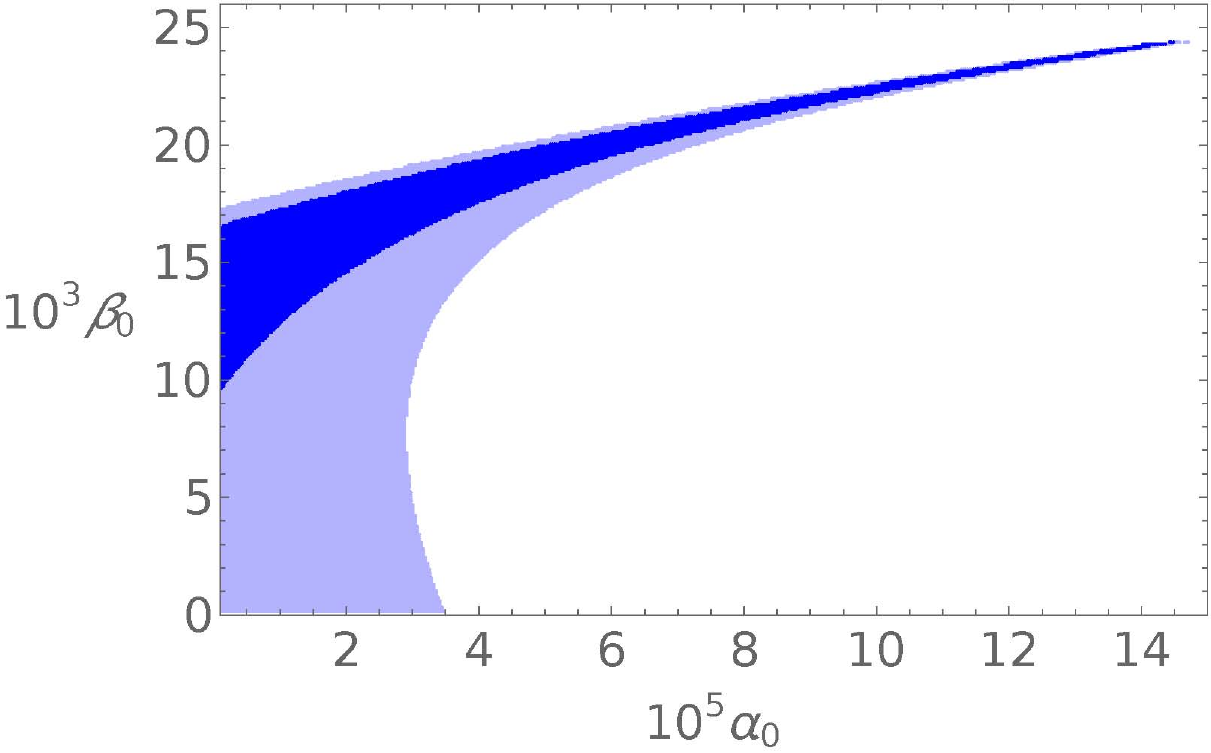} \quad
		\includegraphics[height=4.59cm]{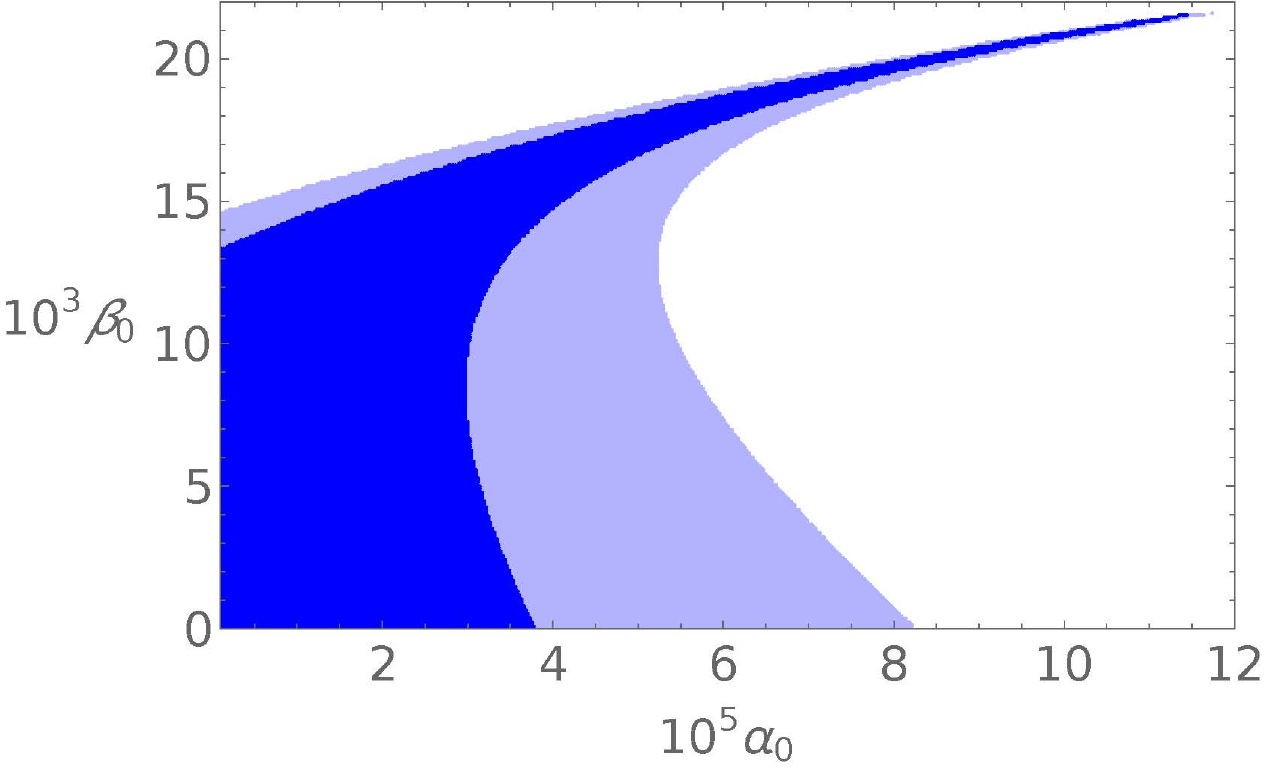}
		\caption{\label{fig:parameter-space} The regions in blue represent the allowed regions
			for the parameters $\alpha _{0}$ and $\beta _{0}$ in $68\%$ and $95\%$ C.L.,
			due to observational data from Planck plus BICEP3/Keck plus BAO \cite%
			{BICEPKeck:2021gln}. In the top graph we have the Plot for $N=52$, while in the
			bottom graph we have it for $N=59$. Note that the constraints for $N=59$ allow a
			larger region for the parameters $\alpha _{0}$ and $\beta _{0}$ in line with
			what we saw in the figure \ref{fig:nsxr-plane}, whose predictions for $N=59$ are
			more within the region of $68\%$ C.L. Note that for large values of $\alpha
			_{0}$ and $\beta _{0}$, around $\alpha _{0}=1.5\times 10^{-4}$ and $\beta
			_{0}=2.5\times 10^{-2}$ ($N=52$) and $\alpha _{0}=2.2\times 10^{-4}$ and $%
			\beta _{0}=1.2\times 10^{-2}$ ($N=59$), the predicted regions for the
			parameters converge to an asymptotic region. In this region, the values of $\alpha _{0}$ and $\beta
			_{0}$ suggest to keep a constraint. \qquad \qquad \qquad \qquad \qquad \qquad \qquad \qquad \qquad \qquad \qquad \qquad}		
	\end{center}
\end{figure}

Another plot developed, figure \ref{fig:parameter-space}, is the parameter space $%
\alpha _{0}\times \beta _{0}$ allowed by the observations. In the top graph of figure \ref{fig:parameter-space}, we
have the Plot for $N=52$, while in the bottom graph of figure \ref{fig:parameter-space}, we have it for $N=59$. The blue
regions represent the allowed regions for the $\alpha _{0}$ and $\beta _{0}$
parameters in $68\%$ and $95\%$ C.L.. Note that the Plot for $N=52$ gives us
a smaller region for the parameters if we compare it to the Plot for $N=59$.
In addition, we can see two regions on each of the Plots. One is
an approximated rectangular region completely within
the $95\%$ C.L. (for $N=52$, the sides correspond to $\alpha _{0}=2.8\times
10^{-5}$ and $\beta _{0}=1.7\times 10^{-2}$, and for $N=59$, $\alpha
_{0}=5.3\times 10^{-5}$ and $\beta _{0}=1.5\times 10^{-2}$), where the
parameters $\alpha _{0}$ and $\beta _{0}$ do not keep a dependency between
them, being able to assume any values independently. In this region of independence between the model parameters, we reproduced the results obtained in Refs. \cite{Huang:2013hsb,Castellanos:2018dub} for each model separately. The other region is the
asymptotic one for large values of $\alpha _{0}$ and $\beta _{0}$, whose occurrence suggests a dependence $\beta _{0} = \beta _{0} (\alpha _{0})$ between the parameters.	

\newpage

\section{Final comments \label{sec:comentarios-finais}}

The Starobinsky model is one of the most competitive candidates for
describing physical inflation. In addition to having a well-grounded
theoretical motivation, it better fits the recent observations \cite%
{Akrami:2018odb,BICEPKeck:2021gln}. Motivated by the success of such a
model, we propose to investigate inflation based on the higher-order
gravitational action characterized by the inclusion of all terms up to the
second-order correction involving only the scalar curvature, namely, the
terms $R^{2}$, $R^{3}$, and $R\square R$. In this sense, our proposed model
has two additional dimensionless parameters, $\alpha _{0}$, and $\beta _{0}$%
, whose values represent deviations from Starobinsky.

Unlike Ref. \cite{Cuzinatto:2018vjt}, whose multi-field treatment used to address the
term $R\square R$ gives us an inflation described by a scalar and a
vector field, here, when passing from the original frame to the representation in the Einstein
frame, the model is described through the dynamics of two scalar fields $%
\chi $ and $\lambda $, where only one of them is associated with a canonical
kinetic term, and whose potential is $V\left( \chi ,\lambda \right) $ given
in Eq. (\ref{eq:potencial-modelo}). The study of inflation in a Friedmann
background, through the analysis of the critical points and phase space of
the model, is essential to verify the existence of an attractor region
associated with the occurrence of an inflationary regime and to know if such
a regime has a graceful exit. We took as a basis the study of particular
cases developed in Refs. \cite{Cuzinatto:2018vjt,PhysRevD.105.063504},
which deal with the Starobinsky$+R\square R$ and Starobinsky$+R^{3}$
extensions. We saw that there is an attractor line near $\chi _{t} \simeq 0$%
, corresponding to the slow-roll inflation, for any value of $\alpha
_{0}<10^{-3}$ and $\beta _{0}<3\times 10^{-2}$. Furthermore, the occurrence
of such a physical inflation regime essentially depends on the initial
conditions for the $\chi $ field. If they are such that the $\chi $ field is
to the right of the critical point $P_{c}$, the value of $\chi $ increases
indefinitely, and inflation never ends. On the other hand, the occurrence of
a consistent physical inflationary regime that has a graceful exit
essentially requires that the initial conditions be such that $\chi <\chi
_{c}$, i.e., that it is to the left of the critical point $P_{c}$. Finally,
we conclude the background analysis with the study of inflation considering
the slow-roll approximation. By defining the slow-roll factor $\delta $,
which in our analysis is responsible for controlling the slow-roll
approximation order, we obtain all relevant quantities, such as $\varepsilon
$ and $\eta $, in the slow-roll leading order.

There is considerable literature about multi-field inflation models, which
we took into account to develop the analysis at the perturbative level \cite%
{Wands:2007bd,RevModPhys.78.537}. The equations of motion for the
scalar perturbations were obtained using the spatially flat gauge. By
writing the equations in the slow-roll leading order approximation, we saw
that the scalar perturbations of the metric are sub-dominant concerning the
perturbations $\delta \chi $ and $\delta \lambda $. At
this point, we performed a correct decomposition of the perturbations in the
tangent (adiabatic perturbations) and orthogonal (isocurvature
perturbations) directions to the phase space background trajectories. This
way, adiabatic $Q_{\sigma }$ and isocurvature $Q_{s}$ perturbations are completely separated. Such
a decomposition allows us to consistently establish the curvature
perturbation, which led us to obtain observational constraints different
from those obtained in Ref. \cite{Cuzinatto:2018vjt}. The action written in terms of $Q_{\sigma }$ and $Q_{s}$ makes it clear that there
are irremediably ghost-type instabilities in the model since the kinetic
terms have opposite signs. Next, we write the equations of motion in a
Mukhanov-Sasaki form in order to study their solutions. We obtained the
exact solutions for the perturbations through a linear combination of the
Hankel functions. Their analysis leads us to conclude that the isocurvature
perturbation associated with the ghost field is negligible after inflation and that the adiabatic one has the same behavior as in the case of a single-field inflation. All previous results were obtained considering the slow-roll approximation. Thus, a question that remains is whether the suppression of isocurvature perturbation holds beyond the slow-roll regime. This issue will be addressed in a further work.

Finally, we confront our model with recent observations from the Planck
satellite, BICEP3/Keck and BAO \cite{Akrami:2018odb,BICEPKeck:2021gln}, making use of a constraint on the number of $e$-folds $N$ of inflation ($%
52\leq N\leq 59$) based on reheating modeling \cite{PhysRevD.105.063504}.
For that, we made two types of Plots, namely, the usual $n_{s}\times
r_{0.002}$ plane and the parameter space $\alpha _{0}\times \beta _{0}$. In
this analysis, we have three parameters: $\alpha _{0}$, $\beta _{0}$, and $N$%
. Thus, to build the Plots, we set one of the parameters and vary the
others. Fixing the parameter $\beta _{0}$, we observe that the region
predicted by the model in the $n_{s}\times r_{0.002}$ plane shifts to the
left and slightly downwards. On the other hand, fixing the parameter $\alpha
_{0}$, we notice that the predicted region shifts to the right and slightly
upwards. By setting $\alpha _{0}=0$, we get the Starobinsky$+R\square R$
model. In this context, we saw that inconsistency in establishing the
curvature perturbation in Ref. \cite{Cuzinatto:2018vjt} led them to obtain
values higher than ours for the tensor-to-scalar ratio. In turn, by fixing
the number of $e$-folds $N$, we construct the parameter space $\alpha
_{0}\times \beta _{0}$ constrained by the observations. In general, the
model predictions are more in agreement with the observations for a number
of $e$-folds $N = 59$. Our analysis, conservatively, restrict the
parameters to maximum values of $\alpha _{0}\sim 10^{-4}$ and $\beta
_{0}\sim 10^{-2}$. It is also worth pointing out the behavior of the $\alpha
_{0}\times \beta _{0}$ parameter space. The $R^{3}$ and $R\square R$ terms are
second-order correction terms on energy scales and, therefore, should
contribute similarly to inflation. In this sense, the joint effect of such
terms is reflected in the plot $\alpha _{0}\times \beta _{0}$. In fact,
there is a considerable region in which the parameters do not depend on each
other and which we can associate with the models separately discussed in
Refs. \cite{Huang:2013hsb,PhysRevD.105.063504,Castellanos:2018dub,Cuzinatto:2018vjt}. However,
there is an asymptotic region for large values of $\alpha _{0}$ and $\beta
_{0}$, where such parameters seem to keep a constraint. In this particular region, a change in one of the parameters necessarily implies a change in the other, so that the possibility of a dependence $ \beta _{0} = \beta _{0} (\alpha _{0}) $ is something to be investigated. This is a topic that the authors will address in a future research.

\acknowledgments

G. Rodrigues-da-Silva thanks CAPES/UFRN-RN (Brazil) for financial support and L. G.
Medeiros acknowledges CNPq-Brazil (Grant No. 307901/2022-0) for partial financial support.

\appendix
\section{Determination of $\protect\chi _{t}$ and $\protect%
	\lambda $ \label{Ap6}}

The equations of motion of the model can be explicitly written in terms of
the fields and their first derivatives as%
\begin{gather}
\chi _{tt}+3h\chi _{t}-\frac{\beta _{0}}{6}e^{-\chi }\lambda _{t}{}^{2}+%
\frac{1}{3}e^{-2\chi }\lambda \left[ \left( 2-e^{\chi }\right) +\lambda +%
\frac{2}{3}\alpha _{0}\lambda ^{2}\right] =0,  \label{eq:ap-eq-qui} \\
\beta _{0}\left[ \lambda _{tt}-\left( \chi _{t}-3h\right) \lambda _{t}\right]
-\left[ 1-e^{-\chi }\left( 1+\lambda +\alpha _{0}\lambda ^{2}\right) \right]
=0,  \label{eq:ap-eq-lambda}
\end{gather}%
where%
\begin{equation}
h^{2}=\frac{1}{4}\chi _{t}^{2}-\frac{1}{12}\beta _{0}e^{-\chi }\lambda
_{t}^{2}+\frac{1}{6}e^{-2\chi }\lambda \left[ \left( e^{\chi }-1\right) -%
\frac{1}{2}\lambda -\frac{1}{3}\alpha _{0}\lambda ^{2}\right] .
\end{equation}

By studying the case where the model parameters behave as $\beta _{0}\sim
\delta \equiv e^{-\chi }$ and $\alpha _{0}\sim \delta ^{2}$, we assume the
quantities $\chi _{t}$ and $\lambda $ as follows, respectively,%
\begin{equation}
\chi _{t} \simeq c_{1}\delta +\beta _{0}\sum\limits_{n=0}^{\infty
}b_{n}\left( \beta _{0}\delta ^{-1}\right) ^{n}+\alpha _{0}\delta
^{-1}\sum\limits_{n=0}^{\infty }d_{n}\left( \beta _{0}\delta ^{-1}\right)
^{n},  \label{eq:chi-t_ansataz}
\end{equation}%
and%
\begin{equation}
\lambda \simeq \delta ^{-1}+\sum\limits_{n=0}^{\infty }g_{n}\left( \beta
_{0}\delta ^{-1}\right) ^{n}+\alpha _{0}\delta ^{-2}\sum_{n=0}^{\infty
}j_{n}\left( \beta _{0}\delta ^{-1}\right) ^{n},  \label{eq:lambda_ansataz}
\end{equation}%
where we explore several possibilities of construction of the quantities $%
\chi _{t}$ (in first-order) and $\lambda $ (up to zero-order). In addition
to the terms involving $\beta _{0}$ and $\alpha _{0}$ separately, we observe
the need to introduce crossed terms due to the non-linearity of gravitation.
We also notice fine limits when we take the particular cases $\alpha
_{0}\rightarrow 0$ (Starobinsky$+R\square R$) and $\beta _{0}\rightarrow 0$
(Starobinsky+$R^{3}$).

By finding the coefficients of the series Eqs. (\ref{eq:chi-t_ansataz}) and (%
\ref{eq:lambda_ansataz}), we substitute them in Eqs. (\ref{eq:ap-eq-qui})
and (\ref{eq:ap-eq-lambda}) and solve the corresponding systems of
equations. After a long calculation, we obtain that, for consistency with
the Starobinsky case, $g_{0}\simeq -1$, as well as finding the following $n$%
th coefficients of the series%
\begin{equation}
b_{n}\simeq -\frac{2}{3^{n+\frac{5}{2}}},\text{ for }n\geq 0,
\end{equation}%
\begin{equation}
d_{n}\simeq \frac{2}{3^{n+\frac{5}{2}}},\text{ for }n\geq 0,
\end{equation}%
\begin{equation}
g_{n}\simeq \frac{1}{3^{n}},\text{ for }n\geq 1,
\end{equation}%
\begin{equation}
j_{n}\simeq -\frac{1}{3^{n+1}},\text{ for }n\geq 1,
\end{equation}%
as well as%
\begin{equation}
c_{1}\simeq -\frac{2}{3^{\frac{3}{2}}}\text{ \ and \ }j_{0}\simeq -1.
\end{equation}

In possession of the $n$th coefficients, we can substitute them in the
quantities $\chi _{t}$ and $\lambda $. In this case, for $\chi _{t}$,%
\begin{equation}
\chi _{t}\simeq -\frac{2}{3^{\frac{3}{2}}}\delta \left[ 1+\frac{1}{3}\beta
_{0}\delta ^{-1}\sum\limits_{n=0}^{\infty }\left( \frac{\beta _{0}\delta
	^{-1}}{3}\right) ^{n}-\frac{1}{3}\alpha _{0}\delta
^{-2}\sum\limits_{n=0}^{\infty }\left( \frac{\beta _{0}\delta ^{-1}}{3}%
\right) ^{n}\right] ,
\end{equation}%
that converges with%
\begin{equation}
\frac{\beta _{0}\delta ^{-1}}{3}<1.
\end{equation}%
Thus,%
\begin{equation}
\chi _{t}\simeq -\frac{2\sqrt{3}}{3}\frac{\delta }{3-\beta _{0}\delta ^{-1}}%
\left( 1-\frac{1}{3}\alpha _{0}\delta ^{-2}\right) .
\end{equation}%
In turn, for $\lambda $, we have%
\begin{equation}
\lambda \simeq \delta ^{-1}-\frac{3-2\beta _{0}\delta ^{-1}}{3-\beta
	_{0}\delta ^{-1}}-\alpha _{0}\delta ^{-2}\left( 1+\frac{\frac{1}{3}\beta
	_{0}\delta ^{-1}}{3-\beta _{0}\delta ^{-1}}\right) .
\end{equation}%
The previous relation recovers the Starobinsky result%
\begin{equation}
\lambda \simeq \delta ^{-1}-1,
\end{equation}%
and Starobinsky+$R^{3}$, namely,%
\begin{equation}
\lambda \simeq \delta ^{-1}-1-\alpha _{0}\delta ^{-2}.
\end{equation}

\section{Coefficients in the action up to second-order for the perturbations $Q_{\sigma }$ and $Q_{s}$ \label{Ap:coefficients}}

\begin{widetext}

In this appendix, we present the non-trivial coefficients in the second-order
perturbed action for the adiabatic $Q_{\sigma }$ and isocurvature $Q_{s}$
perturbations, Eq. (\ref{eq:acao-qs-geral}).

For $C_{Q_{\sigma }^{2}}$, we have%
\begin{gather}
C_{Q_{\sigma }^{2}}=\left( \frac{1}{8\dot{\sigma}^{4}}\right) \left\{ 6\beta
_{0}^{3}e^{-3\chi }\dot{\lambda}^{6}-55\beta _{0}^{2}e^{-2\chi }\dot{\lambda}%
^{4}\dot{\chi}^{2}-162\dot{\chi}^{6}\right. +  \notag \\
+\left. 12\beta _{0}e^{-\chi }\left[ -\dot{\chi}^{2}\ddot{\lambda}^{2}+2\dot{%
	\lambda}\dot{\chi}\ddot{\lambda}\left( \dot{\chi}^{2}+\ddot{\chi}\right) +%
\dot{\lambda}^{2}\left( 13\dot{\chi}^{4}-2\dot{\chi}^{2}\ddot{\chi}-\ddot{%
	\chi}^{2}\right) \right] \right\} +  \notag \\
-\frac{1}{4}H^{-1}\left( \frac{4\kappa _{0}e^{-2\chi }}{9}\right) \left\{ 3%
\left[ -1+e^{\chi }-\lambda \left( 1+\alpha _{0}\lambda \right) \right] \dot{%
	\lambda}+\lambda \left[ 6-3e^{\chi }+\lambda \left( 3+2\alpha _{0}\lambda
\right) \right] \dot{\chi}\right\} +  \notag \\
+\left( -\frac{1}{4}H^{-1}\right) ^{2}\left( e^{-2\chi }\dot{\sigma}%
^{4}\right) +  \notag \\
-\left( \frac{\kappa _{0}e^{-2\chi }}{6\dot{\sigma}^{2}}\right) \left\{
\left( 3+6\alpha _{0}\lambda \right) \dot{\lambda}^{2}+6\left[ e^{\chi
}-2\left( 1+\lambda +\alpha _{0}\lambda ^{2}\right) \right] \dot{\chi}\dot{%
	\lambda}+\lambda \left( 12-3e^{\chi }+6\lambda +4\alpha _{0}\lambda
^{2}\right) \dot{\chi}^{2}\right\} ,
\end{gather}%
for $C_{Q_{s}^{2}}$,%
\begin{gather}
C_{Q_{s}^{2}}=\left( \frac{1}{24\dot{\sigma}^{4}}\right) \left\{ 2\beta
_{0}^{3}e^{-3\chi }\dot{\lambda}^{6}-27\dot{\chi}^{6}-6\beta
_{0}^{2}e^{-2\chi }\dot{\lambda}^{3}\left[ -4\dot{\chi}\ddot{\lambda}+\dot{%
	\lambda}\left( 5\dot{\chi}^{2}+4\ddot{\chi}\right) \right] \right. +  \notag
\\
\left. +36\beta _{0}e^{-\chi }\left[ \dot{\chi}^{2}\ddot{\lambda}^{2}-2\dot{%
	\chi}\dot{\lambda}\ddot{\lambda}\left( \dot{\chi}^{2}+\ddot{\chi}\right) +%
\dot{\lambda}^{2}\left( 2\dot{\chi}^{4}+2\dot{\chi}^{2}\ddot{\chi}+\ddot{\chi%
}^{2}\right) \right] \right\} +  \notag \\
+\left( \frac{\kappa _{0}}{18\beta _{0}e^{\chi }\dot{\sigma}^{2}}\right)
\left\{ 2\beta _{0}^{2}e^{-2\chi }\lambda \left[ 6+\lambda \left( 3+2\alpha
_{0}\lambda \right) \right] \dot{\lambda}^{2}+9\dot{\chi}\left[ 2\beta _{0}%
\dot{\lambda}+3\left( 1+2\alpha _{0}\lambda \right) \dot{\chi}\right]
\right. +  \notag \\
\left. -3\beta _{0}e^{-\chi }\dot{\lambda}\left[ \beta _{0}\lambda \dot{%
	\lambda}+12\left( 1+\lambda +\alpha _{0}\lambda ^{2}\right) \dot{\chi}\right]
\right\} ,
\end{gather}%
for $C_{Q_{\sigma }Q_{s}}$,

\begin{gather}
C_{Q_{\sigma }Q_{s}}=\left( \frac{\sqrt{3\beta _{0}e^{-\chi }}\dot{\chi}}{2%
	\dot{\sigma}^{4}}\right) \left\{ -3\dot{\chi}^{2}\left[ -\dot{\chi}\ddot{%
	\lambda}+\dot{\lambda}\left( \dot{\chi}^{2}+\ddot{\chi}\right) \right]
+\beta _{0}e^{-\chi }\dot{\lambda}^{2}\left[ -3\dot{\chi}\ddot{\lambda}+\dot{%
	\lambda}\left( 2\dot{\chi}^{2}+3\ddot{\chi}\right) \right] \right\} +  \notag
\\
-\frac{1}{4}H^{-1}\left( \frac{1}{\dot{\sigma}^{2}}\sqrt{\frac{\beta
		_{0}e^{-\chi }}{3}}\right) \left\{ \beta _{0}^{2}e^{-2\chi }\dot{\lambda}%
^{5}+6\dot{\chi}^{2}\left[ -\dot{\chi}\ddot{\lambda}+\dot{\lambda}\left( 2%
\dot{\chi}^{2}+\ddot{\chi}\right) \right] -3\beta _{0}e^{-\chi }\dot{\lambda}%
^{2}\left[ -2\dot{\chi}\ddot{\lambda}+\dot{\lambda}\left( 3\dot{\chi}^{2}+2%
\ddot{\chi}\right) \right] \right\} +  \notag \\
-\frac{1}{4}H^{-1}\left( \frac{4\kappa _{0}e^{-3\chi }}{9\sqrt{3\beta
		_{0}e^{-\chi }}}\right) \left\{ -\beta _{0}\lambda \left[ 6+\lambda \left(
3+2\alpha _{0}\lambda \right) \right] \dot{\lambda}-9e^{2\chi }\dot{\chi}%
+3e^{\chi }\left[ \beta _{0}\lambda \dot{\lambda}+3\left( 1+\lambda +\alpha
_{0}\lambda ^{2}\right) \dot{\chi}\right] \right\} +  \notag \\
+\left( \frac{\kappa _{0}e^{-3\chi }}{3\dot{\sigma}^{2}\sqrt{3\beta
		_{0}e^{-\chi }}}\right) \left\{ -3\beta _{0}\left[ e^{\chi }-2\left(
1+\lambda +\alpha _{0}\lambda ^{2}\right) \right] \dot{\lambda}^{2}\right. +
\notag \\
\left. \left\{ -3e^{\chi }\left[ 3+\left( 6\alpha _{0}-\beta _{0}\right)
\lambda \right] -2\beta _{0}\lambda \left[ 6+\lambda \left( 3+2\alpha
_{0}\lambda \right) \right] \right\} \dot{\lambda}\dot{\chi}-9e^{\chi }\left[
e^{\chi }-2\left( 1+\lambda +\alpha _{0}\lambda ^{2}\right) \right] \dot{\chi%
}^{2}\right\} ,
\end{gather}%
for $C_{\dot{Q}_{\sigma }Q_{s}}$,%
\begin{equation}
C_{\dot{Q}_{\sigma }Q_{s}}=-\left( \frac{1}{\dot{\sigma}^{2}}\sqrt{\frac{%
		\beta _{0}e^{-\chi }}{3}}\right) \left[ \beta _{0}e^{-\chi }\dot{\lambda}%
^{3}+3\dot{\chi}\ddot{\lambda}-3\dot{\lambda}\left( \dot{\chi}^{2}+\ddot{\chi%
}\right) \right] ,
\end{equation}%
for $C_{Q_{\sigma }\dot{Q}_{s}}$,%
\begin{equation}
C_{Q_{\sigma }\dot{Q}_{s}}=\left( \frac{\sqrt{3\beta _{0}e^{-\chi }}}{\dot{%
		\sigma}^{2}}\right) \left[ \dot{\chi}\ddot{\lambda}-\dot{\lambda}\left( \dot{%
	\chi}^{2}+\ddot{\chi}\right) \right] ,
\end{equation}%
for $C_{Q_{\sigma }\dot{Q}_{\sigma }}$,%
\begin{eqnarray}
C_{Q_{\sigma }\dot{Q}_{\sigma }} &=&\left( \frac{\beta _{0}e^{-\chi }\dot{%
		\lambda}^{2}}{2\dot{\sigma}^{2}}\right) \dot{\chi}+  \notag \\
&&-\frac{1}{4H}\left( \frac{1}{\dot{\sigma}^{2}}\right) \left\{ \beta
_{0}e^{-\chi }\left[ \beta _{0}e^{-\chi }\left( 1+\beta _{0}e^{-\chi
}\right) \dot{\lambda}^{4}-12\beta _{0}e^{-\chi }\dot{\lambda}^{2}\dot{\chi}%
^{2}+18\dot{\chi}^{4}\right] \right\} .
\end{eqnarray}%
and finally, for $C_{Q_{s}\dot{Q}_{s}}$,%
\begin{equation}
C_{Q_{s}\dot{Q}_{s}}=\frac{1}{2}\left( -1+\frac{\beta _{0}e^{-\chi }\dot{%
		\lambda}^{2}}{\dot{\sigma}^{2}}\right) \dot{\chi}.
\end{equation}

\end{widetext}

\section{Restriction for $N$ based on a reheating modeling
	\label{Ap:vinculo-reaquecimento}}

In Ref. \cite{PhysRevD.105.063504}, we saw that in the particular case of $%
\beta _{0}=0$, the uncertainty in the number of $e$-folds $N_{k}$ for the
reference scale $k=0.002$\emph{\ }$Mpc^{-1}$, defines the interval%
\begin{equation}
52\leq N_{k}\leq 59.  \label{range}
\end{equation}%
This result was obtained through a very general modeling of the reheating
phase considering that at least the fields of the standard model of
particles are present during this phase.

The basic equations of the performed modeling are \cite{PhysRevD.105.063504}

\begin{align}
	N_{re}& =\frac{4}{3\left( w_{a}-\frac{1}{3}\right) }\left\{ N_{k}+\ln \left( 
	\frac{\rho _{e}^{1/4}}{H_{k}}\right) +\right.  \notag \\
	& \left. +\ln \left[ \left( \frac{k}{a_{0}T_{0}}\right) \left( \frac{30}{\pi
		^{2}}\right) ^{\frac{1}{4}}\left( \frac{g_{re}}{g_{0}^{4}}\right) ^{\frac{1}{%
			12}}\right] \right\} ,  \label{Nre}
\end{align}%
\begin{equation}
	T_{re}=\left( \frac{30\rho _{e}}{g_{re}\pi ^{2}}\right) ^{\frac{1}{4}}\exp %
	\left[ -\frac{3}{4}\left( 1+w_{a}\right) N_{re}\right] ,  \label{Tre}
\end{equation}%
where $N_{re}$ is the number of $e$-folds of the reheating, $T_{re}$ is the
temperature of the reheating, $w_{a}$ is the average of the effective equation
of state during the reheating, $\rho _{e}$ is the energy density at the end of
inflation, $T_{0}$ is the CMB temperature in the present day and $g_{re}$ and $%
g_{0} $ are the relativistic degrees of freedom in reheating and in the present
day, respectively. The range (\ref{range}) is obtained by imposing the
bounds $N_{re}\geq 0$ and $T_{re}\geq T_{re}^{\left( \min \right) }$, where $%
T_{re}^{\left( \min \right) }$ is determined from the decay of the inflaton
field in the matter fields \cite{PhysRevD.105.063504}.

Based on the previous equations and results, it is relatively simple to
conclude that the range obtained in (\ref{range}) also applies to $\beta
_{0}\neq 0$. The main point is that at the end and after inflation where $%
\chi <1\Rightarrow \delta \sim 1$, the proposed model behaves essentially
like the Starobinsky model.\footnote{%
	This occurs because we consider $\alpha _{0}<10^{-3}$ and $\beta
	_{0}<3\times 10^{-2}$.} Thus, the only term present in Eqs. (\ref{Nre}) and
(\ref{Tre}), which may have some relevance when $\beta _{0}\neq 0$ is the
term $H_{k}$. However, by (\ref{h2 aprox}), we see that $H_{k}$ weakly
depends on $\beta _{0}$ even taking into account slow-roll first-order
corrections. With this, we conclude that for our model, it is licit to
consider the range for the number of $e$-folds $N$ as given by Eq. (\ref%
{range}).

\bibliographystyle{apsrev4-2}
\bibliography{references-inflation-R3}

\begin{thebibliography}{76}%
\makeatletter
\providecommand \@ifxundefined [1]{%
 \@ifx{#1\undefined}
}%
\providecommand \@ifnum [1]{%
 \ifnum #1\expandafter \@firstoftwo
 \else \expandafter \@secondoftwo
 \fi
}%
\providecommand \@ifx [1]{%
 \ifx #1\expandafter \@firstoftwo
 \else \expandafter \@secondoftwo
 \fi
}%
\providecommand \natexlab [1]{#1}%
\providecommand \enquote  [1]{``#1''}%
\providecommand \bibnamefont  [1]{#1}%
\providecommand \bibfnamefont [1]{#1}%
\providecommand \citenamefont [1]{#1}%
\providecommand \href@noop [0]{\@secondoftwo}%
\providecommand \href [0]{\begingroup \@sanitize@url \@href}%
\providecommand \@href[1]{\@@startlink{#1}\@@href}%
\providecommand \@@href[1]{\endgroup#1\@@endlink}%
\providecommand \@sanitize@url [0]{\catcode `\\12\catcode `\$12\catcode
  `\&12\catcode `\#12\catcode `\^12\catcode `\_12\catcode `\%12\relax}%
\providecommand \@@startlink[1]{}%
\providecommand \@@endlink[0]{}%
\providecommand \url  [0]{\begingroup\@sanitize@url \@url }%
\providecommand \@url [1]{\endgroup\@href {#1}{\urlprefix }}%
\providecommand \urlprefix  [0]{URL }%
\providecommand \Eprint [0]{\href }%
\providecommand \doibase [0]{https://doi.org/}%
\providecommand \selectlanguage [0]{\@gobble}%
\providecommand \bibinfo  [0]{\@secondoftwo}%
\providecommand \bibfield  [0]{\@secondoftwo}%
\providecommand \translation [1]{[#1]}%
\providecommand \BibitemOpen [0]{}%
\providecommand \bibitemStop [0]{}%
\providecommand \bibitemNoStop [0]{.\EOS\space}%
\providecommand \EOS [0]{\spacefactor3000\relax}%
\providecommand \BibitemShut  [1]{\csname bibitem#1\endcsname}%
\let\auto@bib@innerbib\@empty
\bibitem [{\citenamefont {Clifton}\ \emph {et~al.}(2012)\citenamefont
  {Clifton}, \citenamefont {Ferreira}, \citenamefont {Padilla},\ and\
  \citenamefont {Skordis}}]{Clifton:2011jh}%
  \BibitemOpen
  \bibfield  {author} {\bibinfo {author} {\bibfnamefont {T.}~\bibnamefont
  {Clifton}}, \bibinfo {author} {\bibfnamefont {P.~G.}\ \bibnamefont
  {Ferreira}}, \bibinfo {author} {\bibfnamefont {A.}~\bibnamefont {Padilla}},\
  and\ \bibinfo {author} {\bibfnamefont {C.}~\bibnamefont {Skordis}},\ }\href
  {https://doi.org/10.1016/j.physrep.2012.01.001} {\bibfield  {journal}
  {\bibinfo  {journal} {Phys. Rept.}\ }\textbf {\bibinfo {volume} {513}},\
  \bibinfo {pages} {1} (\bibinfo {year} {2012})},\ \Eprint
  {https://arxiv.org/abs/1106.2476} {arXiv:1106.2476 [astro-ph.CO]}
  \BibitemShut {NoStop}%
\bibitem [{\citenamefont {Hehl}\ \emph {et~al.}(1976)\citenamefont {Hehl},
  \citenamefont {von~der Heyde}, \citenamefont {Kerlick},\ and\ \citenamefont
  {Nester}}]{RevModPhys.48.393}%
  \BibitemOpen
  \bibfield  {author} {\bibinfo {author} {\bibfnamefont {F.~W.}\ \bibnamefont
  {Hehl}}, \bibinfo {author} {\bibfnamefont {P.}~\bibnamefont {von~der Heyde}},
  \bibinfo {author} {\bibfnamefont {G.~D.}\ \bibnamefont {Kerlick}},\ and\
  \bibinfo {author} {\bibfnamefont {J.~M.}\ \bibnamefont {Nester}},\ }\href
  {https://doi.org/10.1103/RevModPhys.48.393} {\bibfield  {journal} {\bibinfo
  {journal} {Rev. Mod. Phys.}\ }\textbf {\bibinfo {volume} {48}},\ \bibinfo
  {pages} {393} (\bibinfo {year} {1976})}\BibitemShut {NoStop}%
\bibitem [{\citenamefont {Horndeski}(1974)}]{Horndeski:1974wa}%
  \BibitemOpen
  \bibfield  {author} {\bibinfo {author} {\bibfnamefont {G.~W.}\ \bibnamefont
  {Horndeski}},\ }\href {https://doi.org/10.1007/BF01807638} {\bibfield
  {journal} {\bibinfo  {journal} {Int. J. Theor. Phys.}\ }\textbf {\bibinfo
  {volume} {10}},\ \bibinfo {pages} {363} (\bibinfo {year} {1974})}\BibitemShut
  {NoStop}%
\bibitem [{\citenamefont {Gottlober}\ \emph {et~al.}(1990)\citenamefont
  {Gottlober}, \citenamefont {Schmidt},\ and\ \citenamefont
  {Starobinsky}}]{Gottlober:1989ww}%
  \BibitemOpen
  \bibfield  {author} {\bibinfo {author} {\bibfnamefont {S.}~\bibnamefont
  {Gottlober}}, \bibinfo {author} {\bibfnamefont {H.~J.}\ \bibnamefont
  {Schmidt}},\ and\ \bibinfo {author} {\bibfnamefont {A.~A.}\ \bibnamefont
  {Starobinsky}},\ }\href {https://doi.org/10.1088/0264-9381/7/5/018}
  {\bibfield  {journal} {\bibinfo  {journal} {Class. Quant. Grav.}\ }\textbf
  {\bibinfo {volume} {7}},\ \bibinfo {pages} {893} (\bibinfo {year}
  {1990})}\BibitemShut {NoStop}%
\bibitem [{\citenamefont {{Sotiriou}}\ and\ \citenamefont
  {{Faraoni}}(2010)}]{RevModPhys.82.451}%
  \BibitemOpen
  \bibfield  {author} {\bibinfo {author} {\bibfnamefont {T.~P.}\ \bibnamefont
  {{Sotiriou}}}\ and\ \bibinfo {author} {\bibfnamefont {V.}~\bibnamefont
  {{Faraoni}}},\ }\href {https://doi.org/10.1103/RevModPhys.82.451} {\bibfield
  {journal} {\bibinfo  {journal} {Rev. Mod. Phys.}\ }\textbf {\bibinfo {volume}
  {82}},\ \bibinfo {pages} {451} (\bibinfo {year} {2010})}\BibitemShut
  {NoStop}%
\bibitem [{\citenamefont {Nojiri}\ and\ \citenamefont
  {Odintsov}(2011)}]{Nojiri:2010wj}%
  \BibitemOpen
  \bibfield  {author} {\bibinfo {author} {\bibfnamefont {S.}~\bibnamefont
  {Nojiri}}\ and\ \bibinfo {author} {\bibfnamefont {S.~D.}\ \bibnamefont
  {Odintsov}},\ }\href {https://doi.org/10.1016/j.physrep.2011.04.001}
  {\bibfield  {journal} {\bibinfo  {journal} {Phys. Rept.}\ }\textbf {\bibinfo
  {volume} {505}},\ \bibinfo {pages} {59} (\bibinfo {year} {2011})},\ \Eprint
  {https://arxiv.org/abs/1011.0544} {arXiv:1011.0544 [gr-qc]} \BibitemShut
  {NoStop}%
\bibitem [{\citenamefont {De~Felice}\ and\ \citenamefont
  {Tsujikawa}(2010)}]{DeFelice:2010aj}%
  \BibitemOpen
  \bibfield  {author} {\bibinfo {author} {\bibfnamefont {A.}~\bibnamefont
  {De~Felice}}\ and\ \bibinfo {author} {\bibfnamefont {S.}~\bibnamefont
  {Tsujikawa}},\ }\href {https://doi.org/10.12942/lrr-2010-3} {\bibfield
  {journal} {\bibinfo  {journal} {Living Rev. Rel.}\ }\textbf {\bibinfo
  {volume} {13}},\ \bibinfo {pages} {3} (\bibinfo {year} {2010})},\ \Eprint
  {https://arxiv.org/abs/1002.4928} {arXiv:1002.4928 [gr-qc]} \BibitemShut
  {NoStop}%
\bibitem [{\citenamefont {Capozziello}\ and\ \citenamefont
  {De~Laurentis}(2011)}]{Capozziello:2011et}%
  \BibitemOpen
  \bibfield  {author} {\bibinfo {author} {\bibfnamefont {S.}~\bibnamefont
  {Capozziello}}\ and\ \bibinfo {author} {\bibfnamefont {M.}~\bibnamefont
  {De~Laurentis}},\ }\href {https://doi.org/10.1016/j.physrep.2011.09.003}
  {\bibfield  {journal} {\bibinfo  {journal} {Phys. Rept.}\ }\textbf {\bibinfo
  {volume} {509}},\ \bibinfo {pages} {167} (\bibinfo {year} {2011})},\ \Eprint
  {https://arxiv.org/abs/1108.6266} {arXiv:1108.6266 [gr-qc]} \BibitemShut
  {NoStop}%
\bibitem [{\citenamefont {Cuzinatto}\ \emph {et~al.}(2016)\citenamefont
  {Cuzinatto}, \citenamefont {de~Melo}, \citenamefont {Medeiros},\ and\
  \citenamefont {Pompeia}}]{Cuzinatto:2016ehv}%
  \BibitemOpen
  \bibfield  {author} {\bibinfo {author} {\bibfnamefont {R.~R.}\ \bibnamefont
  {Cuzinatto}}, \bibinfo {author} {\bibfnamefont {C.~A.~M.}\ \bibnamefont
  {de~Melo}}, \bibinfo {author} {\bibfnamefont {L.~G.}\ \bibnamefont
  {Medeiros}},\ and\ \bibinfo {author} {\bibfnamefont {P.~J.}\ \bibnamefont
  {Pompeia}},\ }\href {https://doi.org/10.1103/PhysRevD.93.124034,
  10.1103/PhysRevD.98.029901} {\bibfield  {journal} {\bibinfo  {journal} {Phys.
  Rev.}\ }\textbf {\bibinfo {volume} {D93}},\ \bibinfo {pages} {124034}
  (\bibinfo {year} {2016})},\ \bibinfo {note} {[Erratum: Phys.
  Rev.D98,no.2,029901(2018)]},\ \Eprint {https://arxiv.org/abs/1603.01563}
  {arXiv:1603.01563 [gr-qc]} \BibitemShut {NoStop}%
\bibitem [{\citenamefont {Nojiri}\ \emph {et~al.}(2017)\citenamefont {Nojiri},
  \citenamefont {Odintsov},\ and\ \citenamefont {Oikonomou}}]{Nojiri:2017ncd}%
  \BibitemOpen
  \bibfield  {author} {\bibinfo {author} {\bibfnamefont {S.}~\bibnamefont
  {Nojiri}}, \bibinfo {author} {\bibfnamefont {S.~D.}\ \bibnamefont
  {Odintsov}},\ and\ \bibinfo {author} {\bibfnamefont {V.~K.}\ \bibnamefont
  {Oikonomou}},\ }\href {https://doi.org/10.1016/j.physrep.2017.06.001}
  {\bibfield  {journal} {\bibinfo  {journal} {Phys. Rept.}\ }\textbf {\bibinfo
  {volume} {692}},\ \bibinfo {pages} {1} (\bibinfo {year} {2017})},\ \Eprint
  {https://arxiv.org/abs/1705.11098} {arXiv:1705.11098 [gr-qc]} \BibitemShut
  {NoStop}%
\bibitem [{\citenamefont {Stelle}(1978)}]{Stelle:1977ry}%
  \BibitemOpen
  \bibfield  {author} {\bibinfo {author} {\bibfnamefont {K.~S.}\ \bibnamefont
  {Stelle}},\ }\href {https://doi.org/10.1007/BF00760427} {\bibfield  {journal}
  {\bibinfo  {journal} {Gen. Rel. Grav.}\ }\textbf {\bibinfo {volume} {9}},\
  \bibinfo {pages} {353} (\bibinfo {year} {1978})}\BibitemShut {NoStop}%
\bibitem [{\citenamefont {Nelson}(2010)}]{Nelson:2010ig}%
  \BibitemOpen
  \bibfield  {author} {\bibinfo {author} {\bibfnamefont {W.}~\bibnamefont
  {Nelson}},\ }\href {https://doi.org/10.1103/PhysRevD.82.104026} {\bibfield
  {journal} {\bibinfo  {journal} {Phys. Rev.}\ }\textbf {\bibinfo {volume}
  {D82}},\ \bibinfo {pages} {104026} (\bibinfo {year} {2010})},\ \Eprint
  {https://arxiv.org/abs/1010.3986} {arXiv:1010.3986 [gr-qc]} \BibitemShut
  {NoStop}%
\bibitem [{\citenamefont {L\"u}\ \emph {et~al.}(2015)\citenamefont {L\"u},
  \citenamefont {Perkins}, \citenamefont {Pope},\ and\ \citenamefont
  {Stelle}}]{PhysRevD.92.124019}%
  \BibitemOpen
  \bibfield  {author} {\bibinfo {author} {\bibfnamefont {H.}~\bibnamefont
  {L\"u}}, \bibinfo {author} {\bibfnamefont {A.}~\bibnamefont {Perkins}},
  \bibinfo {author} {\bibfnamefont {C.~N.}\ \bibnamefont {Pope}},\ and\
  \bibinfo {author} {\bibfnamefont {K.~S.}\ \bibnamefont {Stelle}},\ }\href
  {https://doi.org/10.1103/PhysRevD.92.124019} {\bibfield  {journal} {\bibinfo
  {journal} {Phys. Rev. D}\ }\textbf {\bibinfo {volume} {92}},\ \bibinfo
  {pages} {124019} (\bibinfo {year} {2015})}\BibitemShut {NoStop}%
\bibitem [{\citenamefont {Lu}\ \emph {et~al.}(2015)\citenamefont {Lu},
  \citenamefont {Perkins}, \citenamefont {Pope},\ and\ \citenamefont
  {Stelle}}]{Lu:2015cqa}%
  \BibitemOpen
  \bibfield  {author} {\bibinfo {author} {\bibfnamefont {H.}~\bibnamefont
  {Lu}}, \bibinfo {author} {\bibfnamefont {A.}~\bibnamefont {Perkins}},
  \bibinfo {author} {\bibfnamefont {C.~N.}\ \bibnamefont {Pope}},\ and\
  \bibinfo {author} {\bibfnamefont {K.~S.}\ \bibnamefont {Stelle}},\ }\href
  {https://doi.org/10.1103/PhysRevLett.114.171601} {\bibfield  {journal}
  {\bibinfo  {journal} {Phys. Rev. Lett.}\ }\textbf {\bibinfo {volume} {114}},\
  \bibinfo {pages} {171601} (\bibinfo {year} {2015})},\ \Eprint
  {https://arxiv.org/abs/1502.01028} {arXiv:1502.01028 [hep-th]} \BibitemShut
  {NoStop}%
\bibitem [{\citenamefont {Kokkotas}\ \emph {et~al.}(2017)\citenamefont
  {Kokkotas}, \citenamefont {Konoplya},\ and\ \citenamefont
  {Zhidenko}}]{Kokkotas:2017zwt}%
  \BibitemOpen
  \bibfield  {author} {\bibinfo {author} {\bibfnamefont {K.}~\bibnamefont
  {Kokkotas}}, \bibinfo {author} {\bibfnamefont {R.~A.}\ \bibnamefont
  {Konoplya}},\ and\ \bibinfo {author} {\bibfnamefont {A.}~\bibnamefont
  {Zhidenko}},\ }\href {https://doi.org/10.1103/PhysRevD.96.064007} {\bibfield
  {journal} {\bibinfo  {journal} {Phys. Rev.}\ }\textbf {\bibinfo {volume}
  {D96}},\ \bibinfo {pages} {064007} (\bibinfo {year} {2017})},\ \Eprint
  {https://arxiv.org/abs/1705.09875} {arXiv:1705.09875 [gr-qc]} \BibitemShut
  {NoStop}%
\bibitem [{\citenamefont {Bueno}\ and\ \citenamefont
  {Cano}(2017)}]{Bueno:2017sui}%
  \BibitemOpen
  \bibfield  {author} {\bibinfo {author} {\bibfnamefont {P.}~\bibnamefont
  {Bueno}}\ and\ \bibinfo {author} {\bibfnamefont {P.~A.}\ \bibnamefont
  {Cano}},\ }\href {https://doi.org/10.1088/1361-6382/aa8056} {\bibfield
  {journal} {\bibinfo  {journal} {Class. Quant. Grav.}\ }\textbf {\bibinfo
  {volume} {34}},\ \bibinfo {pages} {175008} (\bibinfo {year} {2017})},\
  \Eprint {https://arxiv.org/abs/1703.04625} {arXiv:1703.04625 [hep-th]}
  \BibitemShut {NoStop}%
\bibitem [{\citenamefont {Goldstein}\ and\ \citenamefont
  {Mashiyane}(2018)}]{Goldstein:2017rxn}%
  \BibitemOpen
  \bibfield  {author} {\bibinfo {author} {\bibfnamefont {K.}~\bibnamefont
  {Goldstein}}\ and\ \bibinfo {author} {\bibfnamefont {J.~J.}\ \bibnamefont
  {Mashiyane}},\ }\href {https://doi.org/10.1103/PhysRevD.97.024015} {\bibfield
   {journal} {\bibinfo  {journal} {Phys. Rev.}\ }\textbf {\bibinfo {volume}
  {D97}},\ \bibinfo {pages} {024015} (\bibinfo {year} {2018})},\ \Eprint
  {https://arxiv.org/abs/1703.02803} {arXiv:1703.02803 [hep-th]} \BibitemShut
  {NoStop}%
\bibitem [{\citenamefont {Podolsky}\ \emph {et~al.}(2018)\citenamefont
  {Podolsky}, \citenamefont {Svarc}, \citenamefont {Pravda},\ and\
  \citenamefont {Pravdova}}]{Podolsky:2018pfe}%
  \BibitemOpen
  \bibfield  {author} {\bibinfo {author} {\bibfnamefont {J.}~\bibnamefont
  {Podolsky}}, \bibinfo {author} {\bibfnamefont {R.}~\bibnamefont {Svarc}},
  \bibinfo {author} {\bibfnamefont {V.}~\bibnamefont {Pravda}},\ and\ \bibinfo
  {author} {\bibfnamefont {A.}~\bibnamefont {Pravdova}},\ }\href
  {https://doi.org/10.1103/PhysRevD.98.021502} {\bibfield  {journal} {\bibinfo
  {journal} {Phys. Rev.}\ }\textbf {\bibinfo {volume} {D98}},\ \bibinfo {pages}
  {021502} (\bibinfo {year} {2018})},\ \Eprint
  {https://arxiv.org/abs/1806.08209} {arXiv:1806.08209 [gr-qc]} \BibitemShut
  {NoStop}%
\bibitem [{\citenamefont {Rodrigues-da Silva}\ and\ \citenamefont
  {Medeiros}(2020)}]{Rodrigues-da-Silva:2020cpd}%
  \BibitemOpen
  \bibfield  {author} {\bibinfo {author} {\bibfnamefont {G.}~\bibnamefont
  {Rodrigues-da Silva}}\ and\ \bibinfo {author} {\bibfnamefont
  {L.}~\bibnamefont {Medeiros}},\ }\href
  {https://doi.org/10.1103/PhysRevD.101.124061} {\bibfield  {journal} {\bibinfo
   {journal} {Phys. Rev. D}\ }\textbf {\bibinfo {volume} {101}},\ \bibinfo
  {pages} {124061} (\bibinfo {year} {2020})},\ \Eprint
  {https://arxiv.org/abs/2004.04878} {arXiv:2004.04878 [gr-qc]} \BibitemShut
  {NoStop}%
\bibitem [{\citenamefont {Accioly}\ \emph {et~al.}(2017)\citenamefont
  {Accioly}, \citenamefont {Giacchini},\ and\ \citenamefont
  {Shapiro}}]{Accioly:2016qeb}%
  \BibitemOpen
  \bibfield  {author} {\bibinfo {author} {\bibfnamefont {A.}~\bibnamefont
  {Accioly}}, \bibinfo {author} {\bibfnamefont {B.~L.}\ \bibnamefont
  {Giacchini}},\ and\ \bibinfo {author} {\bibfnamefont {I.~L.}\ \bibnamefont
  {Shapiro}},\ }\href {https://doi.org/10.1103/PhysRevD.96.104004} {\bibfield
  {journal} {\bibinfo  {journal} {Phys. Rev.}\ }\textbf {\bibinfo {volume}
  {D96}},\ \bibinfo {pages} {104004} (\bibinfo {year} {2017})},\ \Eprint
  {https://arxiv.org/abs/1610.05260} {arXiv:1610.05260 [gr-qc]} \BibitemShut
  {NoStop}%
\bibitem [{\citenamefont {Giacchini}(2017)}]{GIACCHINI2017306}%
  \BibitemOpen
  \bibfield  {author} {\bibinfo {author} {\bibfnamefont {B.~L.}\ \bibnamefont
  {Giacchini}},\ }\href
  {https://doi.org/https://doi.org/10.1016/j.physletb.2017.01.019} {\bibfield
  {journal} {\bibinfo  {journal} {Physics Letters B}\ }\textbf {\bibinfo
  {volume} {766}},\ \bibinfo {pages} {306 } (\bibinfo {year}
  {2017})}\BibitemShut {NoStop}%
\bibitem [{\citenamefont {Giacchini}\ and\ \citenamefont
  {de~Paula~Netto}(2019)}]{Giacchini:2018gxp}%
  \BibitemOpen
  \bibfield  {author} {\bibinfo {author} {\bibfnamefont {B.~L.}\ \bibnamefont
  {Giacchini}}\ and\ \bibinfo {author} {\bibfnamefont {T.}~\bibnamefont
  {de~Paula~Netto}},\ }\href {https://doi.org/10.1140/epjc/s10052-019-6727-2}
  {\bibfield  {journal} {\bibinfo  {journal} {Eur. Phys. J.}\ }\textbf
  {\bibinfo {volume} {C79}},\ \bibinfo {pages} {217} (\bibinfo {year}
  {2019})},\ \Eprint {https://arxiv.org/abs/1806.05664} {arXiv:1806.05664
  [gr-qc]} \BibitemShut {NoStop}%
\bibitem [{\citenamefont {Berry}\ and\ \citenamefont
  {Gair}(2011)}]{Berry:2011pb}%
  \BibitemOpen
  \bibfield  {author} {\bibinfo {author} {\bibfnamefont {C.~P.~L.}\
  \bibnamefont {Berry}}\ and\ \bibinfo {author} {\bibfnamefont {J.~R.}\
  \bibnamefont {Gair}},\ }\href {https://doi.org/10.1103/PhysRevD.83.104022}
  {\bibfield  {journal} {\bibinfo  {journal} {Phys. Rev. D}\ }\textbf {\bibinfo
  {volume} {83}},\ \bibinfo {pages} {104022} (\bibinfo {year} {2011})},\
  \bibinfo {note} {[Erratum: Phys.Rev.D 85, 089906 (2012)]},\ \Eprint
  {https://arxiv.org/abs/1104.0819} {arXiv:1104.0819 [gr-qc]} \BibitemShut
  {NoStop}%
\bibitem [{\citenamefont {Bhattacharyya}\ and\ \citenamefont
  {Shankaranarayanan}(2017)}]{PhysRevD.96.064044}%
  \BibitemOpen
  \bibfield  {author} {\bibinfo {author} {\bibfnamefont {S.}~\bibnamefont
  {Bhattacharyya}}\ and\ \bibinfo {author} {\bibfnamefont {S.}~\bibnamefont
  {Shankaranarayanan}},\ }\href {https://doi.org/10.1103/PhysRevD.96.064044}
  {\bibfield  {journal} {\bibinfo  {journal} {Phys. Rev. D}\ }\textbf {\bibinfo
  {volume} {96}},\ \bibinfo {pages} {064044} (\bibinfo {year}
  {2017})}\BibitemShut {NoStop}%
\bibitem [{\citenamefont {Zinhailo}(2018)}]{Zinhailo:2018ska}%
  \BibitemOpen
  \bibfield  {author} {\bibinfo {author} {\bibfnamefont {A.~F.}\ \bibnamefont
  {Zinhailo}},\ }\href {https://doi.org/10.1140/epjc/s10052-018-6467-8}
  {\bibfield  {journal} {\bibinfo  {journal} {Eur. Phys. J.}\ }\textbf
  {\bibinfo {volume} {C78}},\ \bibinfo {pages} {992} (\bibinfo {year}
  {2018})},\ \bibinfo {note} {[Eur. Phys. J.78,992(2018)]},\ \Eprint
  {https://arxiv.org/abs/1809.03913} {arXiv:1809.03913 [gr-qc]} \BibitemShut
  {NoStop}%
\bibitem [{\citenamefont {H\"olscher}(2019)}]{Holscher:2018jhm}%
  \BibitemOpen
  \bibfield  {author} {\bibinfo {author} {\bibfnamefont {P.}~\bibnamefont
  {H\"olscher}},\ }\href {https://doi.org/10.1103/PhysRevD.99.064039}
  {\bibfield  {journal} {\bibinfo  {journal} {Phys. Rev. D}\ }\textbf {\bibinfo
  {volume} {99}},\ \bibinfo {pages} {064039} (\bibinfo {year} {2019})},\
  \Eprint {https://arxiv.org/abs/1806.09336} {arXiv:1806.09336 [gr-qc]}
  \BibitemShut {NoStop}%
\bibitem [{\citenamefont {Datta}\ and\ \citenamefont
  {Bose}(2019)}]{Datta:2019npq}%
  \BibitemOpen
  \bibfield  {author} {\bibinfo {author} {\bibfnamefont {S.}~\bibnamefont
  {Datta}}\ and\ \bibinfo {author} {\bibfnamefont {S.}~\bibnamefont {Bose}},\
  }\href@noop {} {\  (\bibinfo {year} {2019})},\ \Eprint
  {https://arxiv.org/abs/1904.01519} {arXiv:1904.01519 [gr-qc]} \BibitemShut
  {NoStop}%
\bibitem [{\citenamefont {Kim}\ \emph {et~al.}(2021)\citenamefont {Kim},
  \citenamefont {Kobakhidze},\ and\ \citenamefont {Picker}}]{Kim:2019sqk}%
  \BibitemOpen
  \bibfield  {author} {\bibinfo {author} {\bibfnamefont {Y.}~\bibnamefont
  {Kim}}, \bibinfo {author} {\bibfnamefont {A.}~\bibnamefont {Kobakhidze}},\
  and\ \bibinfo {author} {\bibfnamefont {Z.~S.~C.}\ \bibnamefont {Picker}},\
  }\href {https://doi.org/10.1140/epjc/s10052-021-09138-0} {\bibfield
  {journal} {\bibinfo  {journal} {Eur. Phys. J. C}\ }\textbf {\bibinfo {volume}
  {81}},\ \bibinfo {pages} {362} (\bibinfo {year} {2021})},\ \Eprint
  {https://arxiv.org/abs/1906.12034} {arXiv:1906.12034 [gr-qc]} \BibitemShut
  {NoStop}%
\bibitem [{\citenamefont {Yamada}\ \emph {et~al.}(2019)\citenamefont {Yamada},
  \citenamefont {Narikawa},\ and\ \citenamefont {Tanaka}}]{Yamada:2019zrb}%
  \BibitemOpen
  \bibfield  {author} {\bibinfo {author} {\bibfnamefont {K.}~\bibnamefont
  {Yamada}}, \bibinfo {author} {\bibfnamefont {T.}~\bibnamefont {Narikawa}},\
  and\ \bibinfo {author} {\bibfnamefont {T.}~\bibnamefont {Tanaka}},\ }\href
  {https://doi.org/10.1093/ptep/ptz103} {\bibfield  {journal} {\bibinfo
  {journal} {PTEP}\ }\textbf {\bibinfo {volume} {2019}},\ \bibinfo {pages}
  {103E01} (\bibinfo {year} {2019})},\ \Eprint
  {https://arxiv.org/abs/1905.11859} {arXiv:1905.11859 [gr-qc]} \BibitemShut
  {NoStop}%
\bibitem [{\citenamefont {Gogoi}\ and\ \citenamefont
  {Dev~Goswami}(2020)}]{Gogoi:2020ypn}%
  \BibitemOpen
  \bibfield  {author} {\bibinfo {author} {\bibfnamefont {D.~J.}\ \bibnamefont
  {Gogoi}}\ and\ \bibinfo {author} {\bibfnamefont {U.}~\bibnamefont
  {Dev~Goswami}},\ }\href {https://doi.org/10.1140/epjc/s10052-020-08684-3}
  {\bibfield  {journal} {\bibinfo  {journal} {Eur. Phys. J. C}\ }\textbf
  {\bibinfo {volume} {80}},\ \bibinfo {pages} {1101} (\bibinfo {year}
  {2020})},\ \Eprint {https://arxiv.org/abs/2006.04011} {arXiv:2006.04011
  [gr-qc]} \BibitemShut {NoStop}%
\bibitem [{\citenamefont {Faria}(2020)}]{Faria:2020kbv}%
  \BibitemOpen
  \bibfield  {author} {\bibinfo {author} {\bibfnamefont {F.~F.}\ \bibnamefont
  {Faria}},\ }\href {https://doi.org/10.1140/epjc/s10052-020-8224-z} {\bibfield
   {journal} {\bibinfo  {journal} {Eur. Phys. J. C}\ }\textbf {\bibinfo
  {volume} {80}},\ \bibinfo {pages} {645} (\bibinfo {year} {2020})},\ \Eprint
  {https://arxiv.org/abs/2007.03637} {arXiv:2007.03637 [gr-qc]} \BibitemShut
  {NoStop}%
\bibitem [{\citenamefont {Ezquiaga}\ \emph {et~al.}(2021)\citenamefont
  {Ezquiaga}, \citenamefont {Hu}, \citenamefont {Lagos},\ and\ \citenamefont
  {Lin}}]{Ezquiaga:2021ler}%
  \BibitemOpen
  \bibfield  {author} {\bibinfo {author} {\bibfnamefont {J.~M.}\ \bibnamefont
  {Ezquiaga}}, \bibinfo {author} {\bibfnamefont {W.}~\bibnamefont {Hu}},
  \bibinfo {author} {\bibfnamefont {M.}~\bibnamefont {Lagos}},\ and\ \bibinfo
  {author} {\bibfnamefont {M.-X.}\ \bibnamefont {Lin}},\ }\href
  {https://doi.org/10.1088/1475-7516/2021/11/048} {\bibfield  {journal}
  {\bibinfo  {journal} {JCAP}\ }\textbf {\bibinfo {volume} {11}}\bibfield
  {number} {\bibinfo  {number} { (11)},\ \bibinfo {pages} {048}},\ }\Eprint
  {https://arxiv.org/abs/2108.10872} {arXiv:2108.10872 [astro-ph.CO]}
  \BibitemShut {NoStop}%
\bibitem [{\citenamefont {Vilhena}\ \emph {et~al.}(2021)\citenamefont
  {Vilhena}, \citenamefont {Medeiros},\ and\ \citenamefont
  {Cuzinatto}}]{PhysRevD.104.084061}%
  \BibitemOpen
  \bibfield  {author} {\bibinfo {author} {\bibfnamefont {S.~G.}\ \bibnamefont
  {Vilhena}}, \bibinfo {author} {\bibfnamefont {L.~G.}\ \bibnamefont
  {Medeiros}},\ and\ \bibinfo {author} {\bibfnamefont {R.~R.}\ \bibnamefont
  {Cuzinatto}},\ }\href {https://doi.org/10.1103/PhysRevD.104.084061}
  {\bibfield  {journal} {\bibinfo  {journal} {Phys. Rev. D}\ }\textbf {\bibinfo
  {volume} {104}},\ \bibinfo {pages} {084061} (\bibinfo {year}
  {2021})}\BibitemShut {NoStop}%
\bibitem [{\citenamefont {Abbott}\ \emph
  {et~al.}(2016{\natexlab{a}})\citenamefont {Abbott} \emph
  {et~al.}}]{Abbott:2016blz}%
  \BibitemOpen
  \bibfield  {author} {\bibinfo {author} {\bibfnamefont {B.~P.}\ \bibnamefont
  {Abbott}} \emph {et~al.} (\bibinfo {collaboration} {LIGO Scientific,
  Virgo}),\ }\href {https://doi.org/10.1103/PhysRevLett.116.061102} {\bibfield
  {journal} {\bibinfo  {journal} {Phys. Rev. Lett.}\ }\textbf {\bibinfo
  {volume} {116}},\ \bibinfo {pages} {061102} (\bibinfo {year}
  {2016}{\natexlab{a}})},\ \Eprint {https://arxiv.org/abs/1602.03837}
  {arXiv:1602.03837 [gr-qc]} \BibitemShut {NoStop}%
\bibitem [{\citenamefont {Abbott}\ \emph
  {et~al.}(2016{\natexlab{b}})\citenamefont {Abbott} \emph
  {et~al.}}]{TheLIGOScientific:2016src}%
  \BibitemOpen
  \bibfield  {author} {\bibinfo {author} {\bibfnamefont {B.~P.}\ \bibnamefont
  {Abbott}} \emph {et~al.} (\bibinfo {collaboration} {LIGO Scientific,
  Virgo}),\ }\href {https://doi.org/10.1103/PhysRevLett.116.221101,
  10.1103/PhysRevLett.121.129902} {\bibfield  {journal} {\bibinfo  {journal}
  {Phys. Rev. Lett.}\ }\textbf {\bibinfo {volume} {116}},\ \bibinfo {pages}
  {221101} (\bibinfo {year} {2016}{\natexlab{b}})},\ \bibinfo {note} {[Erratum:
  Phys. Rev. Lett.121,no.12,129902(2018)]},\ \Eprint
  {https://arxiv.org/abs/1602.03841} {arXiv:1602.03841 [gr-qc]} \BibitemShut
  {NoStop}%
\bibitem [{\citenamefont {Abbott}\ \emph {et~al.}(2017)\citenamefont {Abbott}
  \emph {et~al.}}]{PhysRevLett.119.161101}%
  \BibitemOpen
  \bibfield  {author} {\bibinfo {author} {\bibfnamefont {B.~P.}\ \bibnamefont
  {Abbott}} \emph {et~al.} (\bibinfo {collaboration} {LIGO Scientific
  Collaboration and Virgo Collaboration}),\ }\href
  {https://doi.org/10.1103/PhysRevLett.119.161101} {\bibfield  {journal}
  {\bibinfo  {journal} {Phys. Rev. Lett.}\ }\textbf {\bibinfo {volume} {119}},\
  \bibinfo {pages} {161101} (\bibinfo {year} {2017})}\BibitemShut {NoStop}%
\bibitem [{\citenamefont
  {{Starobinsky}}(1980)}]{doi.org/10.1016/0370-2693(80)90670-X}%
  \BibitemOpen
  \bibfield  {author} {\bibinfo {author} {\bibfnamefont {A.}~\bibnamefont
  {{Starobinsky}}},\ }\href
  {https://doi.org/https://doi.org/10.1016/0370-2693(80)90670-X} {\bibfield
  {journal} {\bibinfo  {journal} {Physics Letters B}\ }\textbf {\bibinfo
  {volume} {91}},\ \bibinfo {pages} {99 } (\bibinfo {year} {1980})}\BibitemShut
  {NoStop}%
\bibitem [{\citenamefont {Starobinsky}(1983)}]{Starobinsky:1983zz}%
  \BibitemOpen
  \bibfield  {author} {\bibinfo {author} {\bibfnamefont {A.~A.}\ \bibnamefont
  {Starobinsky}},\ }\href@noop {} {\bibfield  {journal} {\bibinfo  {journal}
  {Sov. Astron. Lett.}\ }\textbf {\bibinfo {volume} {9}},\ \bibinfo {pages}
  {302} (\bibinfo {year} {1983})}\BibitemShut {NoStop}%
\bibitem [{\citenamefont {Akrami}\ \emph {et~al.}(2018)\citenamefont {Akrami}
  \emph {et~al.}}]{Akrami:2018odb}%
  \BibitemOpen
  \bibfield  {author} {\bibinfo {author} {\bibfnamefont {Y.}~\bibnamefont
  {Akrami}} \emph {et~al.} (\bibinfo {collaboration} {Planck}),\ }\href@noop {}
  {\  (\bibinfo {year} {2018})},\ \Eprint {https://arxiv.org/abs/1807.06211}
  {arXiv:1807.06211 [astro-ph.CO]} \BibitemShut {NoStop}%
\bibitem [{\citenamefont {Ade}\ \emph {et~al.}(2021)\citenamefont {Ade} \emph
  {et~al.}}]{BICEPKeck:2021gln}%
  \BibitemOpen
  \bibfield  {author} {\bibinfo {author} {\bibfnamefont {P.~A.~R.}\
  \bibnamefont {Ade}} \emph {et~al.} (\bibinfo {collaboration} {BICEP/Keck}),\
  }\href {https://doi.org/10.1103/PhysRevLett.127.151301} {\bibfield  {journal}
  {\bibinfo  {journal} {Phys. Rev. Lett.}\ }\textbf {\bibinfo {volume} {127}},\
  \bibinfo {pages} {151301} (\bibinfo {year} {2021})},\ \Eprint
  {https://arxiv.org/abs/2110.00483} {arXiv:2110.00483 [astro-ph.CO]}
  \BibitemShut {NoStop}%
\bibitem [{\citenamefont {Martin}\ \emph {et~al.}(2014)\citenamefont {Martin},
  \citenamefont {Ringeval},\ and\ \citenamefont {Vennin}}]{MARTIN201475}%
  \BibitemOpen
  \bibfield  {author} {\bibinfo {author} {\bibfnamefont {J.}~\bibnamefont
  {Martin}}, \bibinfo {author} {\bibfnamefont {C.}~\bibnamefont {Ringeval}},\
  and\ \bibinfo {author} {\bibfnamefont {V.}~\bibnamefont {Vennin}},\ }\href
  {https://doi.org/https://doi.org/10.1016/j.dark.2014.01.003} {\bibfield
  {journal} {\bibinfo  {journal} {Physics of the Dark Universe}\ }\textbf
  {\bibinfo {volume} {5-6}},\ \bibinfo {pages} {75 } (\bibinfo {year}
  {2014})},\ \bibinfo {note} {hunt for Dark Matter}\BibitemShut {NoStop}%
\bibitem [{\citenamefont {Huang}(2014)}]{Huang:2013hsb}%
  \BibitemOpen
  \bibfield  {author} {\bibinfo {author} {\bibfnamefont {Q.-G.}\ \bibnamefont
  {Huang}},\ }\href {https://doi.org/10.1088/1475-7516/2014/02/035} {\bibfield
  {journal} {\bibinfo  {journal} {JCAP}\ }\textbf {\bibinfo {volume} {02}},\
  \bibinfo {pages} {035}},\ \Eprint {https://arxiv.org/abs/1309.3514}
  {arXiv:1309.3514 [hep-th]} \BibitemShut {NoStop}%
\bibitem [{\citenamefont {Cheong}\ \emph {et~al.}(2020)\citenamefont {Cheong},
  \citenamefont {Lee},\ and\ \citenamefont {Park}}]{Cheong:2020rao}%
  \BibitemOpen
  \bibfield  {author} {\bibinfo {author} {\bibfnamefont {D.~Y.}\ \bibnamefont
  {Cheong}}, \bibinfo {author} {\bibfnamefont {H.~M.}\ \bibnamefont {Lee}},\
  and\ \bibinfo {author} {\bibfnamefont {S.~C.}\ \bibnamefont {Park}},\ }\href
  {https://doi.org/10.1016/j.physletb.2020.135453} {\bibfield  {journal}
  {\bibinfo  {journal} {Phys. Lett. B}\ }\textbf {\bibinfo {volume} {805}},\
  \bibinfo {pages} {135453} (\bibinfo {year} {2020})},\ \Eprint
  {https://arxiv.org/abs/2002.07981} {arXiv:2002.07981 [hep-ph]} \BibitemShut
  {NoStop}%
\bibitem [{\citenamefont {Rodrigues-da Silva}\ \emph
  {et~al.}(2022)\citenamefont {Rodrigues-da Silva}, \citenamefont
  {Bezerra-Sobrinho},\ and\ \citenamefont {Medeiros}}]{PhysRevD.105.063504}%
  \BibitemOpen
  \bibfield  {author} {\bibinfo {author} {\bibfnamefont {G.}~\bibnamefont
  {Rodrigues-da Silva}}, \bibinfo {author} {\bibfnamefont {J.}~\bibnamefont
  {Bezerra-Sobrinho}},\ and\ \bibinfo {author} {\bibfnamefont {L.~G.}\
  \bibnamefont {Medeiros}},\ }\href
  {https://doi.org/10.1103/PhysRevD.105.063504} {\bibfield  {journal} {\bibinfo
   {journal} {Phys. Rev. D}\ }\textbf {\bibinfo {volume} {105}},\ \bibinfo
  {pages} {063504} (\bibinfo {year} {2022})}\BibitemShut {NoStop}%
\bibitem [{\citenamefont {Sebastiani}\ \emph {et~al.}(2014)\citenamefont
  {Sebastiani}, \citenamefont {Cognola}, \citenamefont {Myrzakulov},
  \citenamefont {Odintsov},\ and\ \citenamefont
  {Zerbini}}]{Sebastiani:2013eqa}%
  \BibitemOpen
  \bibfield  {author} {\bibinfo {author} {\bibfnamefont {L.}~\bibnamefont
  {Sebastiani}}, \bibinfo {author} {\bibfnamefont {G.}~\bibnamefont {Cognola}},
  \bibinfo {author} {\bibfnamefont {R.}~\bibnamefont {Myrzakulov}}, \bibinfo
  {author} {\bibfnamefont {S.~D.}\ \bibnamefont {Odintsov}},\ and\ \bibinfo
  {author} {\bibfnamefont {S.}~\bibnamefont {Zerbini}},\ }\href
  {https://doi.org/10.1103/PhysRevD.89.023518} {\bibfield  {journal} {\bibinfo
  {journal} {Phys. Rev. D}\ }\textbf {\bibinfo {volume} {89}},\ \bibinfo
  {pages} {023518} (\bibinfo {year} {2014})},\ \Eprint
  {https://arxiv.org/abs/1311.0744} {arXiv:1311.0744 [gr-qc]} \BibitemShut
  {NoStop}%
\bibitem [{\citenamefont {Odintsov}\ and\ \citenamefont
  {Oikonomou}(2018)}]{Odintsov:2017fnc}%
  \BibitemOpen
  \bibfield  {author} {\bibinfo {author} {\bibfnamefont {S.~D.}\ \bibnamefont
  {Odintsov}}\ and\ \bibinfo {author} {\bibfnamefont {V.~K.}\ \bibnamefont
  {Oikonomou}},\ }\href {https://doi.org/10.1016/j.aop.2017.11.026} {\bibfield
  {journal} {\bibinfo  {journal} {Annals Phys.}\ }\textbf {\bibinfo {volume}
  {388}},\ \bibinfo {pages} {267} (\bibinfo {year} {2018})},\ \Eprint
  {https://arxiv.org/abs/1710.01226} {arXiv:1710.01226 [gr-qc]} \BibitemShut
  {NoStop}%
\bibitem [{\citenamefont {Ivanov}\ \emph {et~al.}(2021)\citenamefont {Ivanov},
  \citenamefont {Ketov}, \citenamefont {Pozdeeva},\ and\ \citenamefont
  {Vernov}}]{Ivanov:2021chn}%
  \BibitemOpen
  \bibfield  {author} {\bibinfo {author} {\bibfnamefont {V.~R.}\ \bibnamefont
  {Ivanov}}, \bibinfo {author} {\bibfnamefont {S.~V.}\ \bibnamefont {Ketov}},
  \bibinfo {author} {\bibfnamefont {E.~O.}\ \bibnamefont {Pozdeeva}},\ and\
  \bibinfo {author} {\bibfnamefont {S.~Y.}\ \bibnamefont {Vernov}},\
  }\href@noop {} {\  (\bibinfo {year} {2021})},\ \Eprint
  {https://arxiv.org/abs/2111.09058} {arXiv:2111.09058 [gr-qc]} \BibitemShut
  {NoStop}%
\bibitem [{\citenamefont {Salvio}(2017)}]{Salvio:2017xul}%
  \BibitemOpen
  \bibfield  {author} {\bibinfo {author} {\bibfnamefont {A.}~\bibnamefont
  {Salvio}},\ }\href {https://doi.org/10.1140/epjc/s10052-017-4825-6}
  {\bibfield  {journal} {\bibinfo  {journal} {Eur. Phys. J. C}\ }\textbf
  {\bibinfo {volume} {77}},\ \bibinfo {pages} {267} (\bibinfo {year} {2017})},\
  \Eprint {https://arxiv.org/abs/1703.08012} {arXiv:1703.08012 [astro-ph.CO]}
  \BibitemShut {NoStop}%
\bibitem [{\citenamefont {Salvio}(2019)}]{Salvio:2019wcp}%
  \BibitemOpen
  \bibfield  {author} {\bibinfo {author} {\bibfnamefont {A.}~\bibnamefont
  {Salvio}},\ }\href {https://doi.org/10.1140/epjc/s10052-019-7267-5}
  {\bibfield  {journal} {\bibinfo  {journal} {Eur. Phys. J. C}\ }\textbf
  {\bibinfo {volume} {79}},\ \bibinfo {pages} {750} (\bibinfo {year} {2019})},\
  \Eprint {https://arxiv.org/abs/1907.00983} {arXiv:1907.00983 [hep-ph]}
  \BibitemShut {NoStop}%
\bibitem [{\citenamefont {Anselmi}\ \emph {et~al.}(2020)\citenamefont
  {Anselmi}, \citenamefont {Bianchi},\ and\ \citenamefont
  {Piva}}]{Anselmi:2020lpp}%
  \BibitemOpen
  \bibfield  {author} {\bibinfo {author} {\bibfnamefont {D.}~\bibnamefont
  {Anselmi}}, \bibinfo {author} {\bibfnamefont {E.}~\bibnamefont {Bianchi}},\
  and\ \bibinfo {author} {\bibfnamefont {M.}~\bibnamefont {Piva}},\ }\href
  {https://doi.org/10.1007/JHEP07(2020)211} {\bibfield  {journal} {\bibinfo
  {journal} {JHEP}\ }\textbf {\bibinfo {volume} {07}},\ \bibinfo {pages}
  {211}},\ \Eprint {https://arxiv.org/abs/2005.10293} {arXiv:2005.10293
  [hep-th]} \BibitemShut {NoStop}%
\bibitem [{\citenamefont {Anselmi}(2021)}]{Anselmi:2021dag}%
  \BibitemOpen
  \bibfield  {author} {\bibinfo {author} {\bibfnamefont {D.}~\bibnamefont
  {Anselmi}},\ }\href {https://doi.org/10.1088/1475-7516/2021/07/037}
  {\bibfield  {journal} {\bibinfo  {journal} {JCAP}\ }\textbf {\bibinfo
  {volume} {07}},\ \bibinfo {pages} {037}},\ \Eprint
  {https://arxiv.org/abs/2105.05864} {arXiv:2105.05864 [hep-th]} \BibitemShut
  {NoStop}%
\bibitem [{\citenamefont {Anselmi}\ \emph {et~al.}(2021)\citenamefont
  {Anselmi}, \citenamefont {Fruzza},\ and\ \citenamefont
  {Piva}}]{Anselmi:2021rye}%
  \BibitemOpen
  \bibfield  {author} {\bibinfo {author} {\bibfnamefont {D.}~\bibnamefont
  {Anselmi}}, \bibinfo {author} {\bibfnamefont {F.}~\bibnamefont {Fruzza}},\
  and\ \bibinfo {author} {\bibfnamefont {M.}~\bibnamefont {Piva}},\ }\href
  {https://doi.org/10.1088/1361-6382/ac2b07} {\bibfield  {journal} {\bibinfo
  {journal} {Class. Quant. Grav.}\ }\textbf {\bibinfo {volume} {38}},\ \bibinfo
  {pages} {225011} (\bibinfo {year} {2021})},\ \Eprint
  {https://arxiv.org/abs/2103.01653} {arXiv:2103.01653 [hep-th]} \BibitemShut
  {NoStop}%
\bibitem [{\citenamefont {Berkin}\ and\ \citenamefont
  {Maeda}(1990)}]{Berkin:1990nu}%
  \BibitemOpen
  \bibfield  {author} {\bibinfo {author} {\bibfnamefont {A.~L.}\ \bibnamefont
  {Berkin}}\ and\ \bibinfo {author} {\bibfnamefont {K.-i.}\ \bibnamefont
  {Maeda}},\ }\href {https://doi.org/10.1016/0370-2693(90)90657-R} {\bibfield
  {journal} {\bibinfo  {journal} {Phys. Lett.}\ }\textbf {\bibinfo {volume}
  {B245}},\ \bibinfo {pages} {348} (\bibinfo {year} {1990})}\BibitemShut
  {NoStop}%
\bibitem [{\citenamefont {Asorey}\ \emph {et~al.}(1997)\citenamefont {Asorey},
  \citenamefont {Lopez},\ and\ \citenamefont {Shapiro}}]{Asorey:1996hz}%
  \BibitemOpen
  \bibfield  {author} {\bibinfo {author} {\bibfnamefont {M.}~\bibnamefont
  {Asorey}}, \bibinfo {author} {\bibfnamefont {J.~L.}\ \bibnamefont {Lopez}},\
  and\ \bibinfo {author} {\bibfnamefont {I.~L.}\ \bibnamefont {Shapiro}},\
  }\href {https://doi.org/10.1142/S0217751X97002991} {\bibfield  {journal}
  {\bibinfo  {journal} {Int. J. Mod. Phys.}\ }\textbf {\bibinfo {volume}
  {A12}},\ \bibinfo {pages} {5711} (\bibinfo {year} {1997})},\ \Eprint
  {https://arxiv.org/abs/hep-th/9610006} {arXiv:hep-th/9610006 [hep-th]}
  \BibitemShut {NoStop}%
\bibitem [{\citenamefont {Iihoshi}(2011)}]{Iihoshi:2010pf}%
  \BibitemOpen
  \bibfield  {author} {\bibinfo {author} {\bibfnamefont {M.}~\bibnamefont
  {Iihoshi}},\ }\href {https://doi.org/10.1088/1475-7516/2011/02/022}
  {\bibfield  {journal} {\bibinfo  {journal} {JCAP}\ }\textbf {\bibinfo
  {volume} {1102}},\ \bibinfo {pages} {022}},\ \Eprint
  {https://arxiv.org/abs/1011.3927} {arXiv:1011.3927 [hep-th]} \BibitemShut
  {NoStop}%
\bibitem [{\citenamefont {Modesto}(2016)}]{Modesto:2016ofr}%
  \BibitemOpen
  \bibfield  {author} {\bibinfo {author} {\bibfnamefont {L.}~\bibnamefont
  {Modesto}},\ }\href {https://doi.org/10.1016/j.nuclphysb.2016.06.004}
  {\bibfield  {journal} {\bibinfo  {journal} {Nucl. Phys.}\ }\textbf {\bibinfo
  {volume} {B909}},\ \bibinfo {pages} {584} (\bibinfo {year} {2016})},\ \Eprint
  {https://arxiv.org/abs/1602.02421} {arXiv:1602.02421 [hep-th]} \BibitemShut
  {NoStop}%
\bibitem [{\citenamefont {Cuzinatto}\ \emph
  {et~al.}(2019{\natexlab{a}})\citenamefont {Cuzinatto}, \citenamefont
  {de~Melo}, \citenamefont {Medeiros},\ and\ \citenamefont
  {Pompeia}}]{Cuzinatto:2018chu}%
  \BibitemOpen
  \bibfield  {author} {\bibinfo {author} {\bibfnamefont {R.~R.}\ \bibnamefont
  {Cuzinatto}}, \bibinfo {author} {\bibfnamefont {C.~A.~M.}\ \bibnamefont
  {de~Melo}}, \bibinfo {author} {\bibfnamefont {L.~G.}\ \bibnamefont
  {Medeiros}},\ and\ \bibinfo {author} {\bibfnamefont {P.~J.}\ \bibnamefont
  {Pompeia}},\ }\href {https://doi.org/10.1103/PhysRevD.99.084053} {\bibfield
  {journal} {\bibinfo  {journal} {Phys. Rev.}\ }\textbf {\bibinfo {volume}
  {D99}},\ \bibinfo {pages} {084053} (\bibinfo {year} {2019}{\natexlab{a}})},\
  \Eprint {https://arxiv.org/abs/1806.08850} {arXiv:1806.08850 [gr-qc]}
  \BibitemShut {NoStop}%
\bibitem [{\citenamefont {Cuzinatto}\ \emph
  {et~al.}(2019{\natexlab{b}})\citenamefont {Cuzinatto}, \citenamefont
  {Medeiros},\ and\ \citenamefont {Pompeia}}]{Cuzinatto:2018vjt}%
  \BibitemOpen
  \bibfield  {author} {\bibinfo {author} {\bibfnamefont {R.~R.}\ \bibnamefont
  {Cuzinatto}}, \bibinfo {author} {\bibfnamefont {L.~G.}\ \bibnamefont
  {Medeiros}},\ and\ \bibinfo {author} {\bibfnamefont {P.~J.}\ \bibnamefont
  {Pompeia}},\ }\href {https://doi.org/10.1088/1475-7516/2019/02/055}
  {\bibfield  {journal} {\bibinfo  {journal} {JCAP}\ }\textbf {\bibinfo
  {volume} {1902}},\ \bibinfo {pages} {055}},\ \Eprint
  {https://arxiv.org/abs/1810.08911} {arXiv:1810.08911 [gr-qc]} \BibitemShut
  {NoStop}%
\bibitem [{\citenamefont {Castellanos}\ \emph {et~al.}(2018)\citenamefont
  {Castellanos}, \citenamefont {Sobreira}, \citenamefont {Shapiro},\ and\
  \citenamefont {Starobinsky}}]{Castellanos:2018dub}%
  \BibitemOpen
  \bibfield  {author} {\bibinfo {author} {\bibfnamefont {A.~R.~R.}\
  \bibnamefont {Castellanos}}, \bibinfo {author} {\bibfnamefont
  {F.}~\bibnamefont {Sobreira}}, \bibinfo {author} {\bibfnamefont {I.~L.}\
  \bibnamefont {Shapiro}},\ and\ \bibinfo {author} {\bibfnamefont {A.~A.}\
  \bibnamefont {Starobinsky}},\ }\href
  {https://doi.org/10.1088/1475-7516/2018/12/007} {\bibfield  {journal}
  {\bibinfo  {journal} {JCAP}\ }\textbf {\bibinfo {volume} {1812}}\bibfield
  {number} {\bibinfo  {number} { (12)},\ \bibinfo {pages} {007}},\ }\Eprint
  {https://arxiv.org/abs/1810.07787} {arXiv:1810.07787 [gr-qc]} \BibitemShut
  {NoStop}%
\bibitem [{\citenamefont {Koshelev}\ \emph {et~al.}(2016)\citenamefont
  {Koshelev}, \citenamefont {Modesto}, \citenamefont {Rachwal},\ and\
  \citenamefont {Starobinsky}}]{Koshelev:2016xqb}%
  \BibitemOpen
  \bibfield  {author} {\bibinfo {author} {\bibfnamefont {A.~S.}\ \bibnamefont
  {Koshelev}}, \bibinfo {author} {\bibfnamefont {L.}~\bibnamefont {Modesto}},
  \bibinfo {author} {\bibfnamefont {L.}~\bibnamefont {Rachwal}},\ and\ \bibinfo
  {author} {\bibfnamefont {A.~A.}\ \bibnamefont {Starobinsky}},\ }\href
  {https://doi.org/10.1007/JHEP11(2016)067} {\bibfield  {journal} {\bibinfo
  {journal} {JHEP}\ }\textbf {\bibinfo {volume} {11}},\ \bibinfo {pages}
  {067}},\ \Eprint {https://arxiv.org/abs/1604.03127} {arXiv:1604.03127
  [hep-th]} \BibitemShut {NoStop}%
\bibitem [{\citenamefont {Edholm}(2017)}]{Edholm:2016seu}%
  \BibitemOpen
  \bibfield  {author} {\bibinfo {author} {\bibfnamefont {J.}~\bibnamefont
  {Edholm}},\ }\href {https://doi.org/10.1103/PhysRevD.95.044004} {\bibfield
  {journal} {\bibinfo  {journal} {Phys. Rev. D}\ }\textbf {\bibinfo {volume}
  {95}},\ \bibinfo {pages} {044004} (\bibinfo {year} {2017})},\ \Eprint
  {https://arxiv.org/abs/1611.05062} {arXiv:1611.05062 [gr-qc]} \BibitemShut
  {NoStop}%
\bibitem [{\citenamefont {Diamandis}\ \emph {et~al.}(2017)\citenamefont
  {Diamandis}, \citenamefont {Georgalas}, \citenamefont {Kaskavelis},
  \citenamefont {Lahanas},\ and\ \citenamefont
  {Pavlopoulos}}]{Diamandis:2017ems}%
  \BibitemOpen
  \bibfield  {author} {\bibinfo {author} {\bibfnamefont {G.~A.}\ \bibnamefont
  {Diamandis}}, \bibinfo {author} {\bibfnamefont {B.~C.}\ \bibnamefont
  {Georgalas}}, \bibinfo {author} {\bibfnamefont {K.}~\bibnamefont
  {Kaskavelis}}, \bibinfo {author} {\bibfnamefont {A.~B.}\ \bibnamefont
  {Lahanas}},\ and\ \bibinfo {author} {\bibfnamefont {G.}~\bibnamefont
  {Pavlopoulos}},\ }\href {https://doi.org/10.1103/PhysRevD.96.044033}
  {\bibfield  {journal} {\bibinfo  {journal} {Phys. Rev. D}\ }\textbf {\bibinfo
  {volume} {96}},\ \bibinfo {pages} {044033} (\bibinfo {year} {2017})},\
  \Eprint {https://arxiv.org/abs/1704.07617} {arXiv:1704.07617 [hep-th]}
  \BibitemShut {NoStop}%
\bibitem [{\citenamefont {Sravan~Kumar}\ and\ \citenamefont
  {Modesto}(2018)}]{SravanKumar:2018dlo}%
  \BibitemOpen
  \bibfield  {author} {\bibinfo {author} {\bibfnamefont {K.}~\bibnamefont
  {Sravan~Kumar}}\ and\ \bibinfo {author} {\bibfnamefont {L.}~\bibnamefont
  {Modesto}},\ }\href@noop {} {\  (\bibinfo {year} {2018})},\ \Eprint
  {https://arxiv.org/abs/1810.02345} {arXiv:1810.02345 [hep-th]} \BibitemShut
  {NoStop}%
\bibitem [{\citenamefont {Bezerra-Sobrinho}\ and\ \citenamefont
  {Medeiros}(2022)}]{Bezerra-Sobrinho:2022dkv}%
  \BibitemOpen
  \bibfield  {author} {\bibinfo {author} {\bibfnamefont {J.}~\bibnamefont
  {Bezerra-Sobrinho}}\ and\ \bibinfo {author} {\bibfnamefont {L.~G.}\
  \bibnamefont {Medeiros}},\ }\href@noop {} {\  (\bibinfo {year} {2022})},\
  \Eprint {https://arxiv.org/abs/2202.13308} {arXiv:2202.13308 [gr-qc]}
  \BibitemShut {NoStop}%
\bibitem [{\citenamefont {Salles}\ and\ \citenamefont
  {Shapiro}(2014)}]{Salles:2014rua}%
  \BibitemOpen
  \bibfield  {author} {\bibinfo {author} {\bibfnamefont {F.~d.~O.}\
  \bibnamefont {Salles}}\ and\ \bibinfo {author} {\bibfnamefont {I.~L.}\
  \bibnamefont {Shapiro}},\ }\href {https://doi.org/10.1103/PhysRevD.89.084054}
  {\bibfield  {journal} {\bibinfo  {journal} {Phys. Rev. D}\ }\textbf {\bibinfo
  {volume} {89}},\ \bibinfo {pages} {084054} (\bibinfo {year} {2014})},\
  \bibinfo {note} {[Erratum: Phys.Rev.D 90, 129903 (2014)]},\ \Eprint
  {https://arxiv.org/abs/1401.4583} {arXiv:1401.4583 [hep-th]} \BibitemShut
  {NoStop}%
\bibitem [{\citenamefont {Peter}\ \emph {et~al.}(2018)\citenamefont {Peter},
  \citenamefont {Salles},\ and\ \citenamefont {Shapiro}}]{Peter:2017xxf}%
  \BibitemOpen
  \bibfield  {author} {\bibinfo {author} {\bibfnamefont {P.}~\bibnamefont
  {Peter}}, \bibinfo {author} {\bibfnamefont {F.~D.~O.}\ \bibnamefont
  {Salles}},\ and\ \bibinfo {author} {\bibfnamefont {I.~L.}\ \bibnamefont
  {Shapiro}},\ }\href {https://doi.org/10.1103/PhysRevD.97.064044} {\bibfield
  {journal} {\bibinfo  {journal} {Phys. Rev. D}\ }\textbf {\bibinfo {volume}
  {97}},\ \bibinfo {pages} {064044} (\bibinfo {year} {2018})},\ \Eprint
  {https://arxiv.org/abs/1801.00063} {arXiv:1801.00063 [gr-qc]} \BibitemShut
  {NoStop}%
\bibitem [{\citenamefont {Mukhanov}\ \emph {et~al.}(1992)\citenamefont
  {Mukhanov}, \citenamefont {Feldman},\ and\ \citenamefont
  {Brandenberger}}]{MUKHANOV1992203}%
  \BibitemOpen
  \bibfield  {author} {\bibinfo {author} {\bibfnamefont {V.}~\bibnamefont
  {Mukhanov}}, \bibinfo {author} {\bibfnamefont {H.}~\bibnamefont {Feldman}},\
  and\ \bibinfo {author} {\bibfnamefont {R.}~\bibnamefont {Brandenberger}},\
  }\href {https://doi.org/https://doi.org/10.1016/0370-1573(92)90044-Z}
  {\bibfield  {journal} {\bibinfo  {journal} {Physics Reports}\ }\textbf
  {\bibinfo {volume} {215}},\ \bibinfo {pages} {203} (\bibinfo {year}
  {1992})}\BibitemShut {NoStop}%
\bibitem [{\citenamefont {Bassett}\ \emph {et~al.}(2006)\citenamefont
  {Bassett}, \citenamefont {Tsujikawa},\ and\ \citenamefont
  {Wands}}]{RevModPhys.78.537}%
  \BibitemOpen
  \bibfield  {author} {\bibinfo {author} {\bibfnamefont {B.~A.}\ \bibnamefont
  {Bassett}}, \bibinfo {author} {\bibfnamefont {S.}~\bibnamefont {Tsujikawa}},\
  and\ \bibinfo {author} {\bibfnamefont {D.}~\bibnamefont {Wands}},\ }\href
  {https://doi.org/10.1103/RevModPhys.78.537} {\bibfield  {journal} {\bibinfo
  {journal} {Rev. Mod. Phys.}\ }\textbf {\bibinfo {volume} {78}},\ \bibinfo
  {pages} {537} (\bibinfo {year} {2006})}\BibitemShut {NoStop}%
\bibitem [{\citenamefont {Baumann}(2018)}]{Baumann:2018muz}%
  \BibitemOpen
  \bibfield  {author} {\bibinfo {author} {\bibfnamefont {D.}~\bibnamefont
  {Baumann}},\ }\href {https://doi.org/10.22323/1.305.0009} {\bibfield
  {journal} {\bibinfo  {journal} {PoS}\ }\textbf {\bibinfo {volume}
  {TASI2017}},\ \bibinfo {pages} {009} (\bibinfo {year} {2018})},\ \Eprint
  {https://arxiv.org/abs/1807.03098} {arXiv:1807.03098 [hep-th]} \BibitemShut
  {NoStop}%
\bibitem [{\citenamefont {Wands}(2008)}]{Wands:2007bd}%
  \BibitemOpen
  \bibfield  {author} {\bibinfo {author} {\bibfnamefont {D.}~\bibnamefont
  {Wands}},\ }\href {https://doi.org/10.1007/978-3-540-74353-8_8} {\bibfield
  {journal} {\bibinfo  {journal} {Lect. Notes Phys.}\ }\textbf {\bibinfo
  {volume} {738}},\ \bibinfo {pages} {275} (\bibinfo {year} {2008})},\ \Eprint
  {https://arxiv.org/abs/astro-ph/0702187} {arXiv:astro-ph/0702187}
  \BibitemShut {NoStop}%
\bibitem [{\citenamefont {Gundhi}\ and\ \citenamefont
  {Steinwachs}(2020)}]{Gundhi:2018wyz}%
  \BibitemOpen
  \bibfield  {author} {\bibinfo {author} {\bibfnamefont {A.}~\bibnamefont
  {Gundhi}}\ and\ \bibinfo {author} {\bibfnamefont {C.~F.}\ \bibnamefont
  {Steinwachs}},\ }\href {https://doi.org/10.1016/j.nuclphysb.2020.114989}
  {\bibfield  {journal} {\bibinfo  {journal} {Nucl. Phys. B}\ }\textbf
  {\bibinfo {volume} {954}},\ \bibinfo {pages} {114989} (\bibinfo {year}
  {2020})},\ \Eprint {https://arxiv.org/abs/1810.10546} {arXiv:1810.10546
  [hep-th]} \BibitemShut {NoStop}%
\bibitem [{\citenamefont {Baumann}(2011)}]{Baumann:2009ds}%
  \BibitemOpen
  \bibfield  {author} {\bibinfo {author} {\bibfnamefont {D.}~\bibnamefont
  {Baumann}},\ }in\ \href {https://doi.org/10.1142/9789814327183_0010} {\emph
  {\bibinfo {booktitle} {{Physics of the large and the small, TASI 09,
  proceedings of the Theoretical Advanced Study Institute in Elementary
  Particle Physics, Boulder, Colorado, USA, 1-26 June 2009}}}}\ (\bibinfo
  {year} {2011})\ pp.\ \bibinfo {pages} {523--686},\ \Eprint
  {https://arxiv.org/abs/0907.5424} {arXiv:0907.5424 [hep-th]} \BibitemShut
  {NoStop}%
\bibitem [{\citenamefont {Piattella}(2018)}]{Piattella:2018hvi}%
  \BibitemOpen
  \bibfield  {author} {\bibinfo {author} {\bibfnamefont {O.~F.}\ \bibnamefont
  {Piattella}},\ }\href {https://doi.org/10.1007/978-3-319-95570-4} {\emph
  {\bibinfo {title} {{Lecture Notes in Cosmology}}}},\ UNITEXT for Physics\
  (\bibinfo  {publisher} {Springer},\ \bibinfo {address} {Cham},\ \bibinfo
  {year} {2018})\ \Eprint {https://arxiv.org/abs/1803.00070} {arXiv:1803.00070
  [astro-ph.CO]} \BibitemShut {NoStop}%
\bibitem [{\citenamefont {Ivanov}\ and\ \citenamefont
  {Tokareva}(2016)}]{Ivanov:2016hcm}%
  \BibitemOpen
  \bibfield  {author} {\bibinfo {author} {\bibfnamefont {M.~M.}\ \bibnamefont
  {Ivanov}}\ and\ \bibinfo {author} {\bibfnamefont {A.~A.}\ \bibnamefont
  {Tokareva}},\ }\href {https://doi.org/10.1088/1475-7516/2016/12/018}
  {\bibfield  {journal} {\bibinfo  {journal} {JCAP}\ }\textbf {\bibinfo
  {volume} {12}},\ \bibinfo {pages} {018}},\ \Eprint
  {https://arxiv.org/abs/1610.05330} {arXiv:1610.05330 [hep-th]} \BibitemShut
  {NoStop}%
\bibitem [{\citenamefont {Sbis\`a}(2015)}]{Sbisa:2014pzo}%
  \BibitemOpen
  \bibfield  {author} {\bibinfo {author} {\bibfnamefont {F.}~\bibnamefont
  {Sbis\`a}},\ }\href {https://doi.org/10.1088/0143-0807/36/1/015009}
  {\bibfield  {journal} {\bibinfo  {journal} {Eur. J. Phys.}\ }\textbf
  {\bibinfo {volume} {36}},\ \bibinfo {pages} {015009} (\bibinfo {year}
  {2015})},\ \Eprint {https://arxiv.org/abs/1406.4550} {arXiv:1406.4550
  [hep-th]} \BibitemShut {NoStop}%
\bibitem [{\citenamefont {Gradshteyn}\ and\ \citenamefont
  {Ryzhik}(2007)}]{gradshteyn2007}%
  \BibitemOpen
  \bibfield  {author} {\bibinfo {author} {\bibfnamefont {I.~S.}\ \bibnamefont
  {Gradshteyn}}\ and\ \bibinfo {author} {\bibfnamefont {I.~M.}\ \bibnamefont
  {Ryzhik}},\ }\href@noop {} {\emph {\bibinfo {title} {Table of integrals,
  series, and products}}},\ \bibinfo {edition} {seventh}\ ed.\ (\bibinfo
  {publisher} {Elsevier/Academic Press, Amsterdam},\ \bibinfo {year} {2007})\
  pp.\ \bibinfo {pages} {xlviii+1171},\ \bibinfo {note} {translated from the
  Russian, Translation edited and with a preface by Alan Jeffrey and Daniel
  Zwillinger, With one CD-ROM (Windows, Macintosh and UNIX)}\BibitemShut
  {NoStop}%
\end{thebibliography}%
\end{document}